\documentclass[iop,apj,twocolappendix]{emulateapj}
\usepackage{amsmath,amssymb,amstext}

\usepackage[breaklinks,colorlinks,citecolor=blue,linkcolor=magenta]{hyperref} 

\usepackage{multirow}
\usepackage{booktabs}
\usepackage{enumitem}
\usepackage{rotating}

\usepackage{natbib}
\newcommand\citeapos[2]{\citeauthor{#1}'s (\citeyear{#2})}
\usepackage[percent]{overpic}
\usepackage{xcolor}

\shorttitle{SFHs of the LEGUS dwarf
  galaxies} \shortauthors{Cignoni et al.}

\begin{document}


\title{Star formation histories of the LEGUS dwarf galaxies (III):\\ the non-bursty nature of 23 star forming dwarf galaxies}%

\author{M. Cignoni\altaffilmark{2,3,4}, E. Sacchi\altaffilmark{5},
  M. Tosi\altaffilmark{4}, A. Aloisi\altaffilmark{5},
  D. O. Cook\altaffilmark{6,7}, D. Calzetti\altaffilmark{8},
  J. C. Lee\altaffilmark{7}, E. Sabbi\altaffilmark{5},
  D. A. Thilker\altaffilmark{14} ,A. Adamo\altaffilmark{9} ,
  D. A. Dale\altaffilmark{10}, B. G. Elmegreen\altaffilmark{11},
  J.S. Gallagher III\altaffilmark{15}, E. K. Grebel\altaffilmark{12},
  K. E. Johnson\altaffilmark{16}, M. Messa\altaffilmark{9}, L. J. Smith\altaffilmark{13} and L. Ubeda\altaffilmark{5}}


\altaffiltext{1}{Based on observations with the NASA/ESA Hubble Space Telescope, obtained at the Space Telescope Science Institute, which is operated by AURA Inc., under NASA contract NAS 5-26555}
\altaffiltext{2}{Department of Physics - University of Pisa, Largo Pontecorvo, 3 Pisa, 56127, Italy }
\altaffiltext{3}{INFN, Largo B. Pontecorvo 3, 56127, Pisa, Italy}
\altaffiltext{4}{INAF-Osservatorio di Astrofisica e Scienza dello
  Spazio, Via Gobetti 93/3, 40129, Bologna, Italy}
\altaffiltext{5}{Space Telescope Science Institute, 3700 San Martin Drive, Baltimore, MD, 21218, USA }
\altaffiltext{6}{Department of Physics \& Astronomy, California Institute of Technology,
Pasadena, CA 91101, USA}
\altaffiltext{7}{IPAC/Caltech, Pasadena, CA 91101, USA}
\altaffiltext{8}{Department of Astronomy, University of Massachusetts
  -- Amherst, Amherst, MA 01003, USA}
\altaffiltext{9}{Department of Astronomy, The Oskar Klein Centre, Stockholm University,
Stockholm, Sweden}
\altaffiltext{10}{Department of Physics and Astronomy, University of
  Wyoming, Laramie, WY}
\altaffiltext{11}{IBM Research Division, T.J. Watson Research Center, Yorktown Hts., NY}
\altaffiltext{12}{Astronomisches Rechen-Institut, Zentrum f\"ur
  Astronomie der Universit\"at Heidelberg, M\"onchhofstr.\ 12-–14,
  69120 Heidelberg, Germany}
\altaffiltext{13}{European Space Agency/Space Telescope Science Institute, Baltimore, MD}
\altaffiltext{14}{Department of Physics and Astronomy, The Johns Hopkins University, Baltimore, MD}
\altaffiltext{15}{Dept. of Astronomy, University of Wisconsin--Madison, Madison,
WI}
\altaffiltext{16}{Department of Astronomy, University of Virginia, Charlottesville, VA}

\begin{abstract}
 
  We derive the recent star formation histories of 23 active dwarf
  galaxies using HST observations from the Legacy Extragalactic UV
  Survey (LEGUS). We apply a color-magnitude diagram fitting technique
  using two independent sets of stellar models, PARSEC-COLIBRI and
  MIST. Despite the non-negligible recent activity, none of the 23
  star forming dwarfs show enhancements in the last 100 Myr larger
  than three times the 100-Myr-average. The unweighted mean of the
  individual SFHs in the last 100 Myr is also consistent with a rather
  constant activity, irrespective of the atomic gas fraction. We
  confirm previous results that for dwarf galaxies the CMD-based
  average star formation rates (SFRs) are generally higher than the
  FUV-based SFR. For half of the sample, the 60-Myr-average CMD-based
  SFR is more than two times the FUV SFR. In contrast, we find
  remarkable agreement between the 10-Myr-average CMD-based SFR and
  the H${\alpha}$-based SFR. Finally, using core helium burning stars
  of intermediate mass we study the pattern of star formation spatial
  progression over the past 60 Myr, and speculate on the possible
  triggers and connections of the star formation activity with the
  environment in which these galaxies live.  Approximately half of our
  galaxies show spatial progression of star formation in the last 60
  Myr, and/or very recent diffuse and off-center activity compared to
  RGB stars.

\end{abstract}



\section{Introduction}

What is true for massive large galaxies may not be true for dwarf
galaxies.  Typically, large galaxies have formed a major fraction of
their stars in the first $\sim 3$ Gyr, whereas dwarf galaxies exhibit
a variety of star formation histories (SFHs, see, e.g.,
\citealt{THT09,mq11,weisz11,gallart15}) and specific star formation
rates, ranging from totally inactive, as in present-day dwarf
spheroidal galaxies, to extremely active, as in blue compact dwarfs
(BCDs). This heterogeneity, probably a consequence of the lack of
global mechanisms like spiral arms or other organized gas motions,
leads to the question of how the star formation in one part of a dwarf
inhibits or enhances the formation of stars in another part. A popular
model, the Stochastic Self-Propagating Star Formation (SSPSF,
\citealt{gerola80}), predicts that a burst of star formation in a
region (``cell'') of the galaxy could trigger secondary SF bursts in
other adjacent cells. The newborn massive young stars can disturb the
gas in an adjacent region with stellar winds, ionization and other
energetic activities. The gas then collapses and begins its own
starburst. The process continues until a stage is reached where no
residual gas is in the condition to be affected by the young stars.

Despite the success of the model in producing a coherent physical
scenario and matching several observed properties of star forming
dwarfs (e.g. the apparent random distribution of star forming sites),
the exact understanding of what sparked the initial starburst activity
remains elusive. How starbursts are activated is an important question
for large galaxies as well as dwarfs, and there are probably different
answers for different galaxies. Potential mechanisms for promoting
star formation (SF) in dwarfs can be broadly categorized as either
internally-driven or externally-driven events. BCDs show star forming
regions kinematically decoupled from the rest of the galaxy
(\citealt{koleva14}), suggesting that giant star forming clumps in
dwarf irregular galaxies could spiral-in towards the center (see
\citealt{elme12}), feeding intense star formation. This is because in
such small galaxies, star forming clumps could be sufficiently massive
to exceed a few percent of the galaxy mass enclosed inside their
orbital radii, therefore producing dynamically significant torques on
dark matter halo particles, halo stars, and the surrounding disk to
lose their angular momentum with a timescale of 1 Gyr.  Moreover,
massive triaxial dark haloes or bars made of dark matter
(\citealt{he04}) could promote the migration of a significant fraction
of gas from the periphery to the center of the dwarf (\citealt{bf02}),
eventually igniting the starburst.

External processes must also be important (see
e.g. \citealt{lah19}). Low-mass galaxies are more fragile with respect
to perturbations of various origins (interactions, inflows, mergers),
hence their evolution is most sensitive to various kinds of galaxy
encounters (e.g., distant tidals, close pass-byes, major and minor
mergers). Indeed, the best-studied starburst dwarfs (e.g. IC10,
NGC5253, NGC1569, II Zw 40, and others) seem to be interacting with,
or absorbing, other structures (like gas clouds). For example, there
is evidence for active accretion onto IC10 in the form of two
in-moving streams of neutral gas with velocity gradients of several
km/s/kpc (\citealt{ashley15}). These gradients correspond to accretion
rates of the same order of the star formation rate, suggesting that
the impact of this gas on the disk of this dwarf irregular is what is
triggering the current starburst. NGC5253 is another interesting
object: its highly efficient star formation might be caused by a
streamer of gas (containing $\sim 2 \times 10^6$ M$_{\odot}$ extending
$\sim 200$-300 pc along the minor axis, entering the galaxy at a rate
of $\sim 20$ pc/Myr) force-fed into the star-forming region by the
galactic potential (\citealt{turner15}). The prototypical starburst
dwarf NGC1569 shows copious gas infall (\citealt{stil02},
\citealt{mule05}) and, along the minor axis, hot gas outflow. The
dwarf II Zw 40 appears to be the merger remnant of two colliding
smaller gas rich dwarf galaxies. In general, dwarfs accreting dwarfs
have been observed (see
e.g. \citealt{tully06,md12,rich12,annibali16,sacchi16}) and dwarf
groups have been recently observed by \cite{stier17} using the
panchromatic TiNy Titans (TNT; \citealt{stier15}) survey, a systematic
study of SF in interacting dwarf galaxies. According to this study,
the interaction between dwarf galaxies could be quantitatively
different as compared to more massive counterparts. In fact, both
paired dwarfs and paired massive galaxies show enhanced SF out to
separations of $\sim 100$ kpc (e.g. \citealt{patton13,stier15}), but
the effect in dwarfs is stronger by a factor of 1.3 and involves a
larger fraction of the virial radius.

In this context, the goal of this paper is to answer the following
questions: 1) Is the starburst process characterised by long lived
($>>$ 10 Myr) bursts or short lived (a few Myr) bursts? 2) How does the SF
spatially progress in these galaxies? 3) Is the SF activity causally
connected to the environment? To quantify the strength of the
starburst we choose a birthrate parameter b $= \mathrm{SFR}/<\mathrm{SFR}>$, where
$<\mathrm{SFR}>$ is the average star formation rate (SFR) over the longest
look-back time that can be investigated in this paper (generally
60$-$100 Myr). We call a galaxy with b $> 3$ a starburst
galaxy.

The best place to look for answers is the sample of nearby star
forming galaxies. The high spatial resolution and the high signal to
noise multi-wavelength information which can be achieved for these
systems is crucial to resolve and measure individual stars, and to
draw their color-magnitude diagrams (CMDs). The CMD of a stellar
system is in fact a fundamental tool to investigate its recent and
past SF activity.  In the distance range 4-12 Mpc, the exquisite
spatial resolution of HST is necessary for a conclusive census, even
for the most luminous stars, because they are usually formed in the
dense cores of star-forming clouds, and therefore are found in compact
groups. Another key advantage of HST with respect to ground-based data
is its access to UV wavelengths.  From a more general perspective,
understanding the formation, evolution, and mass distribution of
massive stars is critical for interpreting the integrated light of
distant galaxies, for correctly modeling the chemical evolution of
galaxies, and ultimately to unravel the history of star formation in
the Universe, in terms of stellar mass formed and energy balance. A by
product of our analysis is a robust test for stellar models of
intermediate and massive stars.

While all studies of SFHs performed so far (e.g. \citealt{THT09} and
references therein, \citealt{weisz11}) aimed at covering the longest
possible look-back time and infer the SFH back to the earliest epochs,
in this series of papers (\citealt{cigno18}, hereafter Paper I) we
take a different approach. We use a combination of UV and optical
photometry provided by the Legacy ExtraGalactic UV Survey (LEGUS;
\citealt{calzetti15}) to build and study the UV color-magnitude
diagrams for a sample of 23 star forming dwarf galaxies (hereafter
SFDs; see Table \ref{tab1}) at distances of 4-12 Mpc, and infer, at
the highest temporal resolution possible (a few Myr in the last 20
Myr), the SFHs over the last 100 Myr. All these CMDs allow us to
isolate samples of He-burning stars of intermediate mass, which are
recognized as good chronometers (this age indicator was pioneered by
\citealt{dp} on the dwarf irregular galaxy Sextans A and recently
applied to several other SFDs by, e.g., \citealt{mq11,mq12}). Tracing
these stars allows us to infer how the SF percolates, helping to shed
light on the cause of such SF enhancements.

\setlength\tabcolsep{1.1pt} 
\begin{deluxetable*}{p{1.9cm}p{1.5cm}p{0.8cm}p{0.5cm}p{0.5cm}p{0.5cm}p{0.5cm}p{0.5cm}p{0.5cm}p{1cm}p{1cm}p{1cm}}
\tablecolumns{11}
\tabletypesize{\tiny}
\tablecaption{Properties of the LEGUS dwarf galaxies sample.\label{tab1}}
\tablewidth{0pt}
\tablehead{\colhead{Name \tablenotemark{a}} & 
  \colhead{Morph. \tablenotemark{a}}  &
  \colhead{Dist.\tablenotemark{b}} & \colhead{SFR$_{60}$
    \tablenotemark{c}} & \colhead{SFR$_{10}$ \tablenotemark{d}} &
  \colhead{SFR(FUV)  \tablenotemark{e}} &
  \colhead{SFR(H$_{\alpha}$)  \tablenotemark{f}} & \colhead{E(B$-$V)  \tablenotemark{g}} & \colhead{E(B$-$V)$_{R}$ \tablenotemark{h}} & \colhead{M$_*$\tablenotemark{i}}  & \colhead{M(HI) \tablenotemark{i}} & \colhead{Class  \tablenotemark{l}} 
\\\\
\colhead{} & \colhead{} & \colhead{Mpc} & \colhead{M$_{\odot}$
  yr$^{-1}$kpc$^{-2}$} & \colhead{M$_{\odot}$ yr$^{-1}$kpc$^{-2}$} &
\colhead{M$_{\odot}$ yr$^{-1}$kpc$^{-2}$} & \colhead{M$_{\odot}$
  yr$^{-1}$kpc$^{-2}$} & \colhead{mag}  & \colhead{mag} &
\colhead{M$_{\odot}$} & \colhead{M$_{\odot}$}&\colhead{} \\
\colhead{} & \colhead{} & \colhead{} & \colhead{$\times 10^{-3}$} & \colhead{$\times 10^{-3}$} & \colhead{$\times 10^{-3}$} & \colhead{$\times 10^{-3}$} & \colhead{}  & \colhead{} & \colhead{} & \colhead{}&\colhead{} \\
\colhead{(1)} & \colhead{(2)} & \colhead{(3)} & \colhead{(4)} & \colhead{(5)} & 
\colhead{(6)} & \colhead{(7)} & \colhead{(8)} & \colhead{(9)} & \colhead{(10)} &
\colhead{(11)} & \colhead{(12)}   
\\
}
\startdata
\hline
\\
NGC5253    & Im         &3.32              &51.0&26.1&40.8&52.3&0.100        &0.30       &  2.2E08 & 1.0E08 &SSCt\\
UGC5139    & IABm       &3.83              &1.09& 0.48& 0.66& 0.59&0.050        &0.05       &  2.5E07 &  2.1E08 &OOD\\ 
UGC4459    & Im         &3.96              & 0.48& 0.72& 0.45& 0.71&0.050        &0.05       &  6.8E06 & 6.8E07 &SSC\\
NGC4449    & IBm        &4.01              &42.1&37.3&33.1&40.0&0.060        &0.35       &  1.1E09 & 2.1E09 &OOC\\
UGC0685    & SAm        &4.37              &0.60& 0.19& 0.18& 0.33&0.100        &0.05       &  9.5E07  & 9.7E07 &OSC\\
NGC5238    & SABdm      &4.43              &0.73& 0.53& 0.46& 0.65&0.050        &0.05       &  1.4E08 & 2.9E07  &SSCt\\
NGC3738    & Im         &5.09              &8.83& 5.31& 2.19& 2.11&0.025       &0.25       &   2.4E08 & 1.5E08 &OSC\\
IC4247     & S?         &5.11              &0.46& 0.21& 0.25& 0.08&0.050        &0.05       &  1.2E08 & 4.0E07   &SSC   \\
UGCA281    & Sm         &5.19              &0.40& 0.54& 0.74& 1.55&0.050        &0.10       &  1.9E07 & 8.3E07 &SSC\\
NGC1705    & SA0/BCG    &5.22              &1.70& 3.46& 2.76& 3.21&0.080        &0.04       &  1.3E08 & 9.4E07  &OSC\\
UGC7242    & Scd        &5.67              & 0.36& 0.18& 0.16& 0.20&0.025       &0.05       &  7.8E07 & 5.0E07 &SOC\\
UGC4305    & Im         &6.40               &3.13& 2.42& 2.81& 3.98&0.080        &0.00       &  2.3E08  &  7.3E08 &SOD\\  
UGC1249    & SBm        &6.40               &5.93& 3.67& 1.93& 1.19&0.100        &0.15       &  5.5E08 & 9.9E08 &OOD\\
NGC5474    & SAcd       &6.60               &4.26& 3.73& 2.52& 1.86&0.075       &0.10       &  8.1E08 & 1.3E09  &SOD\\ 
NGC5477    & SAm        &6.70               &1.00& 0.96& 0.57& 0.59&0.025       &0.15       &  4.0E07 & 1.3E08 &SOD\\
NGC4248    &S?          &6.82              &0.48& 0.32& 0.14& 0.34&0.125       &0.10       & 9.8E08  &  6.1E07 &OOC  \\   
UGC0695    & Sc         &7.80               &0.49& 0.41& 0.07& 0.10&0.100        &0.10       &  1.8E08  & 1.1E08  &SSC\\
NGC4656    & SBm        &7.90               &22.6& 8.50& 6.40& 5.52&0.050        &0.15       &  4.0E08 & 2.2E09 &OOD\\
NGC4485    & IBm        &8.80               &9.41& 5.66& 3.08& 3.87&0.100        &0.25       &  3.7E08 & 4.0E08 &OOD\\
ESO486     & S?         &9.09              &0.48& 0.34& 0.49& 0.45&0.025       &0.05       & 7.2E08  & 2.8E08 &SSD\\
IC0559     & Sc         &10.0                &0.30& 0.17& 0.09& 0.15&0.075       &0.05       & 1.4E08 & 3.7E07 &SSC\\ 
NGC3274    & SABd       &10.0                &4.43& 3.21& 1.23& 1.32&0.075       &0.20       &  1.1E08 & 5.5E08 &SOD \\
UGC5340    & Im         &12.7              &0.92& 0.17& 0.38& 0.66&0.025       &0.05       &  1.0E07 & 2.4E08 &OOD\\
\enddata
\tablenotetext{a}{Galaxy name and morphological type as listed  in NED, the NASA Extragalactic Database.}
\tablenotetext{b}{Distances from \cite{sabbi18}.}
\tablenotetext{c}{CMD-based average SFR in the last 60 Myr derived in
  this work (PARSEC-COLIBRI and MIST solutions are equally
  weighted). It should be noted our CMDs generally miss compact star
  forming regions and clusters.}
\tablenotetext{d}{CMD-based average SFR in the last 10 Myr derived in
  this work (PARSEC-COLIBRI and MIST solutions are  equally
  weighted). It should be noted our CMDs generally miss compact star
  forming regions and clusters.}
\tablenotetext{e}{SFR calculated from the GALEX FUV imaging, normalized to the
  area of the HST/WFC3 footprint and corrected for extinction following the
  prescriptions of \cite{hao11}.}
\tablenotetext{f}{SFR calculated from ground-based H$\alpha$,
  normalized to the area of the HST/WFC3 footprint and corrected for
  extinction following the prescriptions of \cite{calzetti07}. }
\tablenotetext{g}{Foreground reddening derived in this work.}
\tablenotetext{h}{Differential reddening derived in this work.}
\tablenotetext{i}{M$_*$  and HI masses are taken from \cite{calzetti15}.}

\tablenotetext{l}{Galaxy classification from this work (see Section
  5). The first letter describes how the HeB stars of different ages
  are arranged with respect to one another:``S'' stands for similar
  distributions, ``O'' for offset. The second letter compares the
  centroids of HeB and RGB stars: ``S'' stands for similar centroids
  (namely closer than 250 pc), ``O'' for offset centroids. The third
  letter describes the sparseness of HeB and RGB stars on the map: if
  HeB stars are far more concentrated than the RGB stars the third
  letter is ``C'' (concentrated), otherwise we assign a letter ``D''
  (diffuse). We also add the suffix ``t'' if the HeB distributions are
  twisted with respect to the RGB ones. }

\end{deluxetable*}


The paper is structured as follows. The galaxy sample used in this work
is presented in Section 2. The method used to extract the SFHs from
the CMDs is outlined in Section 3. We present the SFHs in Section
4. Section 5 shows the spatial distribution of the SF, and Section 6
discusses these results.

\section{Observations}

The observations were performed for the LEGUS survey
(\citealt{calzetti15}), an {\it{HST}} Treasury Program for a
panchromatic photometric survey of 50 nearby (within $\sim 12$ Mpc)
star forming dwarf and spiral galaxies. The goal of the survey is to
investigate scales and modes of SF using the leverage of the UV
imaging.  Scientific objectives and the data reduction are described
in \cite{calzetti15}, while stellar photometry\footnote{The stellar
  photometry catalogs have been publicly released through MAST and are
  available at:
  https://archive.stsci.edu/prepds/legus/dataproducts-public.html} is
described in detail in \cite{sabbi18}. The observations were obtained
with the Wide Field Camera 3 (WFC3) and complemented with archival
data from the Advanced Camera for Surveys (ACS), in a set of broad
bands over the range 0.27 - 0.81 $\mu$m, namely in the filters F275W,
F336W, F438W, F555W and F814W (equivalent to NUV, U, B, V, and I,
respectively). Resolved stellar photometry was performed using the
DOLPHOT package version 2.0 downloaded on 12 December 2014 from the
website http://americano.dolphinsim.com/dolphot/
(\citealt{dolphin2000}). DOLPHOT performs point-spread function (PSF)
fitting on all the flat-fielded and CTE-corrected images (FLC) per
field, simultaneously. The output photometry from DOLPHOT is on the
calibrated VEGAMAG scale.

In order to reject non-stellar objects and to have a clean final
sample of stars for our CMDs, we applied quality cuts. The DOLPHOT
output was filtered to retain only  objects classified as stars with
signal-to-noise $> 5$ in both filters. The list was further culled
using sharpness ($< 0.15$) and crowding ($< 1.3$).

\section{SFH method}

The SFHs of the LEGUS SFDs were determined using the population
synthesis routine SFERA (Star Formation Evolution Recovery Algorithm),
which employs a synthetic CMD method, along the lines pioneered by
\cite{tosi91}. We provide here only a short description of SFERA's
approach, while the procedure details are described in
\cite{cigno15,cigno16,cigno18}.

As a first step, a basis of synthetic CMDs is generated using
isochrones from an adopted library of stellar models. To test the
systematic uncertainties and physical assumptions behind the stellar
models, the synthetic CMDs are generated adopting two independent and
widely used sets, either the PARSEC-COLIBRI \citep{bressan12,marigo17}
or the MIST (MESA Isochrones and Stellar Tracks) isochrones
(\citealt{paxton11,paxton13,paxton15,choi16,dotter16}). The two sets
differ both in the input physics and in the assumptions about the
efficiency of macroscopic mechanisms, like rotation (PARSEC-COLIBRI
models are static, the MIST ones are rotating with
$v/v_{crit}=0.4)$. The differences between the PARSEC-COLIBRI and the
MIST results should be considered as a lower limit to the actual
differences between data and models.

All basic synthetic CMDs are Monte Carlo realizations of all possible
combinations of 9 equal logarithmic age bins between $\log(t) = 5$ and
$8.5$ (corresponding to the age range of our UV data), and 20
metallicity bins\footnote{We adopt the approximation
  [M/H]$=\log(Z/Z_{\odot})$, with $Z_{\odot}=0.0152$ for
  PARSEC-COLIBRI models and $Z_{\odot}=0.00142$ for MIST models.}
between $[M/H] = -2$ and 0.0.  A \cite{kroupa01} initial mass function
(IMF) between 0.1 and 300 M$_{\odot}$ is then used to fully populate
the CMDs.  Unresolved binaries are also considered and 30\% of
synthetic stars are coupled with a stellar companion sampled from the
same IMF. Concerning the distance modulus, given the relatively low
sensitivity of the UV SFH to this parameter (mainly due to the rapid
evolutionary phases populating the UV CMD), we opted to use the
distances (\citealt{sabbi18}) derived using the tip of the red giant
branch (RGB), a robust feature missing in the UV CMDs but clearly
visible in optical CMDs. On the other hand, the UV color is very
sensitive to extinction, therefore extinction distribution and SFH are
simultaneously fitted to the data. Extinction is parametrized with two
free parameters: a foreground value, applied evenly to all stars,
A$_{V}$, and a differential value, dA$_{V}$\footnote{Each star
  receives a random additional extinction between 0 and dA$_{V}$.},
that is applied in addition to A$_{V}$. We adopted the extinction law
of \cite{cardelli89} and assumed a normal total-to-selective
extinction value of R$_{V}=3.1$.

The observational conditions were simulated in the synthetic CMDs by
using the results of extensive artificial star tests. ``Fake'' sources
are injected (one at a time) onto the actual images and the source
detection routine used for our science images is applied to the fields
containing the combined science images and the fake sources. Counting
how many fake stars are lost as a function of magnitude and position
provides the map of the local incompleteness. Each synthetic CMD is
convolved with incompleteness as well as with the photometric errors
derived from the cumulative magnitude distribution of
mag$_{\mathrm{out}}$-mag$_{\mathrm{input}}$ of the fake stars. Since
our targets are young stars, and young stars tend to crowd in
associations and filaments (hence, most of them will be affected by
more significant errors and incompleteness than the average in the
field), we implemented the procedure described in \cite{cigno16},
which is more appropriate for crowded fields. As a first step, fake
stars are evenly distributed over the galaxy field of view (FoV). The
observed density of stellar sources (plagued by incompleteness) is
then corrected for the local incompleteness, restoring the ``true''
profile of the galaxy. As a second step, fake stars are injected
following the reconstructed profile. The resulting incompleteness,
being weighted with the real stars, will be a less biased estimate of
the actual incompleteness suffered by the young stars.

For deriving the SFH and identifying the best-fit model, the basic
synthetic CMDs are linearly combined and the best combination of
coefficients (minimizing a Poissonian likelihood function of the
data-model residuals) is searched with the hybrid-genetic algorithm
SFERA. Our typical approach is to explore the entire age-metallicity
space. However, given the low dependence upon metallicity in the
quoted range of ages ($< 300$ Myr), with the only significant effect
concerning the color extension of the He-burning loop (which shrinks
with increasing metallicity), we opted to use the available
spectroscopic information together with the photometry to provide
additional constraints on the SFH. In particular: 1) the metallicity
in the last 10 Myr was fixed at the literature\footnote{Metallicities
  are taken from Table 1 of \cite{calzetti15}} spectroscopic value
(within the measured errors); 2) we limited the number of free
parameters by imposing that the metallicity was an increasing function
of time. Throughout this work, we adopt the nebular oxygen abundance
derived from HII region observations as a tracer of the overall
metallicity in the last 10 Myr.

The whole process is performed using the PARSEC-COLIBRI and MIST
stellar models independently. At fixed metallicity, the differences
between PARSEC-COLIBRI and MIST synthetic CMDs are indicative of the
different underlying stellar physics. When deriving the SFH in the
next Section, these differences will serve as an indication of the
potential differences between a given model and the ``truth''.

\section{SFHs}

\subsection{Bursting or continuous?}

The SFHs during the last 100 Myr of 23 SFDs are investigated using the
UV CMD, specifically the F336W vs F336W-V\footnote{In some galaxies V
  stands for the F555W band, in others for the F606W one, depending on
  the availability of archival data.} CMD obtained from the LEGUS
survey. This filter combination enables us to focus on the youngest
populations in our galaxies, mostly traced by main sequence (MS) and
core He-burning (HeB) stars more massive than 5 M$_{\odot}$. HeB stars
monotonically fade as the population age increases, providing an
advantage over the MS in the age dating of the regions, since
subsequent generations of HeB stars do not overlap with each other as
they do on the MS. Moreover, the HeB sequence is on average brighter
than the coeval MS turn-off. However, since the MS evolutionary times
are much longer than post-MS times, star-counts along the MS are
statistically more robust. This property becomes crucial when the SFR
is very low and the sample of post-MS stars is poor. For this reason
HeB and MS stars are used in synergy to infer the recent SFHs. The
bluest LEGUS filter available, the F275W filter, would have improved
the age resolution in the last 50 Myr, but exacerbated the effect of
reddening, thus canceling the resolution gain. Moreover, it would have
shortened the look-back time reachable by the photometry due to the
much lower sensitivity of this filter. At any rate, most of the SFHs
available in the literature are derived using wavelengths longer than
the B-passband, hence the implementation of the F336W vs F336W$-$V
(hereafter U vs U$-$V) CMD permits us to investigate the recent SFH with
unprecedented detail.

\begin{figure*}[t]
\centering \includegraphics[width=15cm]{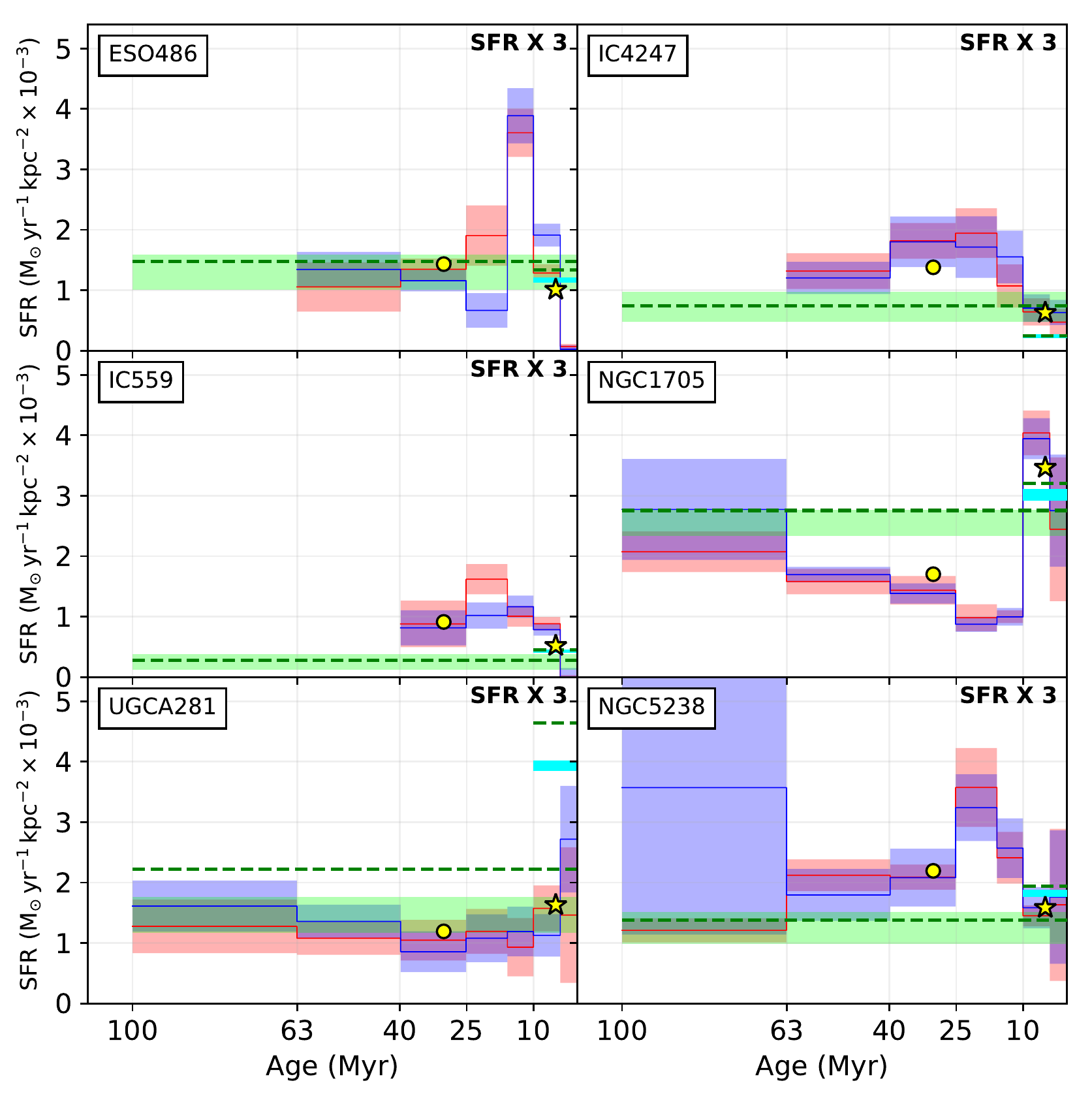}
\caption{Recovered SFHs using PARSEC-COLIBRI (red histogram) and MIST
  (blue histogram) stellar models for the SFDs ESO486, IC4247, IC559,
  NGC1705, UGCA281 and NGC5238. Yellow filled circles and stars
  represent the average SFR density in the last 60 and 10 Myr,
  respectively. Green and cyan horizontal stripes (their width
  reflects the SFR uncertainty) stand for the SFRs derived from FUV
  and H$\alpha$ emission, respectively.  Horizontal dashed lines show
  the same rates corrected for extinction; the corrections utilize
  hybrid star formation rate recipes that include the 24 $\mu m$
  emission to account for the portion that is extinguished by dust
  (\citealt{calzetti07,hao11}). All rates have been normalized to the
  area of the field of view of the HST/WFC3 camera. }
\label{SFHs1} 
\end{figure*}
\begin{figure*}[t]
\centering \includegraphics[width=15cm]{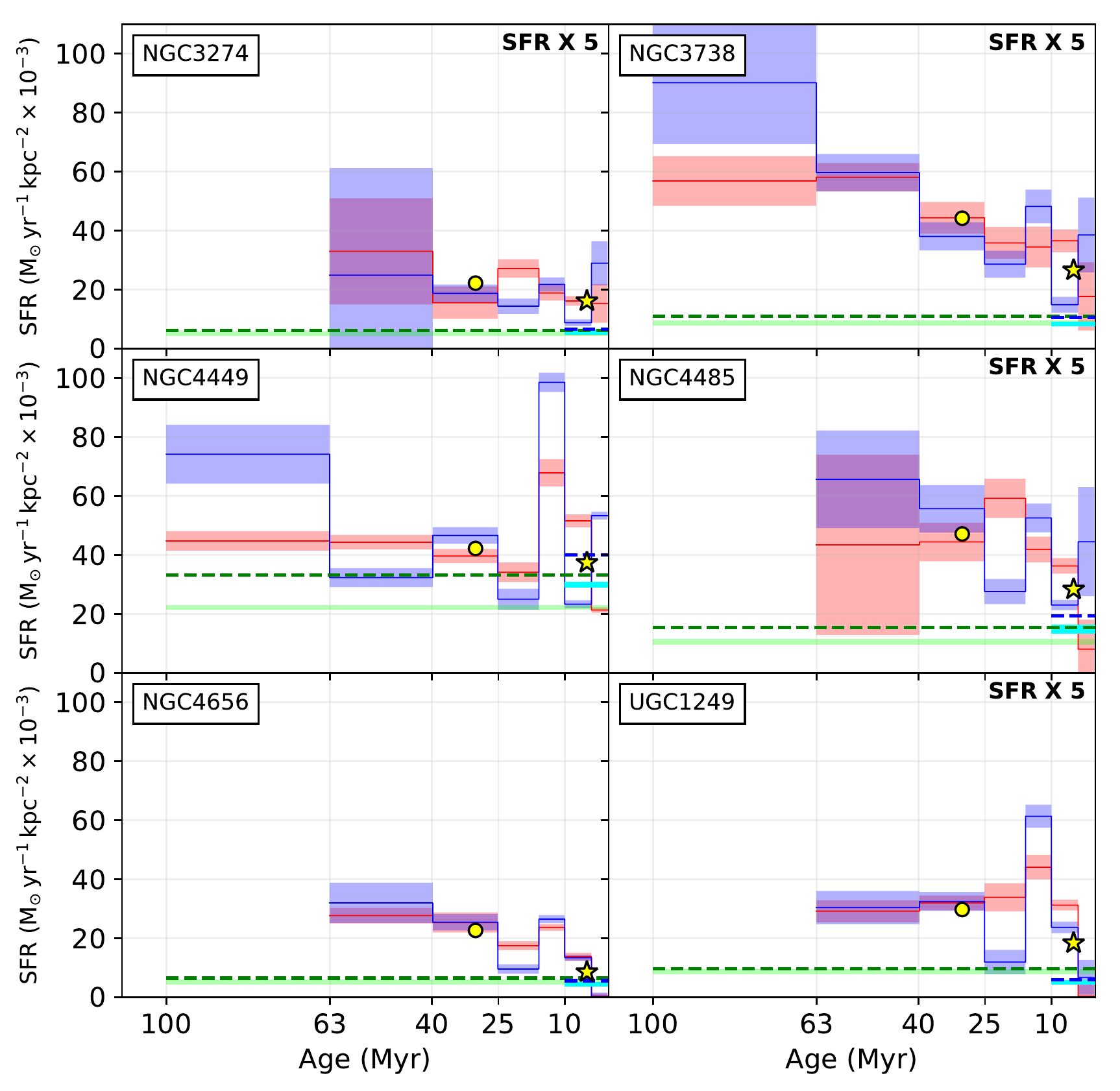}\\
\caption{Same recovered quantities as in Fig. \ref{SFHs1} but for the SFDs NGC3274,
    NGC3738, NGC4449, NGC4485, NGC4656 and UGC1249. }
\label{SFHs2} 
\end{figure*}
\begin{figure*}[t]
\centering \includegraphics[width=15cm]{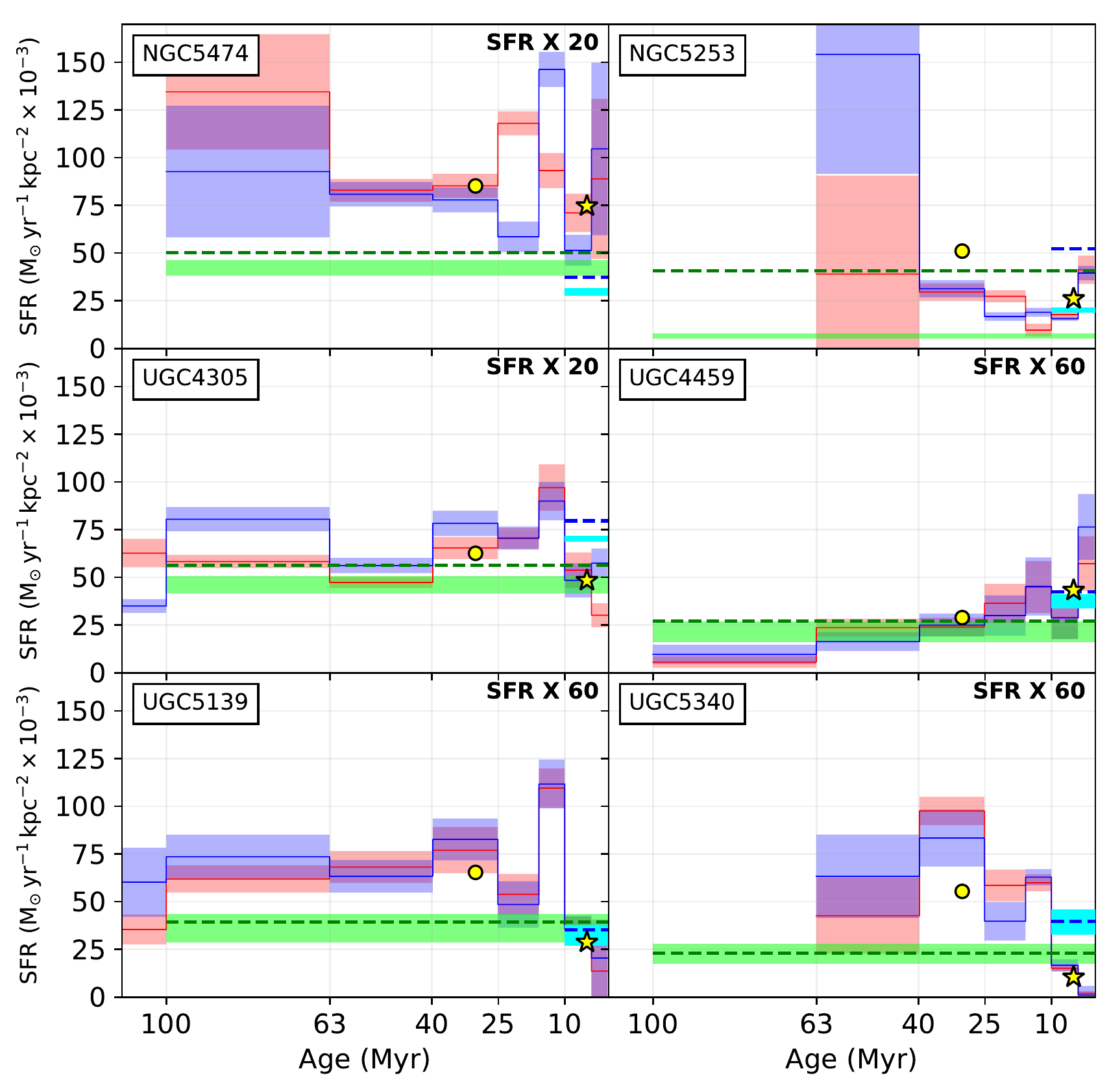}
\caption{Same recovered quantities as in Fig. \ref{SFHs1} but for the SFDs NGC5475,
    NGC5253, UGC4305, UGC4459, UGC5139 and UGC5340. }
\label{SFHs3} 
\end{figure*}
\begin{figure*}[t]
\centering \includegraphics[width=15cm]{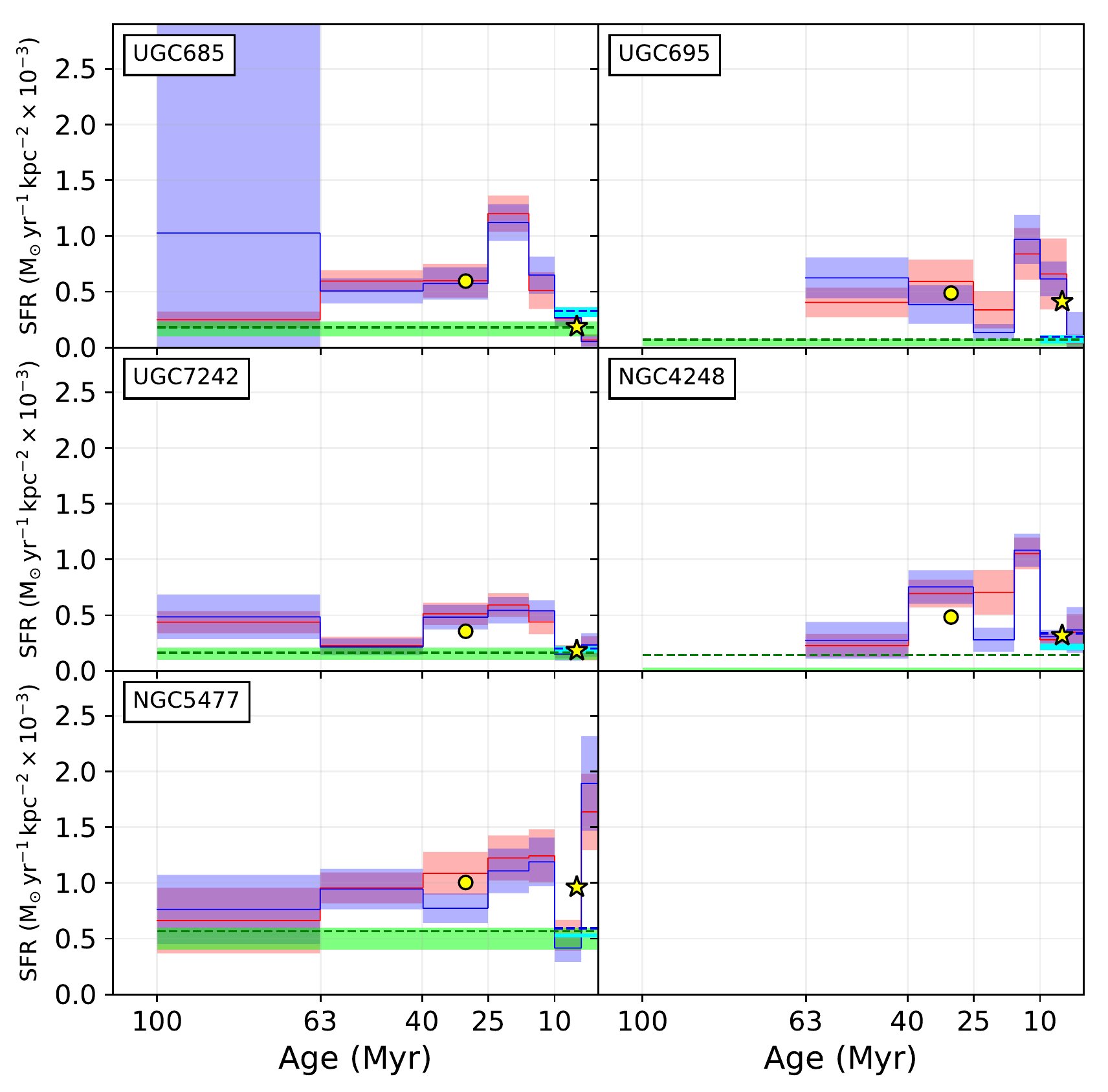}
\caption{Same recovered quantities as in Fig. \ref{SFHs1} but for the SFDs UGC685,
    UGC695, UGC7242, NGC4248 and NGC5477.}
\label{SFHs4} 
\end{figure*}

Figures \ref{SFHs1}, \ref{SFHs2}, \ref{SFHs3}, and \ref{SFHs4} show
the detailed SFHs (red for the PARSEC-COLIBRI solution, blue for the
MIST solution) normalized to the area of the field of view. The
average SFR densities in the last 60 and 10 Myr (PARSEC-COLIBRI and
MIST solutions are equally weighted) are shown with yellow filled
circles and stars, respectively. Horizontal continuous and dashed
lines represent average SFR densities from integrated photometry (FUV
and H$\alpha$), which are extensively discussed in Section
\ref{data-mod}. For each galaxy, Table \ref{tab1} summarizes the
adopted distance (\citealt{sabbi18}), the
average\footnote{i.e. resulting from an equal-weight combination of
  the values obtained with the PARSEC-COLIBRI and MIST solutions.} SFR
density in the last 60 and 10 Myr, the SFR density from FUV and
H$\alpha$ calibrations, the inferred reddening and differential
reddening, the stellar mass and HI mass taken from the literature, and
our new galaxy classification (see Section 5).

The look-back time reached in each galaxy depends on a combination of
factors, such as the galaxy distance, stellar density, and extinction.
A common feature in all targets is the existence of a non-zero (at
1$\sigma$ level) SFR activity at all ages.  While the SFHs present
galaxy-to-galaxy variations, particularly at the youngest epochs, the
general trend is a relatively flat SF as a function of time, with no
major bursts. Even the most ``extreme'' events exceed the average by
only a factor of three. Overall, these results confirm the general
behavior seen in the last 100 Myr of many high resolution optical
studies involving SFDs (see e.g. \citealt{mq10, weisz08}), whose SFHs
are characterised by spikes and lulls a few times higher or lower than
the 100-Myr average SFR. It is important to point out, however, that the
time resolution at older ages gets progressively worse, hence SF
enhancements similar to those detected in the last 30 Myr could be
indistinguishable from a constant activity 100$-$200 Myr ago.  All
targets show SF enhancements in the last 50 Myr, but the SF is not
always continuing to the present time at a similar rate.

\begin{figure*}[t]
\centering \includegraphics[width=8.5cm]{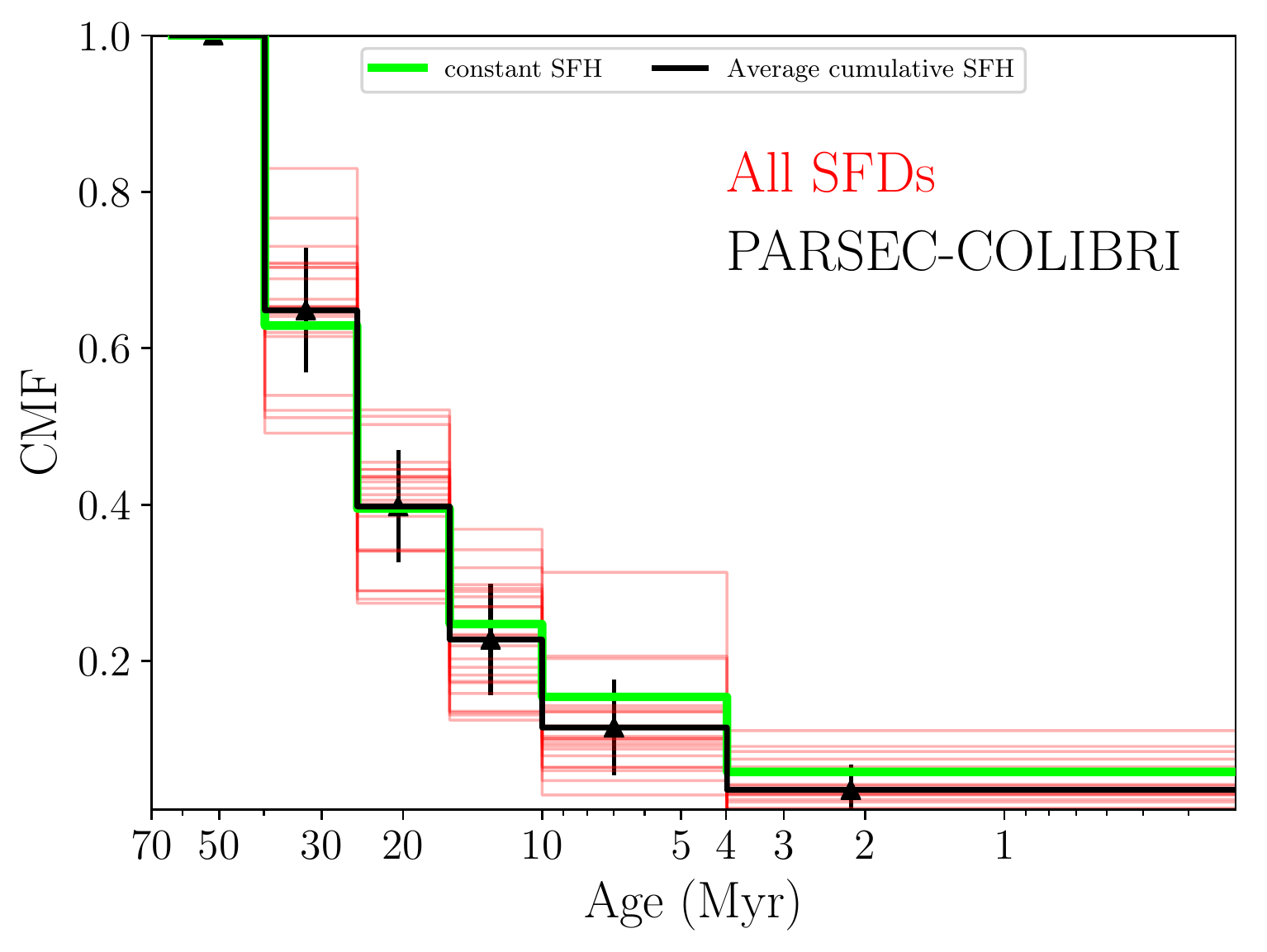}
\centering \includegraphics[width=8.5cm]{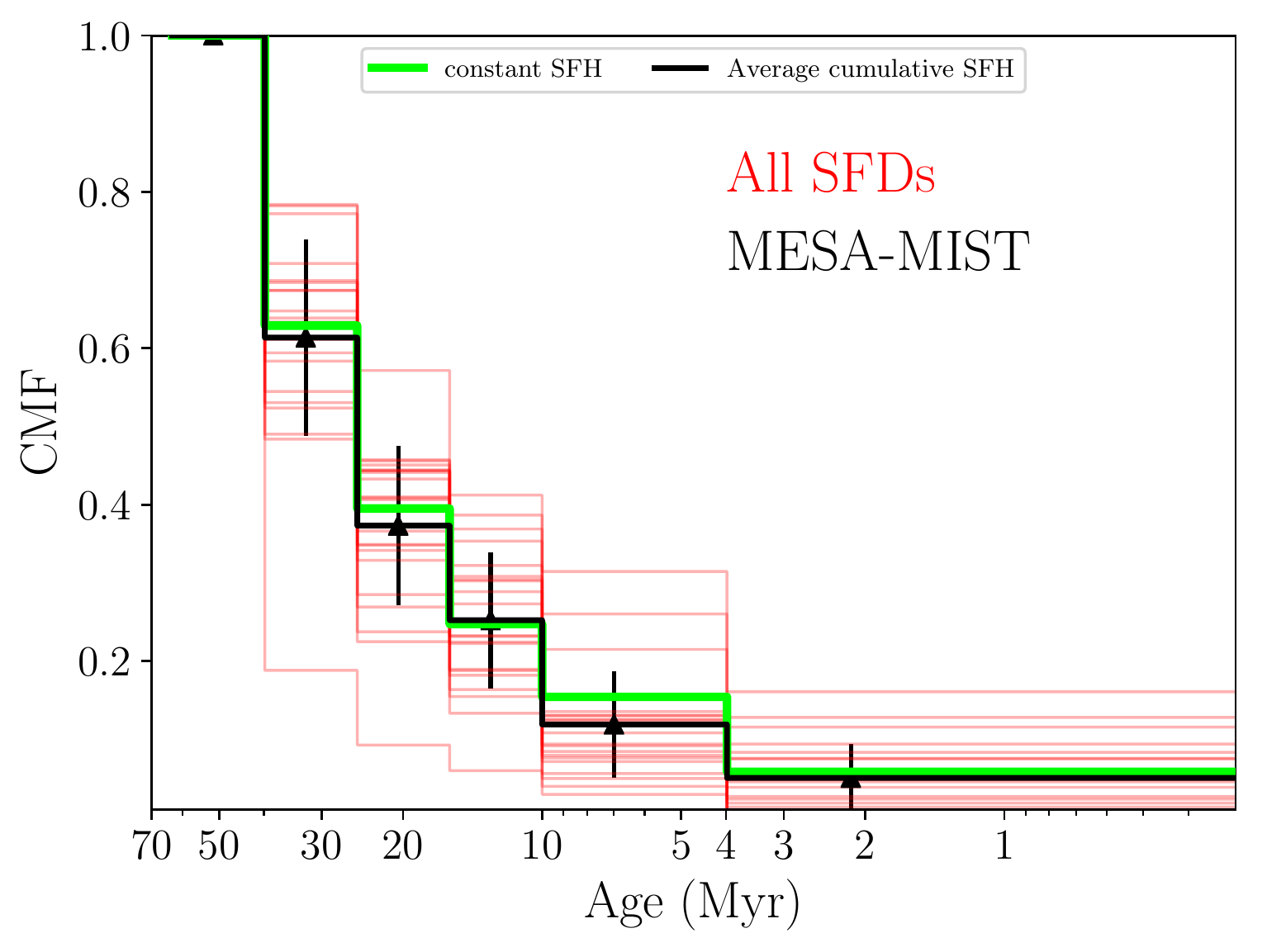}
\caption{Cumulative mass functions (red lines) for all SFHs (in the
  last 60 Myr) derived using PARSEC-COLIBRI (left panel) and MIST
  (right panel) stellar models. The black line is the average
  cumulative SFH (the arithmetic average of all cumulatives for each
  time-step). For comparison, the green line shows the cumulative SFH
  for a constant SF activity. Notice that here the CMF is computed not
  as a function of time but of lookback time.}
\label{cum1_col} 
\end{figure*}

\begin{figure*}[t]
\centering \includegraphics[width=8.5cm]{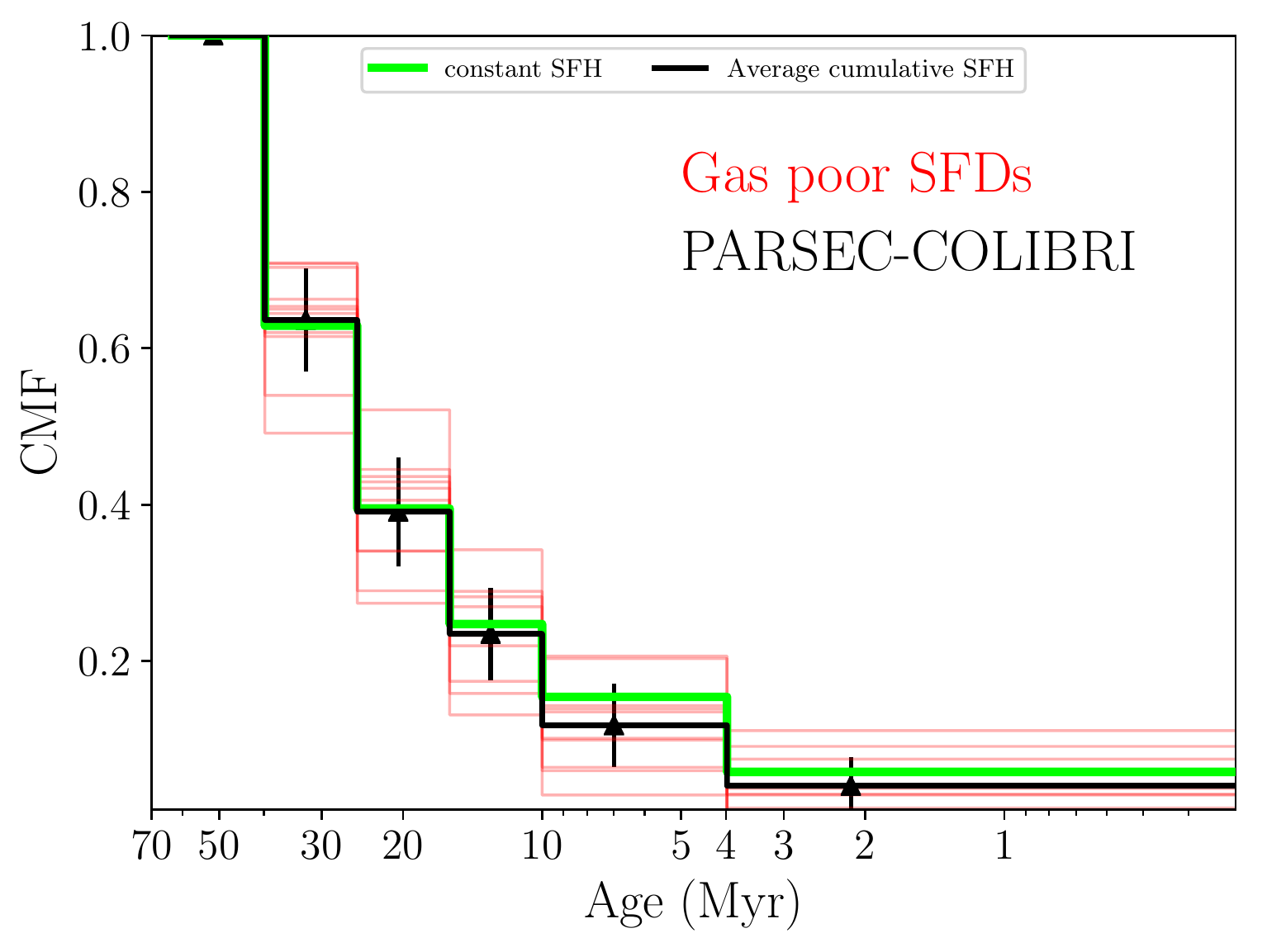}
\centering \includegraphics[width=8.5cm]{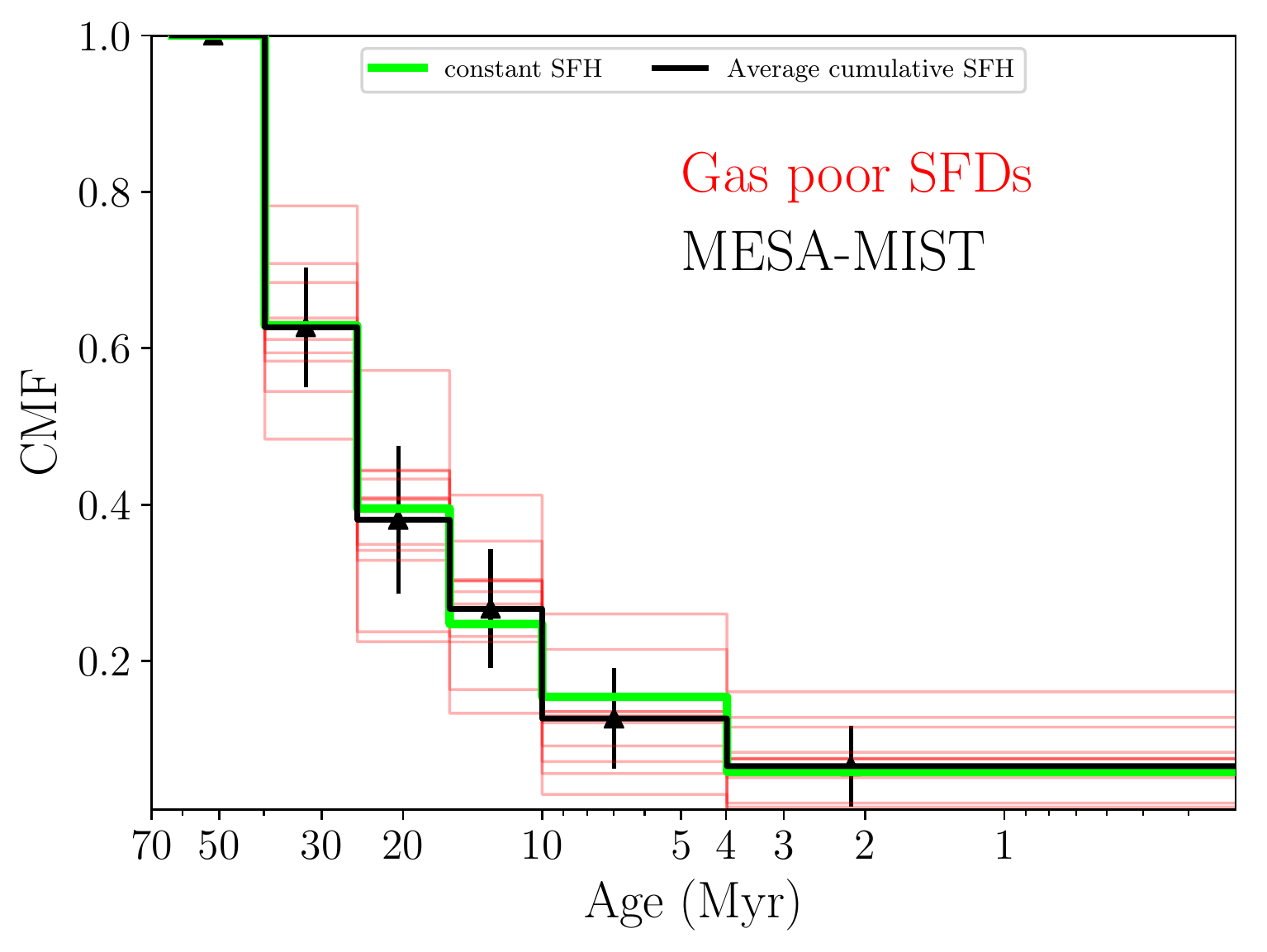}\\
\centering \includegraphics[width=8.5cm]{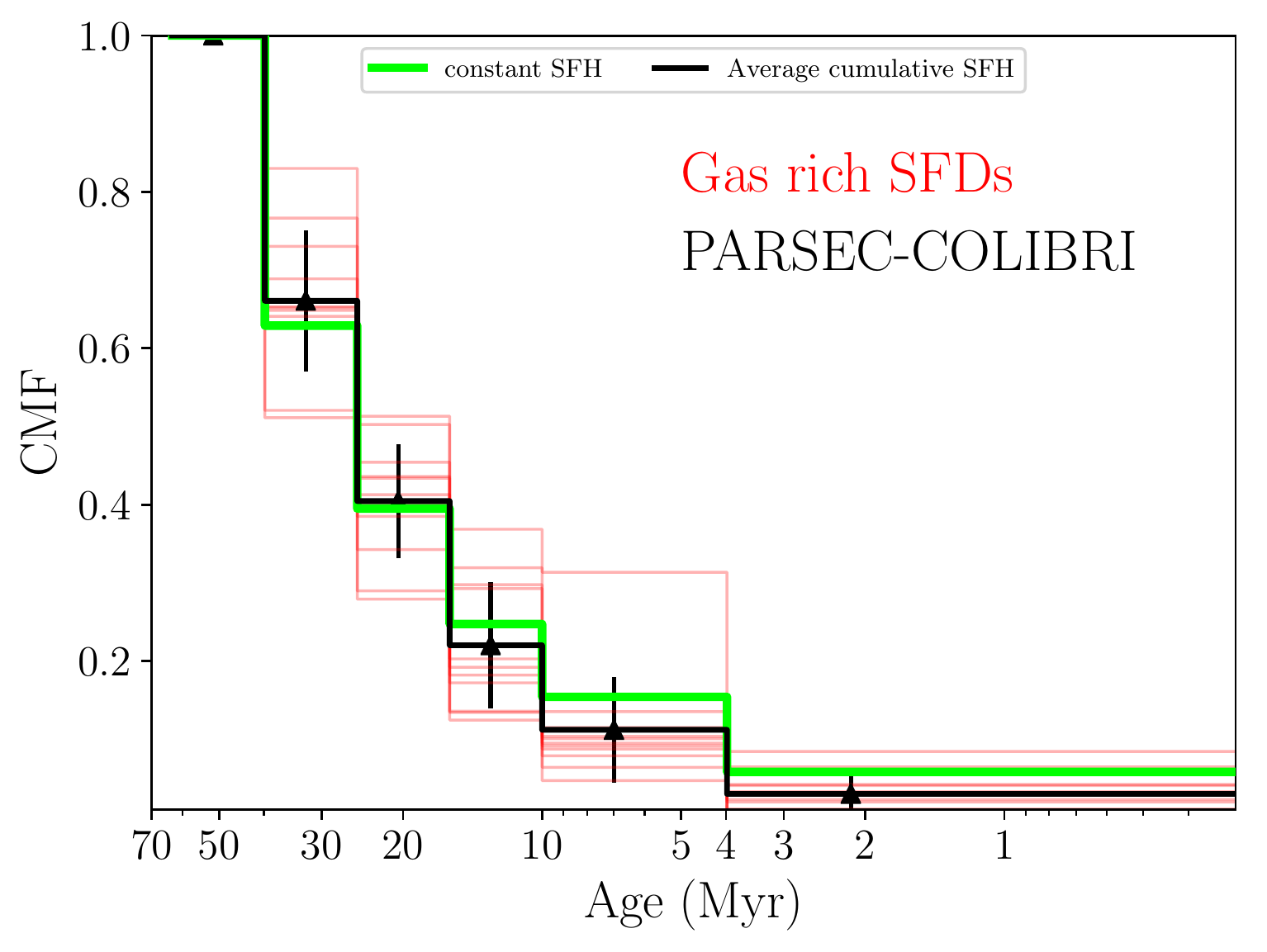}
\centering \includegraphics[width=8.5cm]{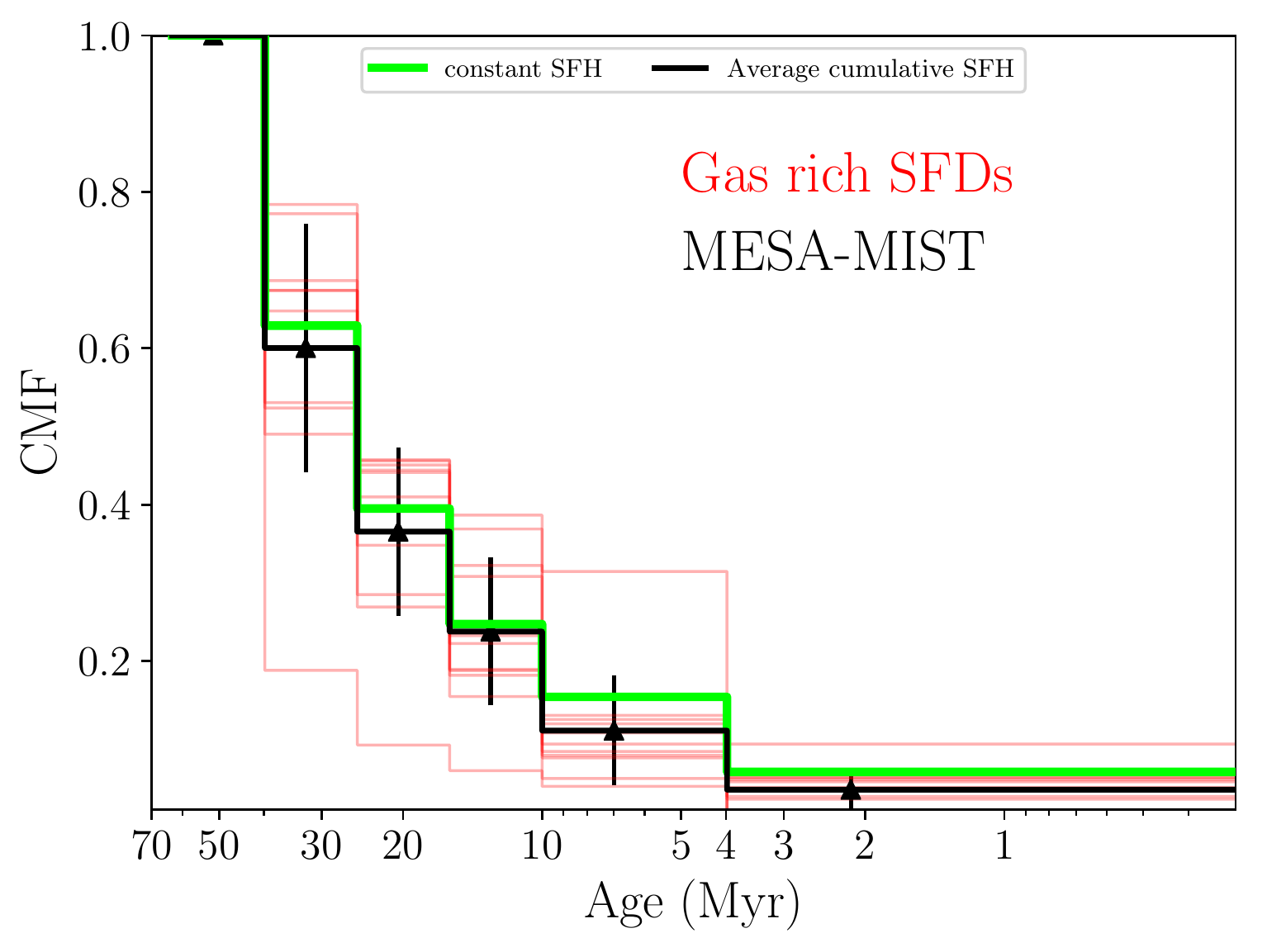}
\caption{Cumulative mass functions for gas poor (top panels) and gas
  rich (bottom panel) SFDs. PARSEC-COLIBRI solutions are on the left
  side, MESA-MIST solutions on the right. Black and green lines have
  the same meaning of those in Fig. \ref{cum1_col}. Notice that here
  the CMF is computed not as a function of time but of lookback
  time.}
\label{cum2_col} 
\end{figure*}

In order to compare the SFHs with each other, we adopt the cumulative
distribution visualization, which gives the fraction of stellar mass
formed up to a given time. This representation allows galaxies of
different masses to be directly compared, and cumulative SFH
measurements are not subject to covariant SFRs in adjacent time
bins. Notice that the stellar mass is summed up as a function of
lookback time and not time, as usually done, to give more emphasis to
the SF behavior in the most recent time bins, which are the focus of
this paper. Figure \ref{cum1_col} shows the cumulative functions of
the recovered SFHs in the last 60 Myr (PARSEC solutions on the left,
MESA-MIST on the right). The black thick line is the average
cumulative SFH, which is the unweighted mean\footnote{Arithmetic
  average.} of the individual cumulative SFHs (the errors are the
standard deviation around the average). This scheme weights all
galaxies equally, so that the resulting SFH indicates what is
``typical'', without the influence of the most massive galaxies that
might dominate a time bin. The green line depicts the cumulative SFH
resulting from a constant SF activity. Independently of the adopted
stellar models, the ``typical'' LEGUS SFD has a SFH that is compatible
with a constant activity in the last 60 Myr. A possible explanation is
that the epoch of the peak activity in each galaxy is randomly
distributed across the sample, hence the average activity is
necessarily flat.

In order to explore the impact of the gas fraction on this result,
Figure \ref{cum2_col} shows the same plot dividing gas poor and gas
rich galaxies. To do this, we consider all galaxies with a ratio
between M$_{\mathrm{HI}}$ and M$_{\mathrm{HI}} +$M$_{\star}$ smaller
than 50\% gas poor, and the others as gas rich (stellar and HI masses
are taken from the literature).  This separation should be taken with
caution since the amount of molecular hydrogen is not accounted for,
and some galaxies in the two samples have similar gas fractions. With
these caveats in mind, the plots suggest that, in the last 60 Myr, gas
poor galaxies experienced a typical SFH that is similar to the typical
SFH of gas rich ones. 


\subsection{Data vs model CMDs}
\label{data-mod}

\begin{figure*}[t]
\centering \includegraphics[width=19cm]{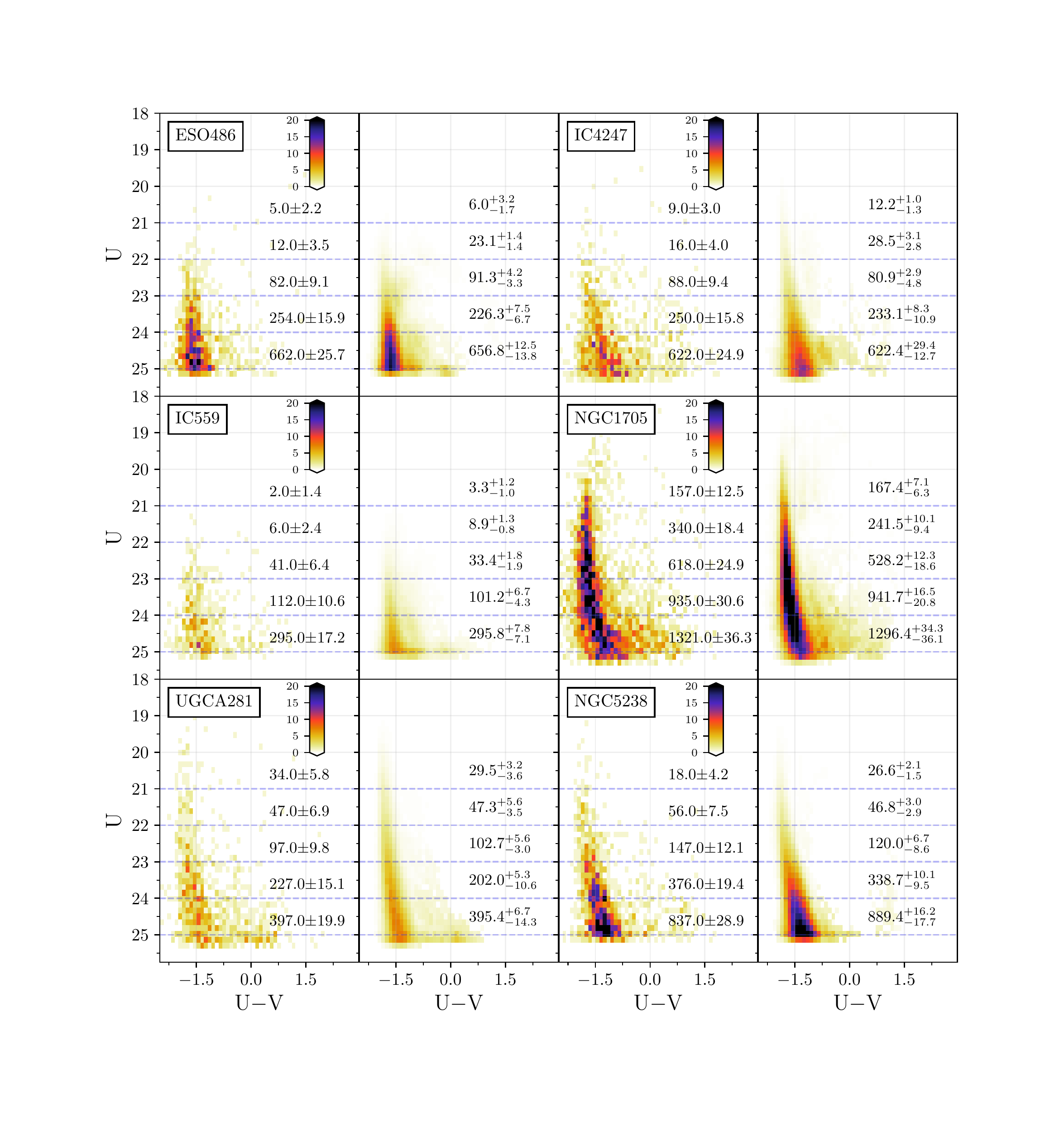}
\caption{ Observational CMDs (on the left) vs synthetic ones (generated from
  PARSEC-COLIBRI and MESA-MIST best SFHs; on the right) for the SFDs ESO486, IC4247,
  IC559, NGC1705, UGCA281 and NGC5238. Observed and predicted
  star-counts in one magnitude bins are also shown.}
\label{cmd1} 
\end{figure*}

\begin{figure*}[t]
\centering \includegraphics[width=19cm]{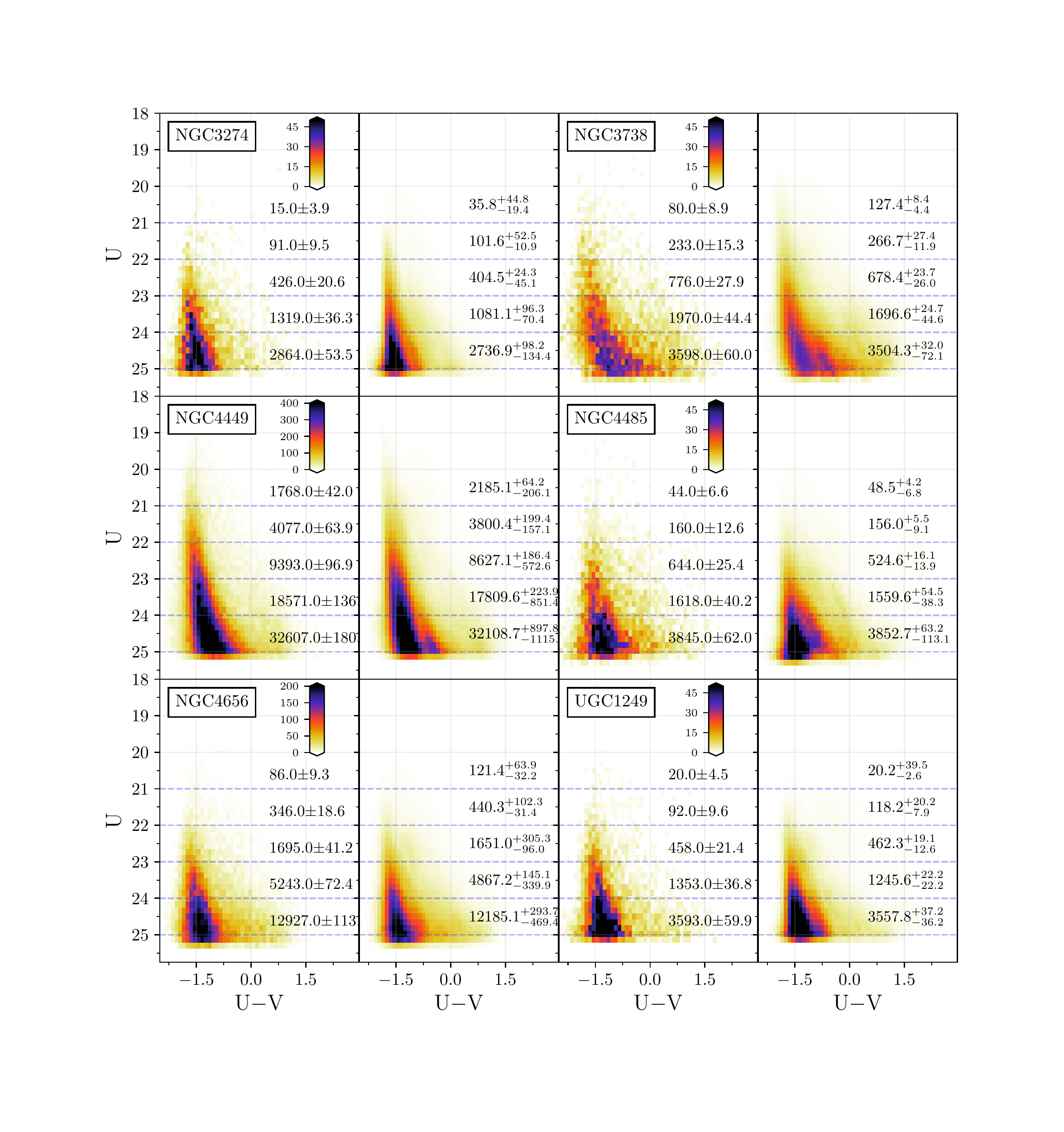}\\
\caption{Same plot as Fig. \ref{cmd1} for the SFDs NGC3274,
    NGC3738, NGC4449, NGC4485, NGC4656 and UGC1249.  }
\label{cmd2} 
\end{figure*}

\begin{figure*}[t]
\centering \includegraphics[width=19cm]{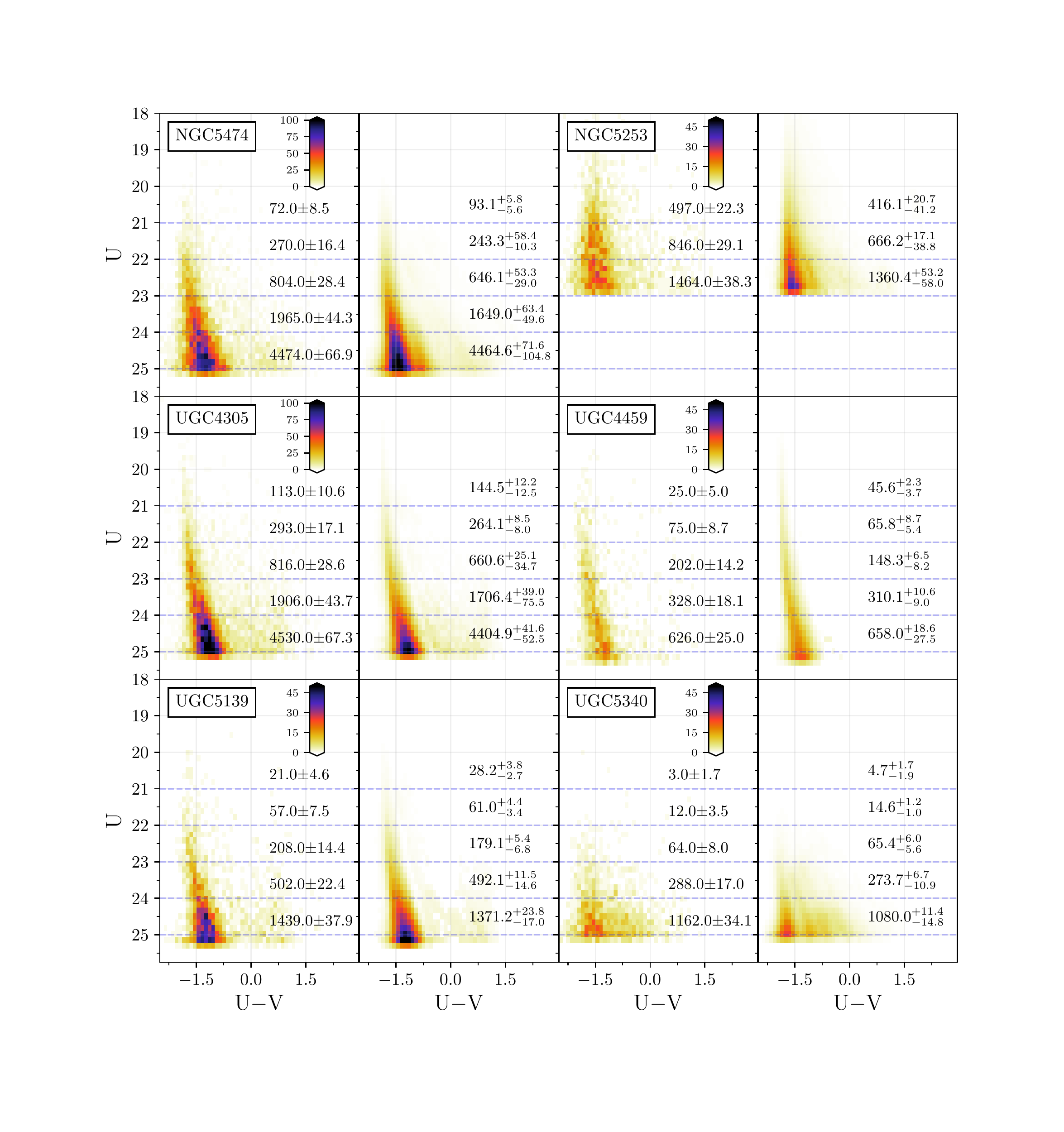}
\caption{Same plot as Fig. \ref{cmd1} for the SFDs NGC5475,
    NGC5253, UGC4305, UGC4459, UGC5139 and UGC5340.}
\label{cmd3} 
\end{figure*}

\begin{figure*}[t]
\centering \includegraphics[width=19cm]{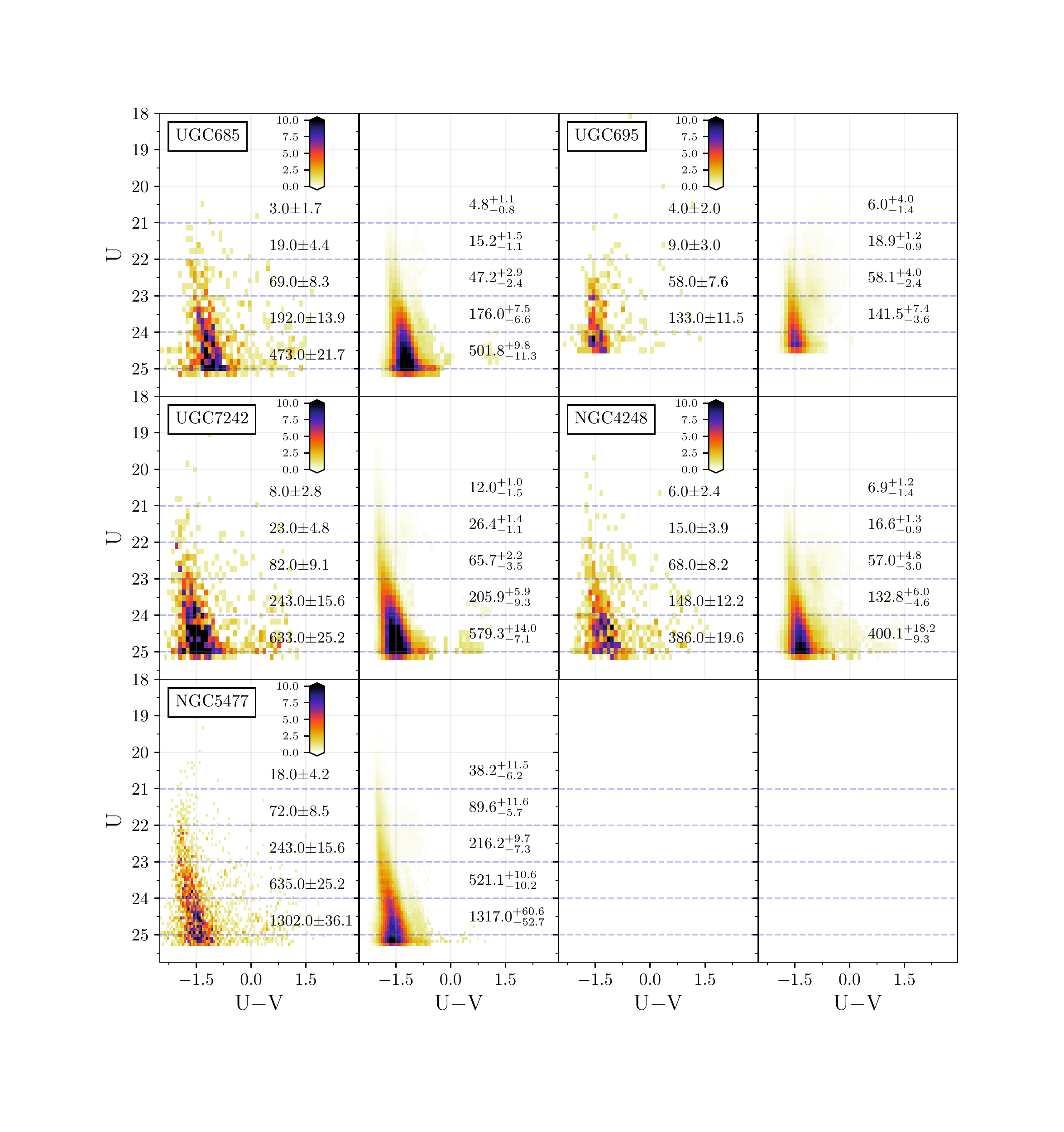}
\caption{Same plot as Fig. \ref{cmd1} for the SFDs UGC685,
    UGC695, UGC7242, NGC4248 and NGC5477.}
\label{cmd4} 
\end{figure*}

The most basic check on the reliability of the recovered SFH is to
compare synthetic and observed CMDs.  In Figures \ref{cmd1},
\ref{cmd2}, \ref{cmd3}, and \ref{cmd4} we compare the observational
Hess diagrams (i.e. the density of points in the CMD) with the
corresponding synthetic ones. The latter are obtained by averaging
several mock CMDs generated from the PARSEC-COLIBRI and MESA-MIST
solutions equally weighted.  This allows us to consider the systematic
errors related to the adopted sets of stellar evolution models.  The
observed ($\pm$ 1 standard deviation) and predicted number (median, 25
and 75 percentiles) of stars are also shown in bins of one
magnitude. Overall, the simulations are in agreement with the
observations, with a few exceptions like NGC4449, NGC5253, NGC3738,
and NGC4656, where significant mismatches between the data and the
models are found, both in terms of star-counts and color spread. These
exceptions are not fully surprising, since these galaxies are the most
crowded ones in the sample, and/or strongly affected by differential
reddening. In fact, when the sample is locally very incomplete, the
reconstructed density profile is based on small numbers of detected
stars, hence it is inherently more uncertain. This problem is
exacerbated in those regions where the incompleteness is also highly
spatially variable. Moreover, in dense systems unresolved clusters
could be contaminating the brighter end of the CMD. Concerning
crowding, NGC5253 is the worst case in the sample: most of its
vigorous SF is concentrated in the central 500 pc (this is also the
part of the galaxy shown in Fig. \ref{cmd3}) and the completeness of
its CMD drops by several magnitudes from the periphery to the central
kpc. Extinction can have a complex distribution, which may depend on
the age of the population (e.g., the youngest stars are likely still
embedded in their dust-rich cocoons), while in our parametrization it
is modeled with two parameters (foreground and differential
extinction) with no dependence on time in the last 200 Myr. Moreover,
the adopted extinction law is \cite{cardelli89} with Rv= 3.1, while
the actual law may vary from galaxy to galaxy (see, e.g.,
\citealt{demarchi16}).

On the side of stellar evolution modelling, stellar models for massive
stars are inherently uncertain (see, e.g.,
\citealt{demink12}). Significant mass-loss may cause a star to lose
over half its mass during its lifetime. This mass loss occurs via
stellar winds and strongly affects the stellar evolution of a star
(however, at low metallicities, we expect a major reduction in wind
feedback from single massive stars; see
e.g. \citealt{rama19}). Moreover, massive stars are often found to be
rapid rotators and have a significant preference to be binaries with
orbital periods of less than a few days
(\citealt{sana11}). Theoretically speaking, about 15-40\% of massive
early-type stars are expected to be products of binary mass transfer
(\citealt{demink14}), i.e. they have accreted mass in a past mass
exchange episode and/or merged with a former binary companion. Binary
mass transfer can produce a surplus of massive stars (adding a tail of
binary products, blue stragglers) or can rejuvenate stars via mass
accretion.

The transition from MS and post-MS is especially problematic in
NGC4449 and NGC3738, with the synthetic CMDs showing a gap between the
MS and the post-MS phase, which is completely absent in the observed
CMD. A similar dip was already noticed by \cite{tang14} in several
dwarf galaxies. These authors showed that this discrepancy is overcome
by extending the overshooting at the base of the convective envelope.

Finally, we cannot exclude that real variations of the IMF are
occurring in some dense star forming regions. In Paper I, we already
found that an IMF flatter than Salpeter's ($s =-2.0$) provides a
better fit to the data of NGC4449, and a similar variation has been
recently found by \cite{sch18} for 30 Doradus, a giant star forming
region in the Large Magellanic Cloud; this system has produced stars
up to very high masses ($\approx$ 200 M$_{\odot}$), with a
statistically significant excess of stars above 30 M$_{\odot}$, and an
IMF shallower than a Salpeter one above 15 M$_{\odot}$.


\subsection{Comparison with integrated FUV and H$\alpha$ rates}

In this section, we compare our SFHs with integrated-light SFR density
measurements using H$\alpha$ and GALEX FUV imaging, respectively.
While H$\alpha$ photons are produced by gas ionized by young hot
stars, tracing the SF in the last few million years, the FUV flux
stems from the photospheres of O- through later-type B-stars, and thus
traces the SF in the last 100 Myr.

The data are taken from the Local Volume Legacy survey (LVL;
\citealt{hao11,dale09,lee11}) where deep GALEX FUV, ground-based
H$\alpha$, and Spitzer 24$\mu m$ imaging were obtained for all of the
galaxies studied here. We perform photometry within the HST WFC3
footprint on the H$\alpha$, FUV, and 24$\mu m$ images, and combine the
resulting fluxes to derive dust-corrected SFRs (flux measurements and
SFR calculations will be described in detail in Cook et al. 2019, in
prep.).  The SFRs are calculated using the prescription of
\cite{murphy11} which assumes a Kroupa IMF. The dust correction is
calculated for H$\alpha$+24$\mu m$ and FUV+24$\mu m$ following the
prescriptions of \cite{calzetti07} and \cite{hao11}, respectively.

Green and cyan horizontal stripes (their width reflects the SFR
uncertainty) in Figures \ref{SFHs1}, \ref{SFHs2}, \ref{SFHs3}, and
\ref{SFHs4} stand for the average SFRs derived from the FUV and from
the H$\alpha$ emission, respectively. Dashed lines show the same rates
corrected for extinction. All rates have been normalized to the area
of the HST/WFC3 field. The average SFR densities in the last 60 and 10
Myr (PARSEC-COLIBRI and MIST solutions are equally weighted) are shown
with yellow filled circles and stars, respectively. Table \ref{tab1}
summarizes all rates for all galaxies.
\begin{figure}[t]
\centering \includegraphics[width=8cm]{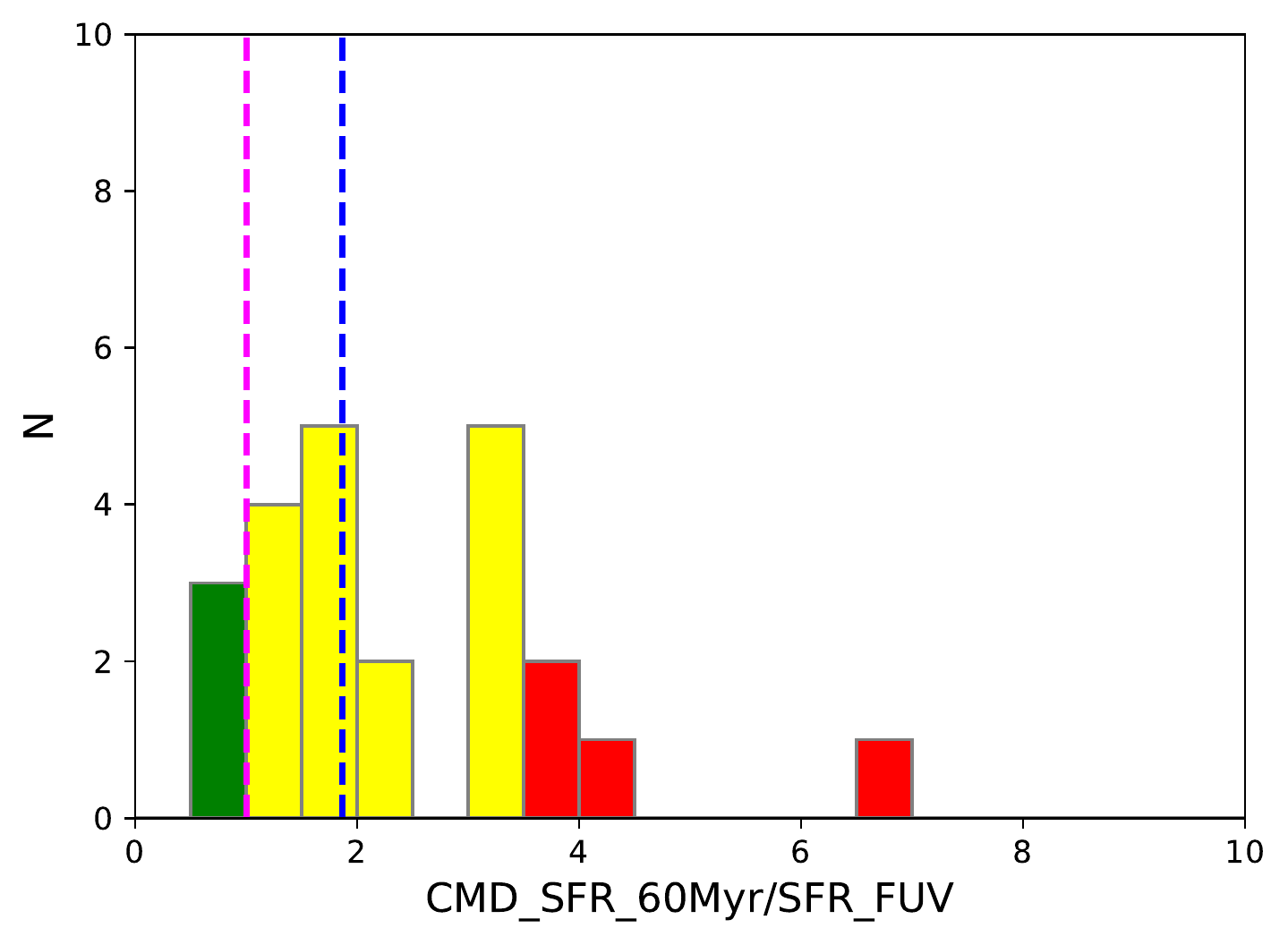}
\caption{Distribution of the ratios between CMD-based SFRs (averaged
  over the last 60 Myr and different stellar models) and FUV SFR
  (extinction corrected) for the entire sample. Blue and magenta
  dashed lines indicate median and unit (ratio$=1$) value,
  respectively. Objects with ratios lower than the 25th percentile are
  in green, higher than the 75th percentile in red and those in
  between in yellow.}
\label{h1} 
\end{figure}
\begin{figure}[t]
\centering \includegraphics[width=8cm]{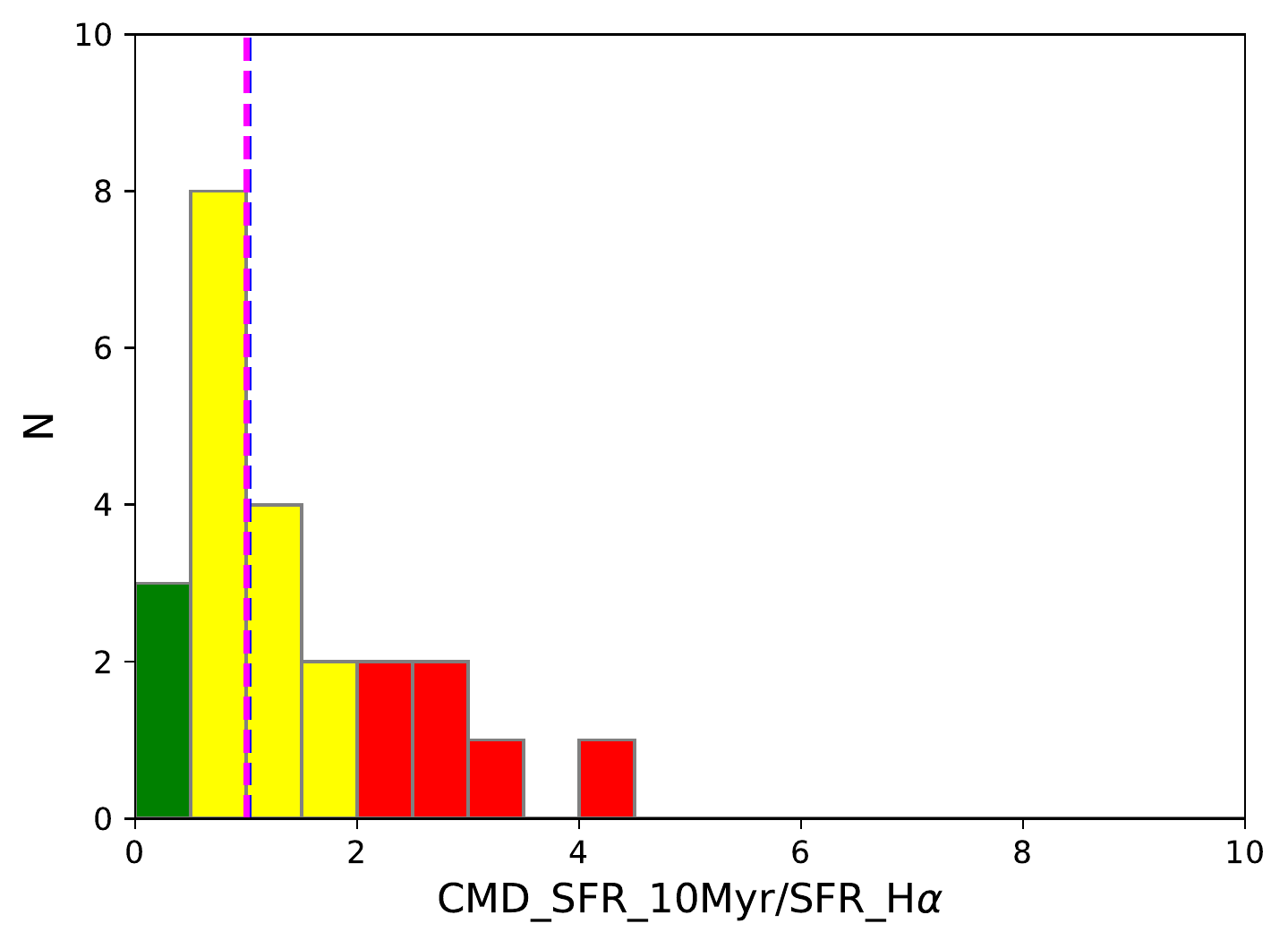}\\
\caption{Distribution of the ratios between CMD-based SFRs (averaged
  over the last 10 Myr) and H$\alpha$ based SFR (extinction corrected)
  for the entire sample. Symbols are the same as in Fig. \ref{h1}. }
\label{h2} 
\end{figure}
For most of the galaxies in the sample, the FUV-based SFRs (corrected
for extinction ) are systematically lower than the CMD-based SFRs by
up to a factor of five. This is even more striking in light of the
fact that, because of crowding, our CMDs generally miss compact star
forming regions and clusters. Fig. \ref{h1} shows the histogram of the
distribution of the ratio between CMD-based SFRs (averaged over the
last 60 Myr and different stellar models) and FUV SFR (corrected for
extinction) for the entire sample. Blue and magenta dashed lines
indicate median and unit (ratio$=1$) values, respectively. Objects
with ratios lower than the 25th percentile are displayed in the
histograms in green, those higher than the 75th percentile are in red
and those in between in yellow.  It is clear from the plot that the
median ($\approx 2$) and unit values are significantly
offset. Moreover, two galaxies show a ratio of 4 or higher.

A similar result was also found by \cite{mq15b} when they compared
optical CMD-based SFRs with integrated FUV counterparts for a sample
of SFDs within 6 Mpc (STARBurst IRregular Dwarf Survey, STARBIRDS;
\citealt{mq15a}). As in our sample, most of their optical CMD-based
SFRs were higher than the FUV-GALEX SFRs by up to a factor of four.

Different explanations for this discrepancy can be envisaged. One
cause could arise from the short time-scale ($60$ Myr) used to average
our rates, in contrast with the FUV which is sensitive to the whole
flux of OB stars. As found by \cite{joh13} for a sample of 50 dwarf
galaxies, the mean age of the stellar population contributing to the
FUV emission is strongly dependent on the SFH of a system, ranging
from $\sim$ 50\% of the FUV flux produced by stars younger than 16 Myr
(for $\approx$ 25\% of the sample galaxies) to 50\% of the flux
produced by stars older than 100 Myr (for $\approx$ 20\% of the sample
galaxies). While this may possibly indicate that some of these
galaxies were much less active prior to 60 Myr ago, such a
synchronised enhancement seems unlikely for all these
systems. Moreover, 1) the CMD-based SFHs of \cite{mq15b} offer a
longer look-back time, 2) if we use the average rate over 100 Myr,
which is indeed available in some of our galaxies, we do not find a
significant difference. Extinction could be also important, since the
FUV is extremely sensitive to the total extinction and extinction
law. However, a systematic difference seems to affect the sample
irrespective of the amount of extinction. Binaries are not considered
in the FUV SFR calibration, while our CMD-based SFRs assume a 30\% of
binaries. However, although this may cause a systematic bias, the
overall effect is probably modest (see, e.g.,
\citealt{cigno16}). Another important point concerns the IMF. In
low-mass systems the upper end of the IMF might be not fully sampled
due to the low SFR. However, if stochasticity was important, we should
see the same discrepancy using H$\alpha$, while the distribution of
the ratio between our CMD-based SFRs in the last 10 Myr and the
H$\alpha$ SFRs (shown in Figure \ref{h2} with the same conventions
adopted in Figure \ref{h1}) peaks very close to 1, and only a couple
of galaxies have a ratio greater than three. As a further support to
this point, \cite{mq15b} found good agreement between NUV fluxes (as
predicted by their CMD-based SFHs) and measured NUV-GALEX fluxes,
hence suggesting that stochasticity has a minor role. Another issue
involves the adopted upper mass limit of the IMF (100M$_{\odot}$ for
the FUV and H$\alpha$ SFR calibrations, 300 M$_{\odot}$ for the
CMD-based SFHs) : although this choice has a negligible effect for the
CMD-based SFRs, the impact on the integrated SFRs could be relevant
(for example, lowering the adopted upper mass limit would increase the
mass to light ratio, hence the inferred SFR). Indeed, some authors
have argued that dwarf galaxies have a mass limit (see
e.g. \citealt{wk05}) lower than the populations of the solar vicinity.

Finally, as pointed out by \cite{mq15b}, a likely cause for the
discrepancy could rest in the stellar evolutionary libraries and/or
stellar atmospheric models used to calibrate the FUV SFR
relation. Interestingly, if models are the culprit, the effect does
not seem related to metallicity, since among the most discrepant
dwarfs in the sample there are UGC1249 and UGC5340, which are rather
metal rich and extremely metal poor, respectively.

\section{SF spatial patterns}

\begin{figure}[t]
\centering \includegraphics[width=8.5cm]{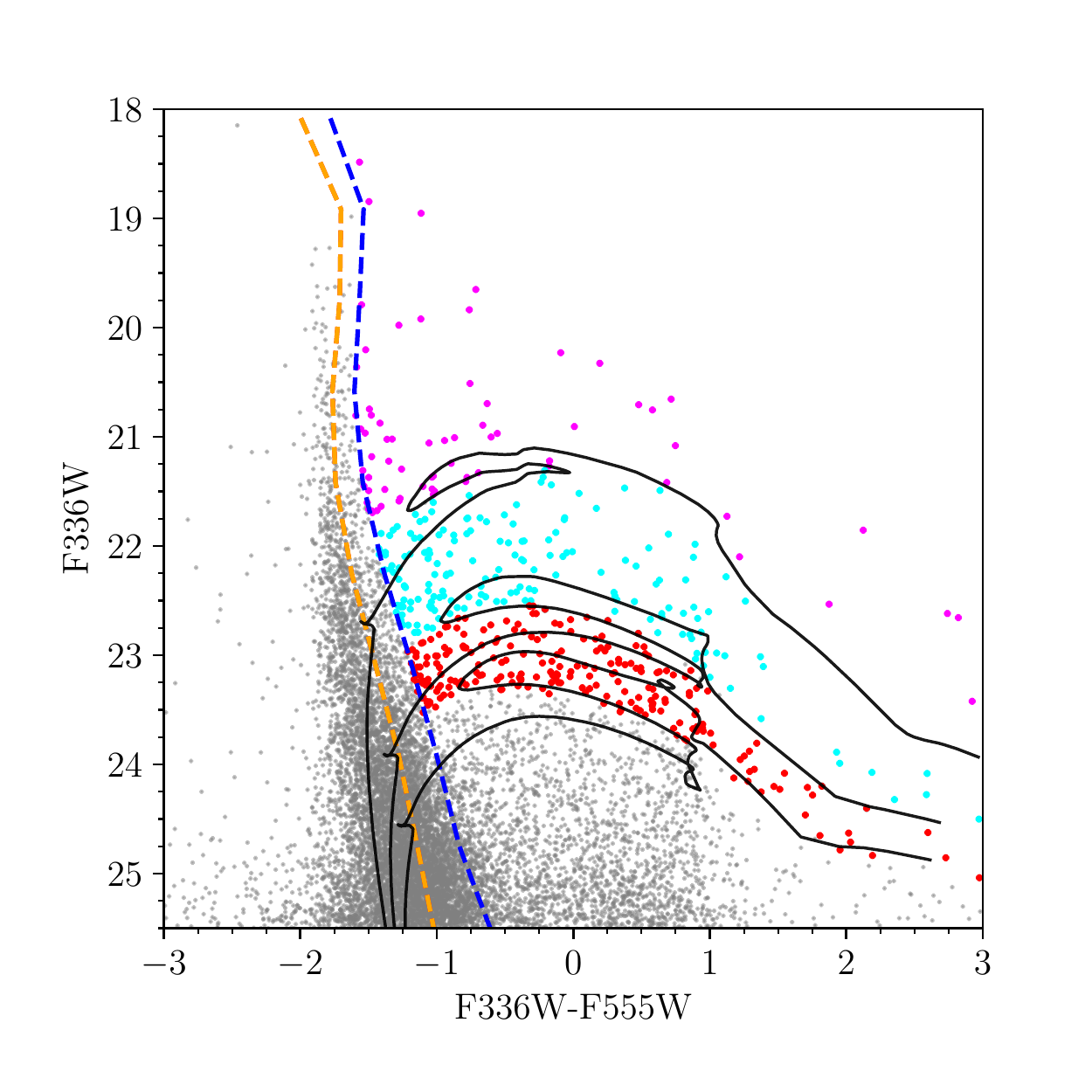}
\caption{HeB selection procedure applied to UGC4305. Red, cyan, and
magenta dots are HeB stars for the age bins $40-60$ Myr, $20-40$ Myr, and
  $0-20$ Myr respectively. The orange and blue dashed lines indicate
  the 50\% and 75\% percentiles of the U-V color distribution. }
\label{clip} 
\end{figure}

The spatial distribution of HeB stars of different ages provides unique
information on how the recent SF has occurred in space and time. In
order to take full advantage of this stellar phase, it is thus
mandatory to select HeB samples as pure as possible. This task is
facilitated in the U vs U-V CMD, since the blue edge of the HeB phase
is well detached from the MS, avoiding confusion between the two
stellar phases unlike in purely optical CMDs. Furthermore, the
temporal resolution is higher than that achievable using MS stars,
because the blue edge of the HeB phase provides a one to one relation
between age and luminosity. On the other hand, there are some
complications due to the more complex morphology of the post-MS
evolution in the U vs U-V CMD (as visible in Fig. \ref{clip}, where
PARSEC-COLIBRI isochrones of the labelled age are overlaid to the CMD
of UGC4305). The HeB phase is not completely horizontal (the red part
is fainter) and its red edge (a mixture of HeB and early AGB stars),
which is promptly visible in the optical CMDs as a red plume above the
RGB tip, is strongly tilted and stretched to the red. Moreover, the
color extension of the theoretical HeB phase is generally too short in
the models, even at the lowest metallicities. Indeed, observations
show a continuity between MS and HeB stars, while a gap between them
is often visible in the models.

For this task we used the PARSEC-COLIBRI isochrones for the
observational metallicities listed in Table 1 of
\cite{calzetti15}. Distance moduli are also taken from the literature
(Table \ref{tab1}), whereas the average extinctions are those
resulting from the synthetic CMD approach.

In order to cope with the mismatch between data and models
for the color of the HeB stars, we opted for the following procedure:

\begin{itemize}
\item we bin the range of F336W magnitudes in intervals of 0.25 mag;
\item for each interval, we choose the value
of the 75\% percentile of the U-V color distribution;
\item we use this value as the corresponding lower bound
of the range of color values allowed for bona-fide HeB stars;
\item we iterate over all magnitude intervals;
\item we finally slice the full HeB sample in terms of age, using the
  PARSEC-COLIBRI HeB models for the assumed spectroscopic metallicity. More
  specifically, since HeB stars of a given age cross twice the
  same color at two different luminosities (the so called ``blue
  loop''), we chose to select ages using the redward excursion (the
  fainter crossing) of each theoretical HeB. 
\end{itemize}
Throughout this paper, we select bona-fide HeB stars from three
different age bins, namely $40-60$ Myr (older epochs are only
accessible for a handful of galaxies in the sample), $20-40$ Myr, and
$0-20$ Myr . Figure \ref{clip} shows an example of such a selection
applied to UGC4305.

We also tested the procedure using larger percentiles (minor chance of
MS contamination) and different models (MESA-MIST). We found that the
major changes regard the statistics of the HeB samples, while the
corresponding spatial distributions are qualitatively similar. Our
samples will be mostly composed of HeB stars at the blue edge, plus minor
contributions of redder HeB (in the UV CMD the central part of the
optical HeB is tilted toward fainter magnitudes and redder colors,
while the red edge is too faint to be detected in the U-band),
sub-giants from the Hertzsprung gap, and MS interlopers (pushed to the
right by differential reddening).

\begin{figure*}[t]
\centering \includegraphics[width=20cm]{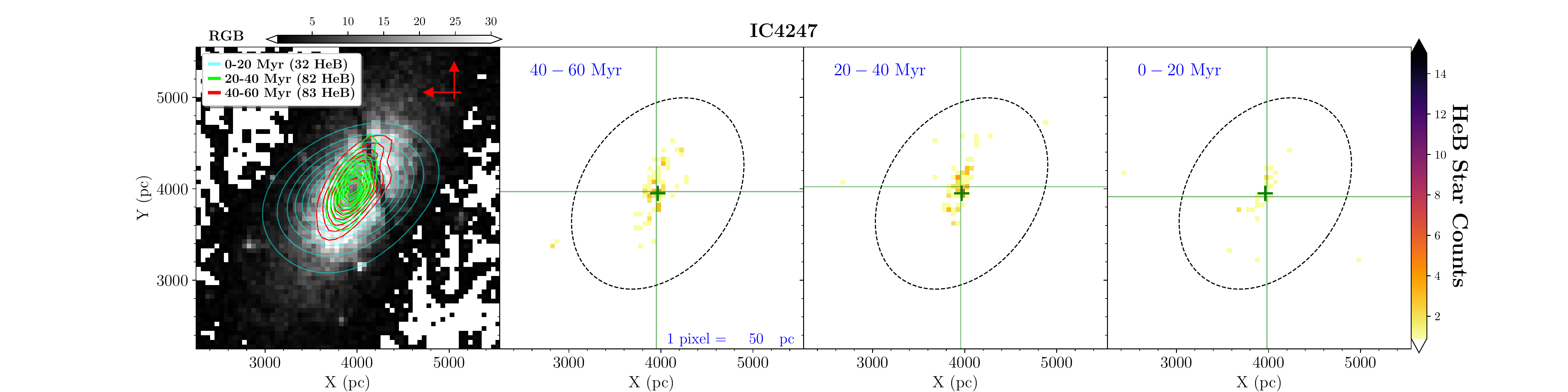}\\
\centering \includegraphics[width=20cm]{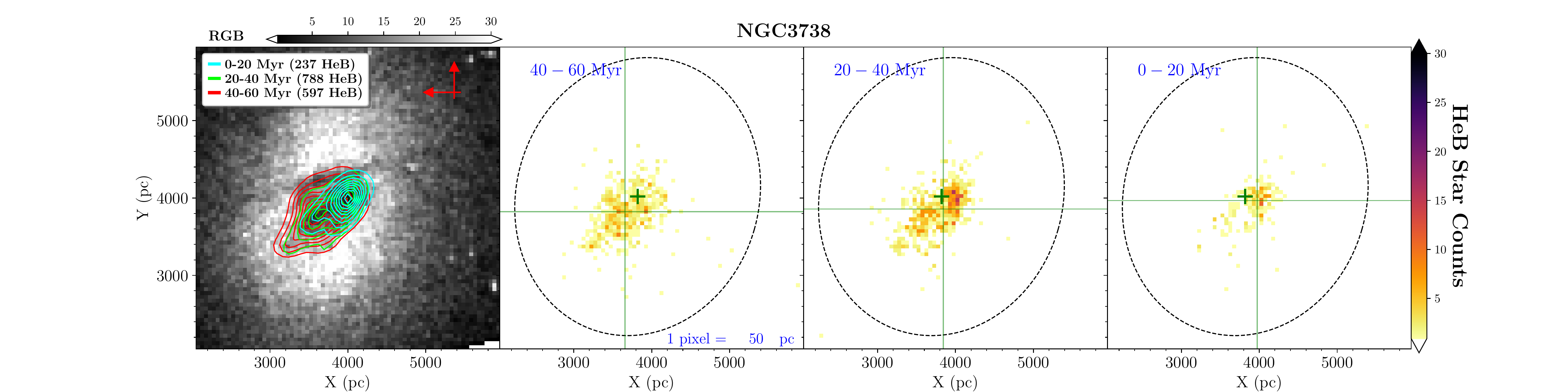}\\
\centering \includegraphics[width=20cm]{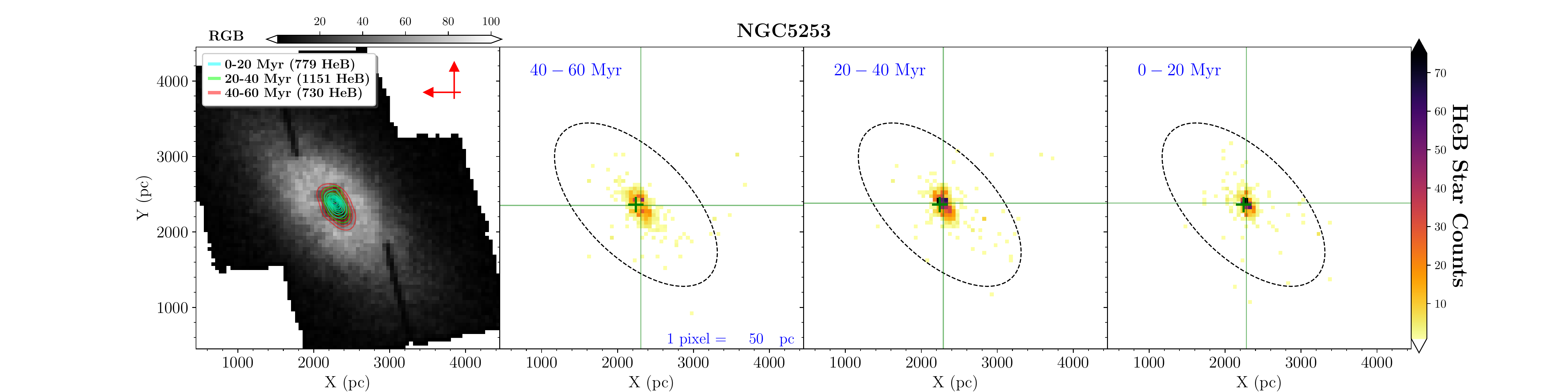}\\
\centering \includegraphics[width=20cm]{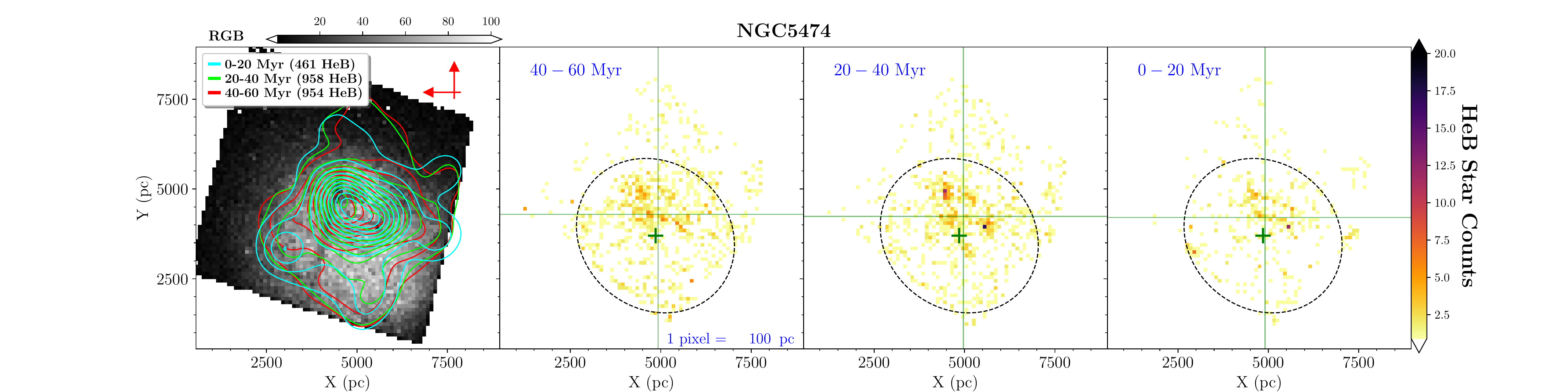}
\caption{From top to bottom, density distributions for the galaxies
  IC4247, NGC3738, NGC5253 and NGC5474. From left to right, the first
  panel of each galaxy shows the RGB density map (with overlaid the
  HeB contour plots estimated with a Gaussian kernel density), while
  the second, third and fourth panels show the density maps of the HeB
  stars with 40-60 Myr, 20-40 Myr and 0-20 Myr, respectively. The
  left-pointing horizontal red arrow shows the East direction, while
  the up-pointing red arrow shows the North direction. The dashed
  ellipses in the HeB density maps show the region where 68\% of the
  RGB stars are contained (assuming that the underlying distribution
  for the RGB stars is a bivariate Gaussian).}
\label{maps1} 
\end{figure*}

\begin{figure*}[t]
\centering \includegraphics[width=20cm]{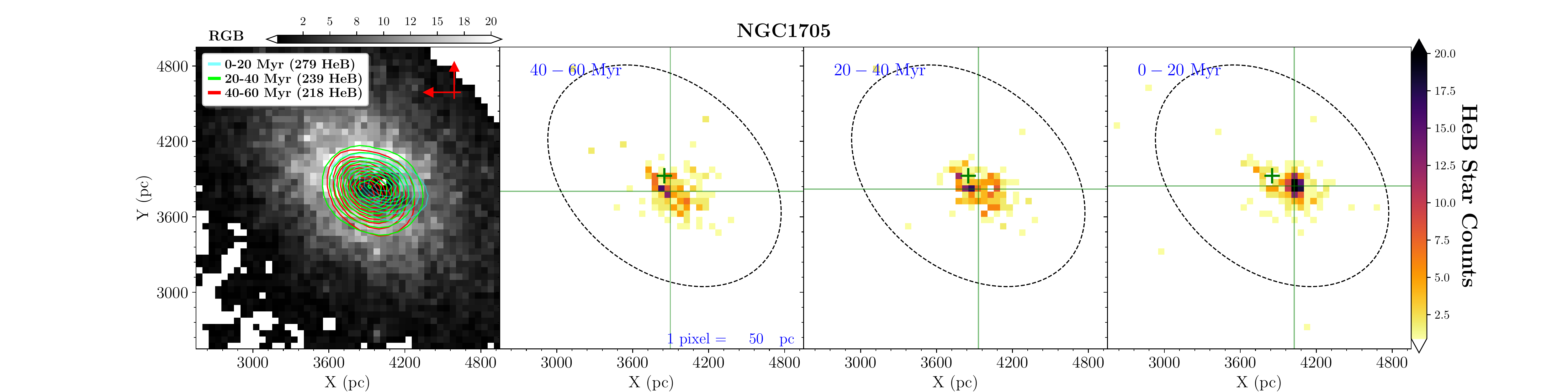}\\
\centering \includegraphics[width=20cm]{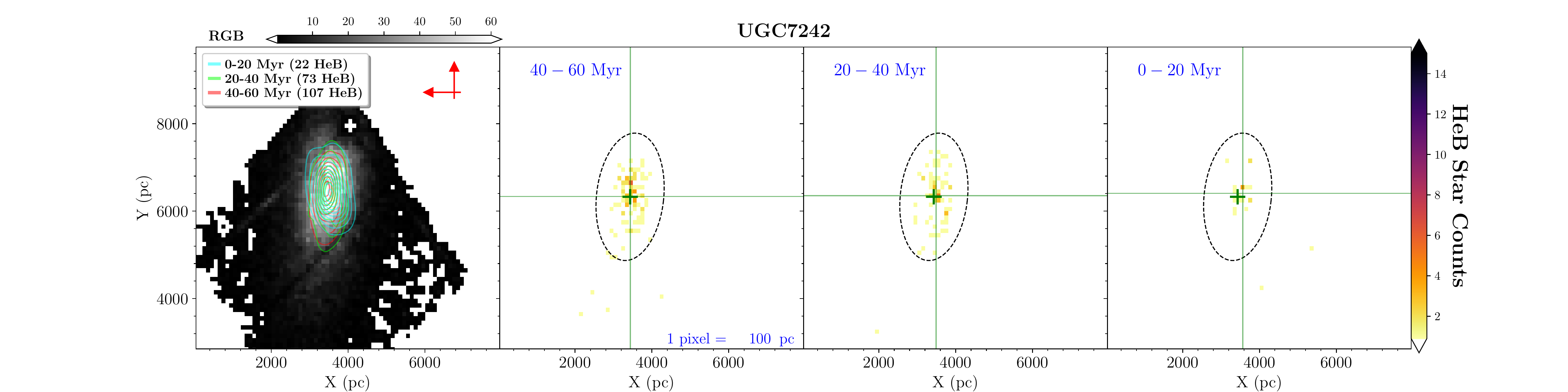}\\
\centering \includegraphics[width=20cm]{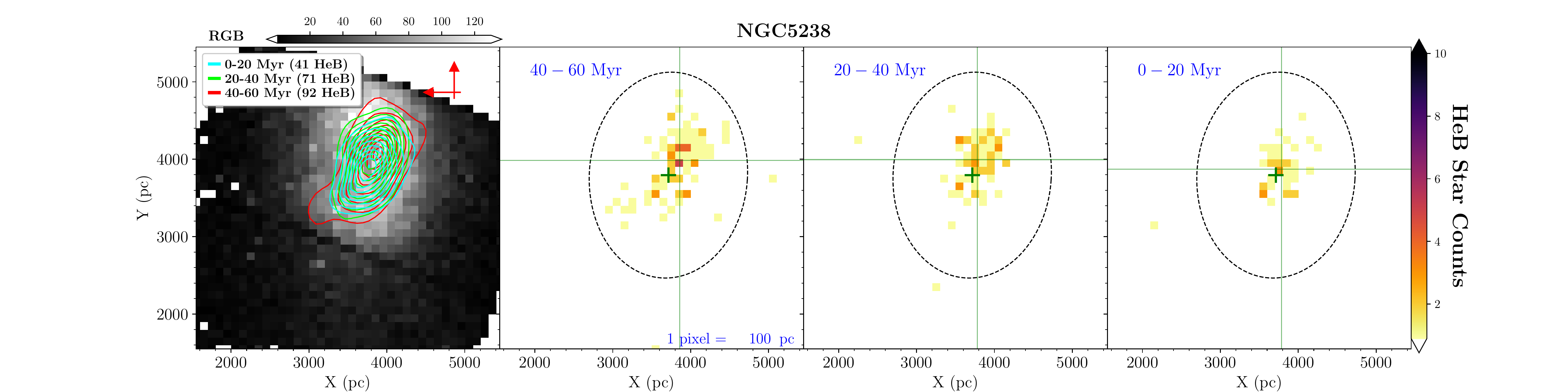}\\
\centering \includegraphics[width=20cm]{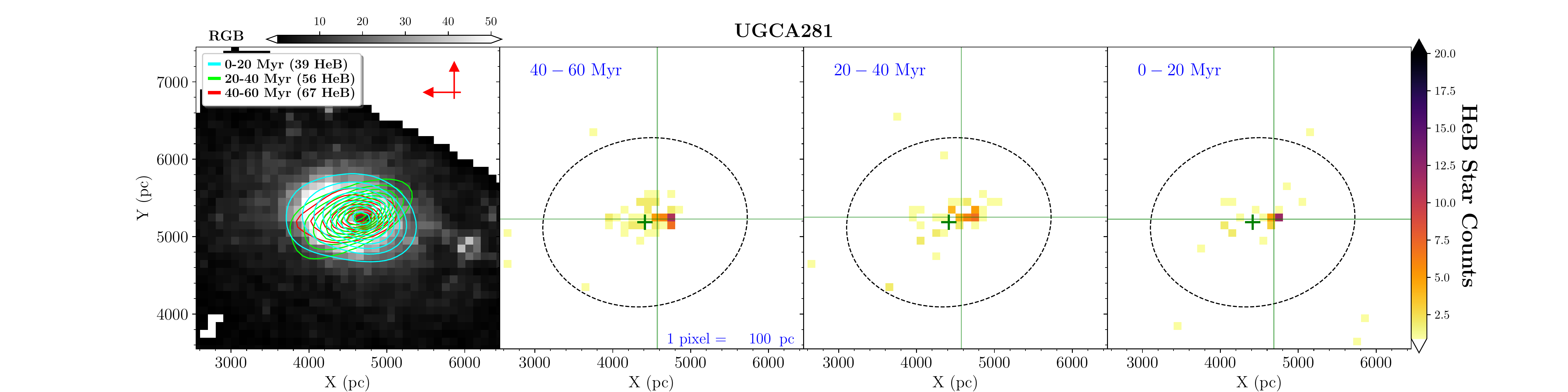}
\caption{Same plot as in Fig. \ref{maps1} for the galaxies NGC1705,
  UGC7242, NGC5238 and UGCA281.  }
\label{maps2} 
\end{figure*}

\begin{figure*}[t]
\centering \includegraphics[width=20cm]{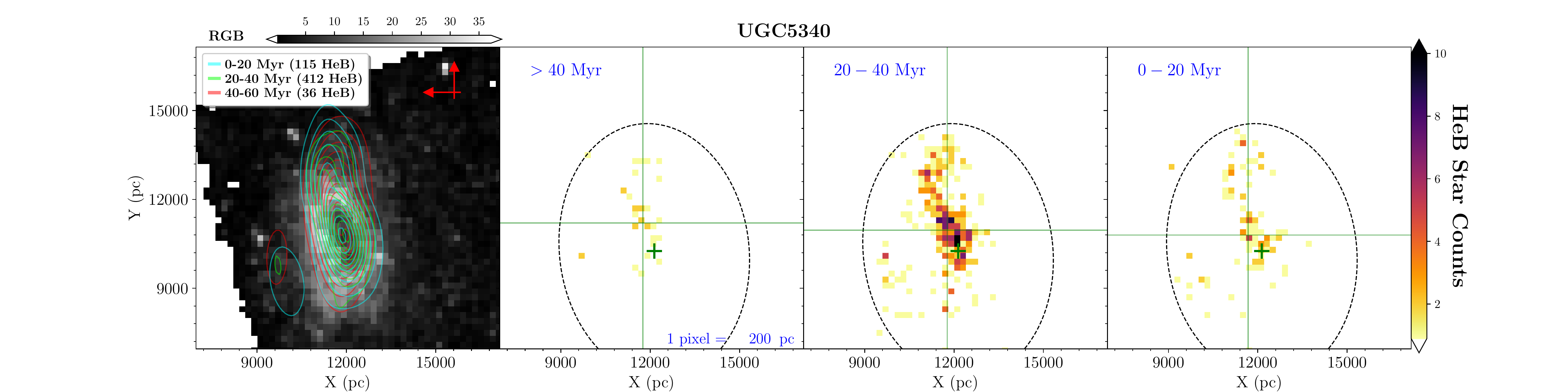}\\
\centering \includegraphics[width=20cm]{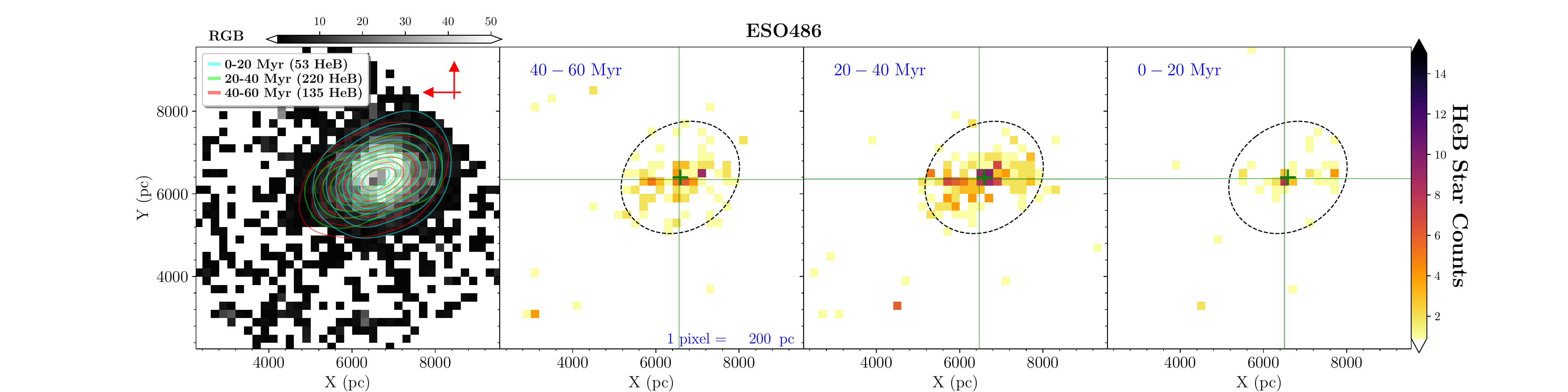}\\
\centering \includegraphics[width=20cm]{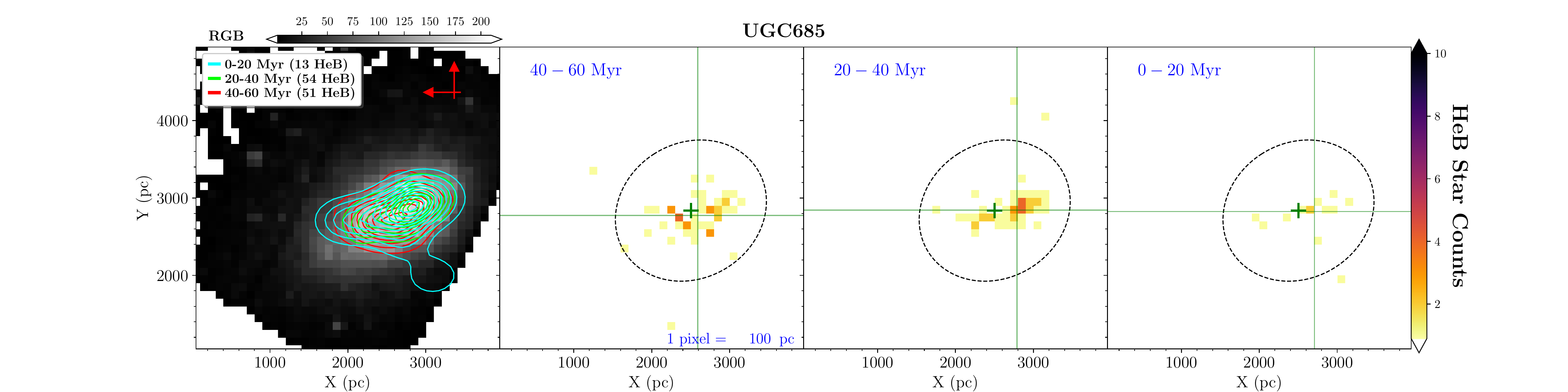}\\
\centering \includegraphics[width=20cm]{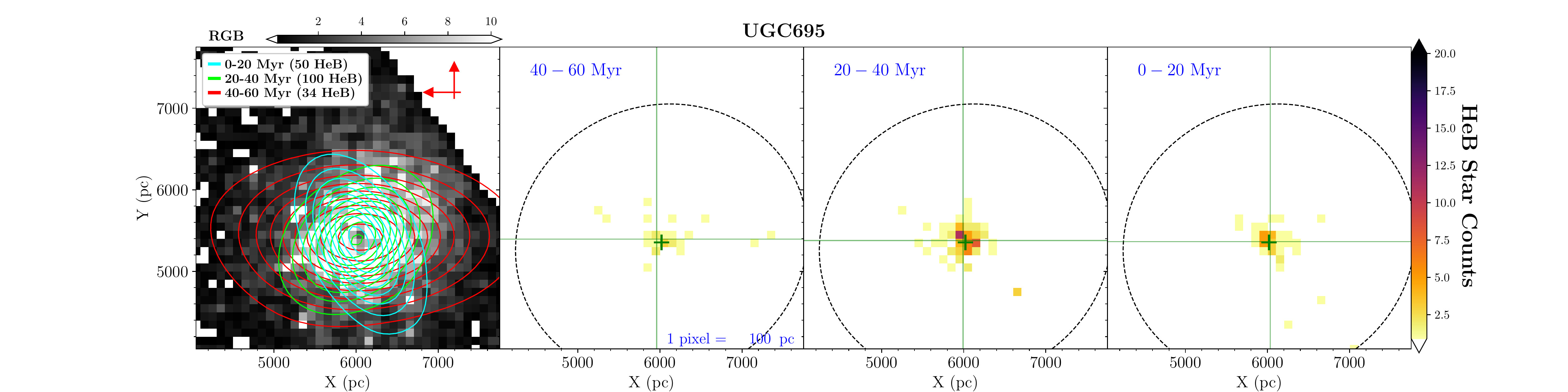}
\caption{Same plot as in Fig. \ref{maps1} for the galaxies UGC5340,
ESO486, UGC685 and UGC695.  }
\label{maps3} 
\end{figure*}

\begin{figure*}[t]
\centering \includegraphics[width=20cm]{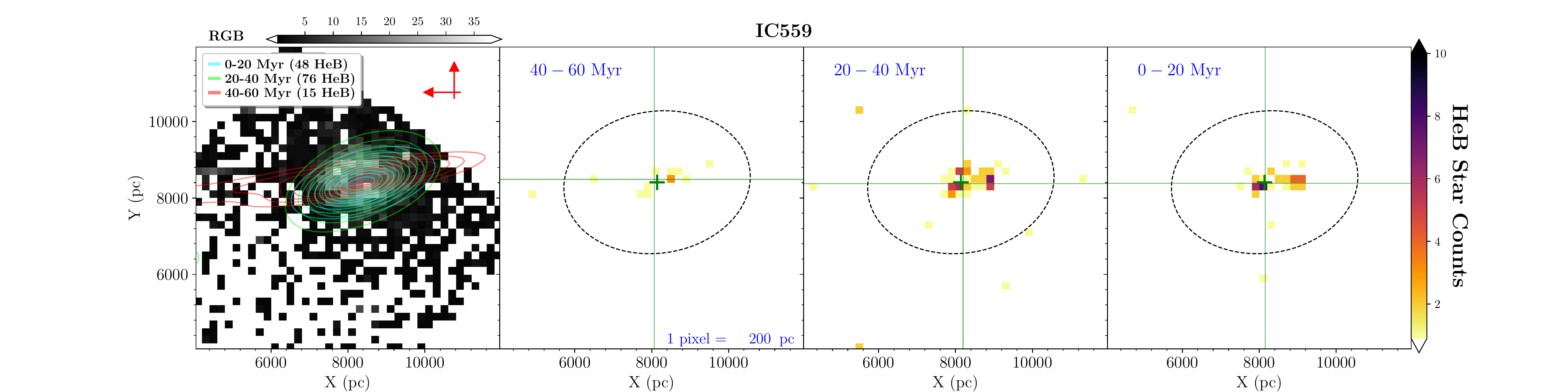}\\
\centering \includegraphics[width=20cm]{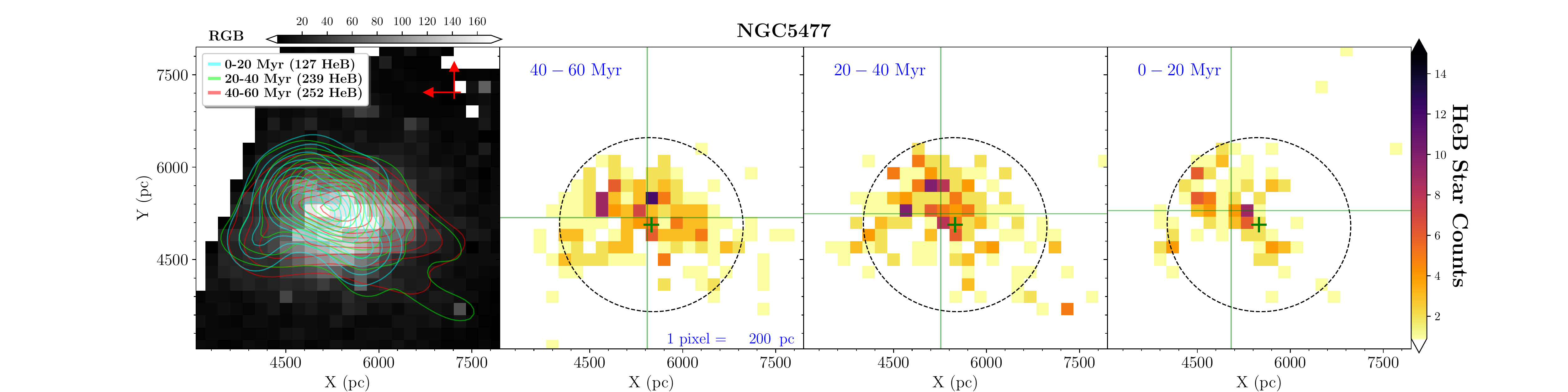}\\
\centering \includegraphics[width=20cm]{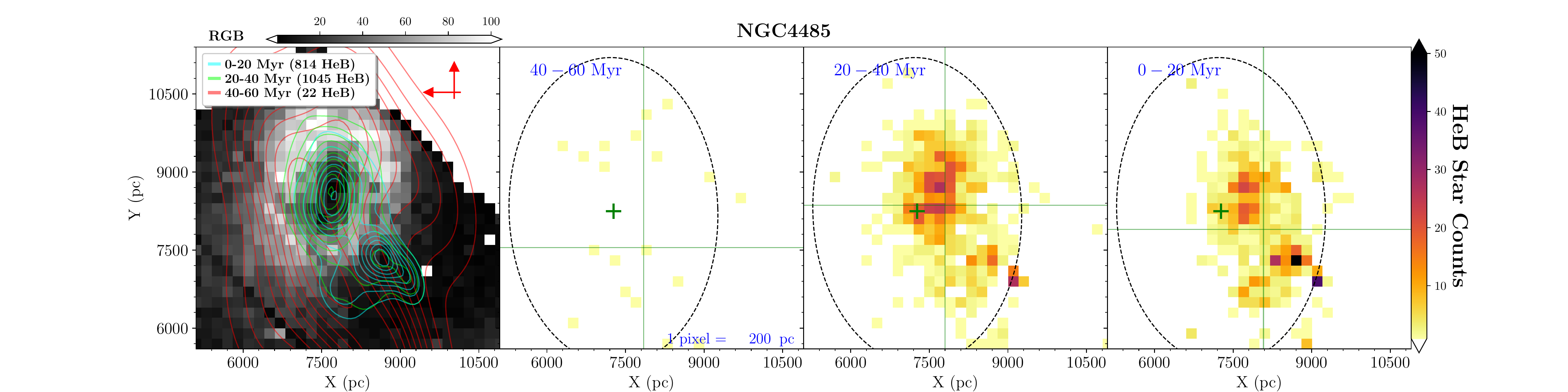}\\
\centering \includegraphics[width=20cm]{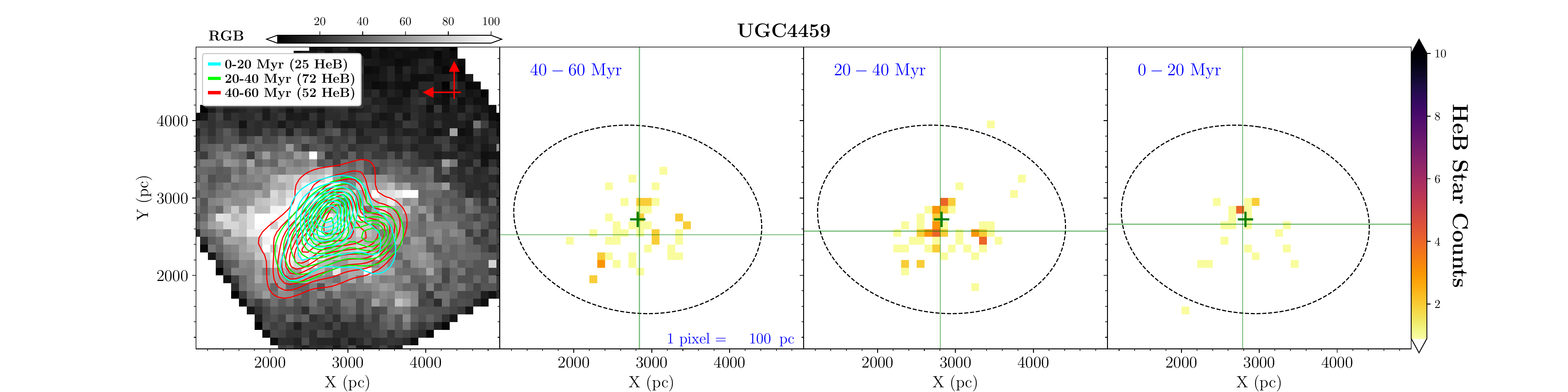}
\caption{Same plot as in Fig. \ref{maps1} for the galaxies IC559,
NGC5477, NGC4485 and UGC4459.}
\label{maps4} 
\end{figure*}

\begin{figure*}[t]
\centering \includegraphics[width=20cm]{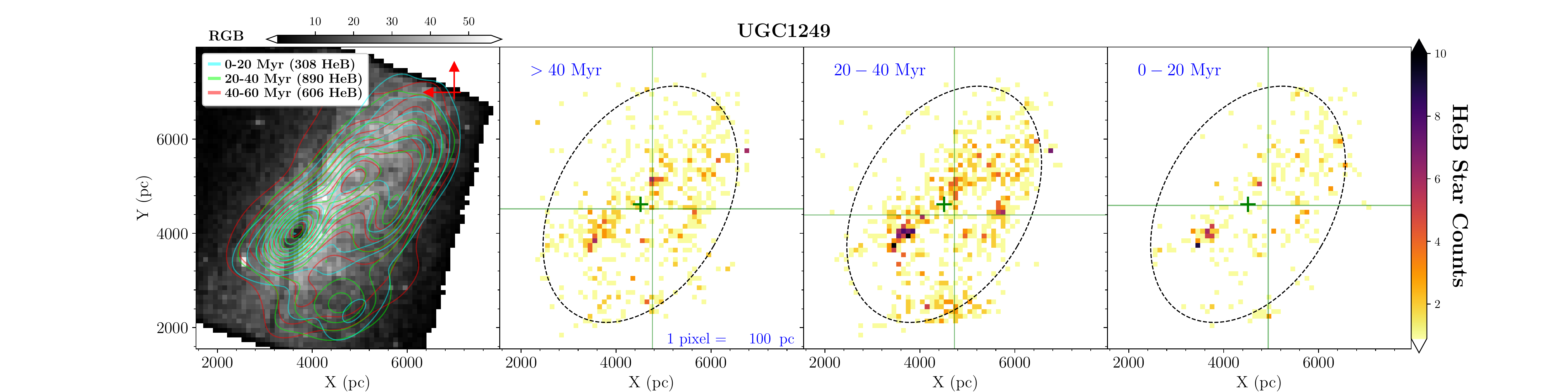}\\
\centering \includegraphics[width=20cm]{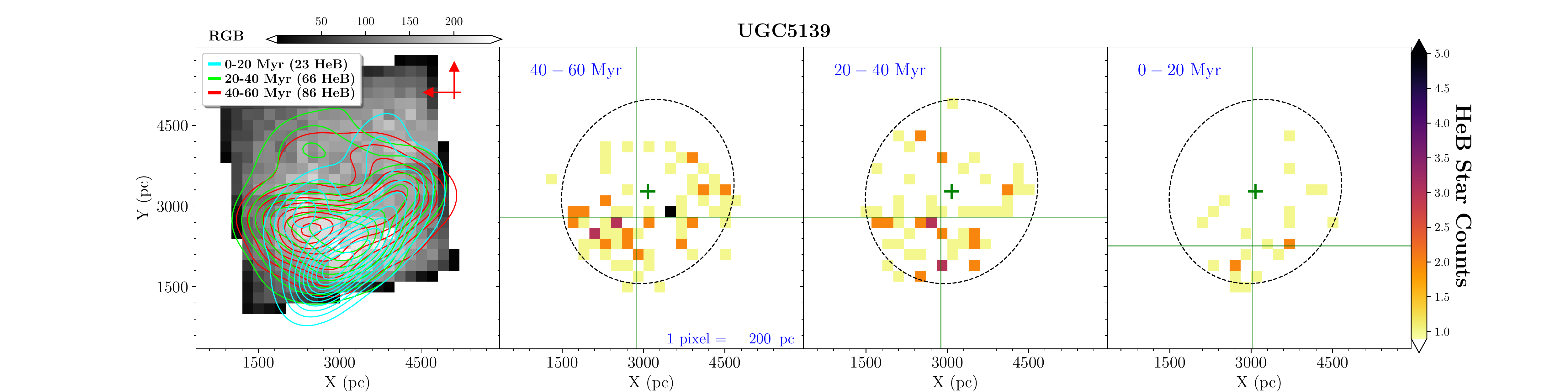}\\
\centering \includegraphics[width=20cm]{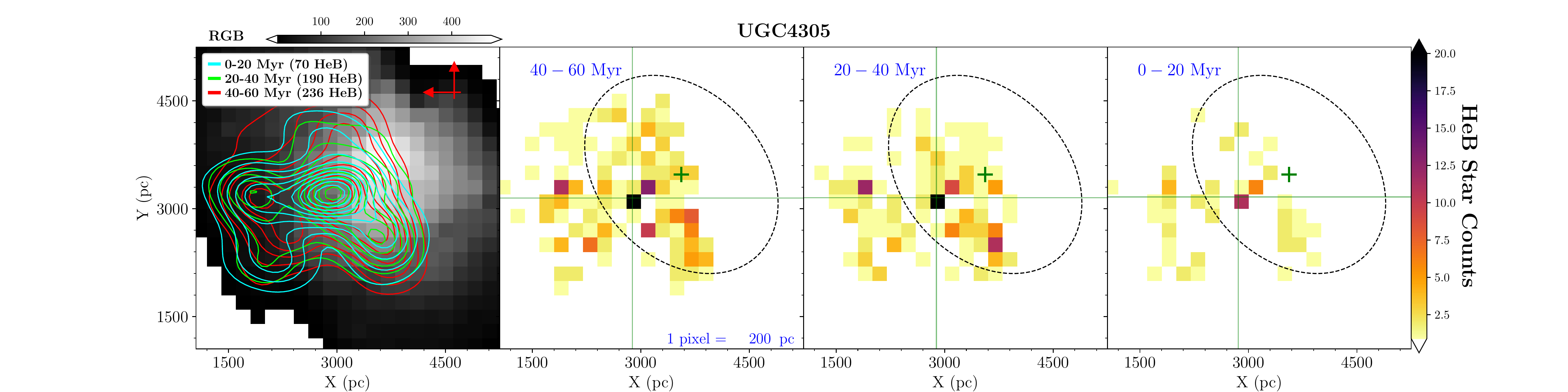}\\
\centering \includegraphics[width=20cm]{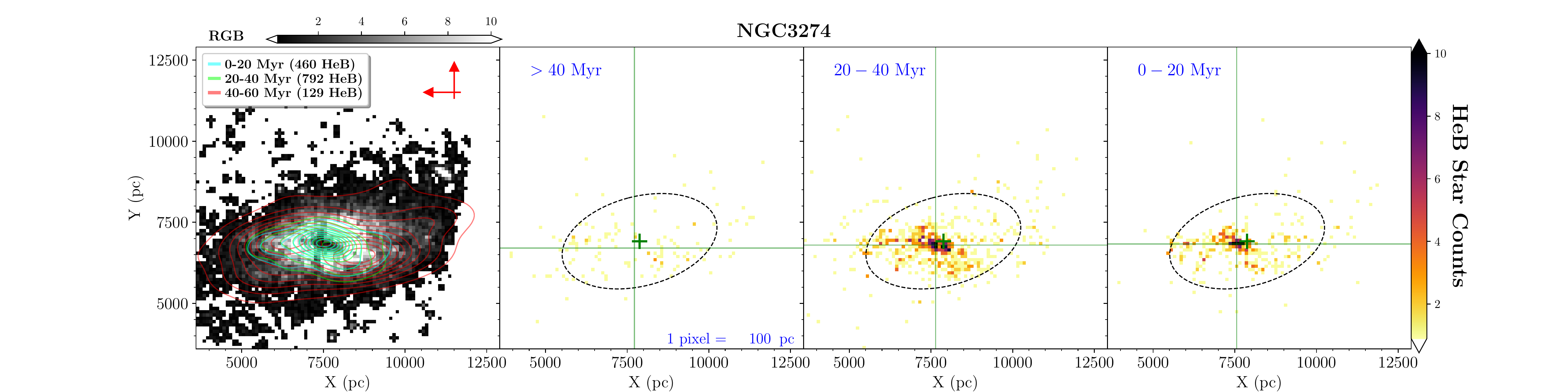}
\caption{Same plot as in Fig. \ref{maps1} for the galaxies UGC1249,
UGC5139, UGC4305 and NGC3274. }
\label{maps5} 
\end{figure*}

\begin{figure*}[t]
\centering \includegraphics[width=20cm]{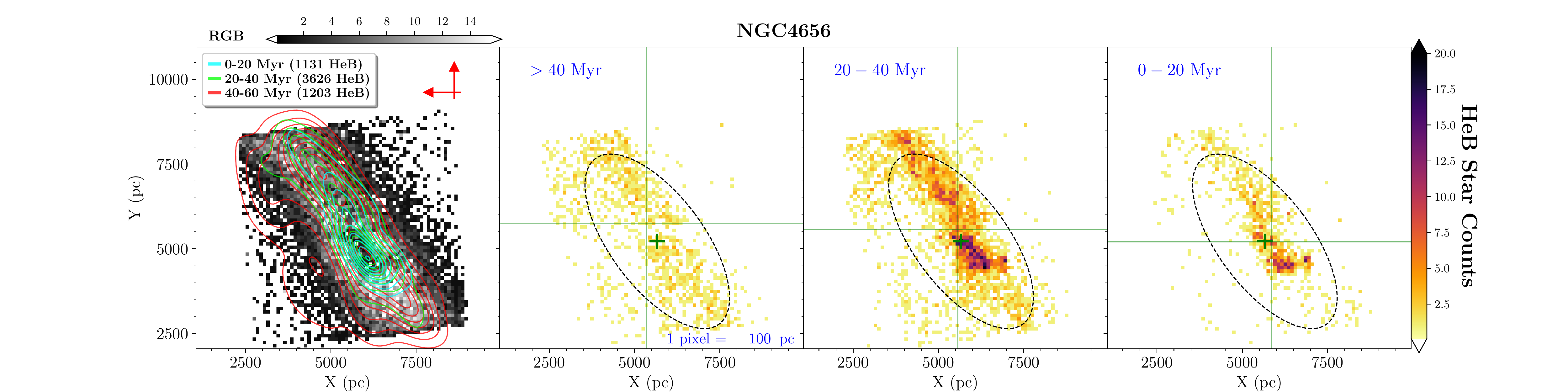}\\
\centering \includegraphics[width=20cm]{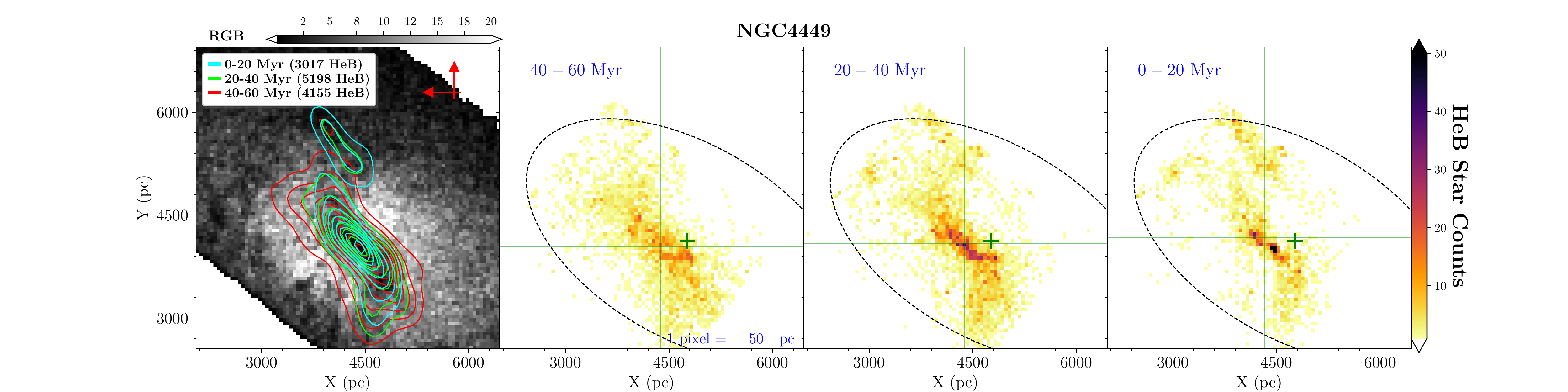}\\
\centering \includegraphics[width=20cm]{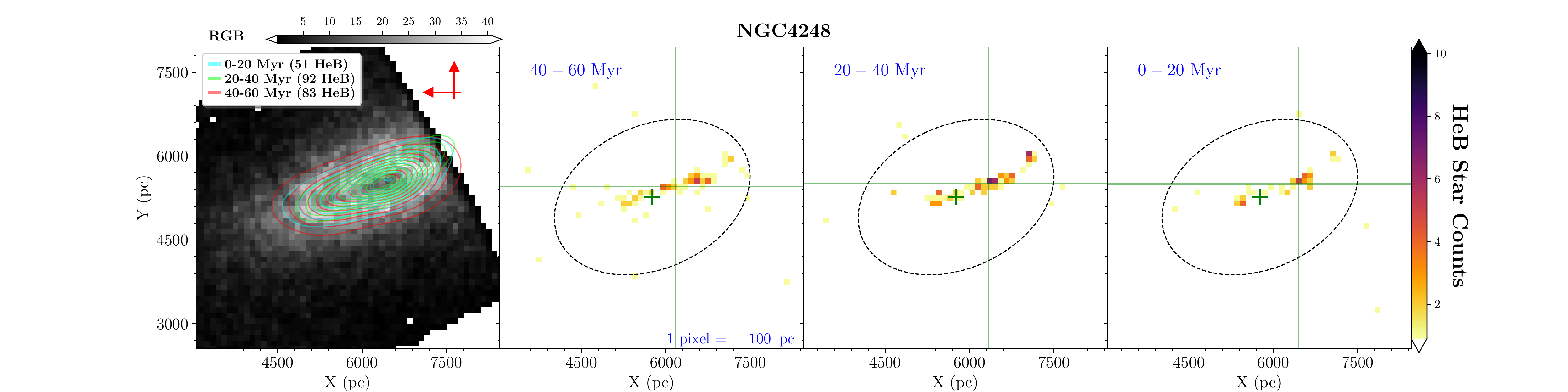}
\caption{Same plot as in Fig. \ref{maps1} for the galaxies NGC4656,
NGC4449, NGC4248. }
\label{maps6} 
\end{figure*}

The HeB spatial distributions for all LEGUS SFDs are shown in
Fig.\ref{maps1}, \ref{maps2}, \ref{maps3}, \ref{maps4}, \ref{maps5},
\ref{maps6}.  For comparison, we also plotted the distribution of RGB
stars as taken from the optical CMDs (thus including also ACS archival
fields, \citealt{sabbi18}). In order to minimize incompleteness
effects, only RGB stars down to 2 mag fainter than the RGB tip are
used. In contrast to the HeB stars, these objects measure the
average activity prior to 1 Gyr ago, so they trace the stellar mass
cumulated in time and space by our galaxies.

For each galaxy, the leftmost panel shows the RGB distribution, while
the HeB distributions for the three age bins are shown in the second,
third. and fourth panel, respectively. To facilitate the comparison,
we also overplot on the RGB distribution a Gaussian kernel estimation
of the HeB stars from the three age bins (red is the oldest, cyan the
youngest; the contour levels are uniformly chosen between 0 and the
highest density point). The dashed ellipses in the HeB density maps
show the region where 68\% of the RGB stars are contained (assuming
that the underlying distribution for the RGB stars is a bivariate
Gaussian), while the little green crosses are their
centroids. Finally, the long green dotted lines locate the centroids
of the HeB stars of the labelled age bin.

It is worth emphasize that the distribution of HeB stars of different
ages reflects the combination of SFR and IMF, since stars of
progressively brighter magnitudes are not only younger, but also more
massive. The same is true for the RGB, which is generally composed of
low mass stars, which are far more frequent (given an IMF) than
intermediate and massive HeB stars younger than 50 Myr. Moreover,
while HeB stars of the age bins 0-20 Myr and 20-40 Myr are not
severely affected by incompleteness in any of our galaxies, the age
bin 40-60 Myr and the RGB stars are frequently lost in very crowded
(and generally young) regions, giving the false impression of a lower
SF activity. For these reasons, the ratio of star counts from
different stellar species should not be taken as a measure of their
SFR ratio. For the same reason, the centroid of the oldest HeB
distribution can be biased, while the RGB centroid is generally less
affected since only minor areas are plagued by significant
incompleteness. Finally, we warn that centroids and 68\% ellipses can
be of limited use in those galaxies that are only partially covered by
LEGUS observations, since boundary effects can have a significant
impact on the geometry of the galaxy. Moreover, many dwarfs in general
have irregular shapes and clumpy structures.

Despite these caveats, the age uncertainty (due to photometric errors)
of HeB stars is generally low, $\sim 20$\% at 20 Myr (higher in very
reddened galaxies), so the distribution of HeB stars from different
epochs reliably traces the direction where the SF is proceeding.

A first inspection of these figures reveals intriguing features and
some similarities:

\begin{itemize}
\item RGB stars generally occupy broader regions and their
  distribution is smoother. This is a non-surprising consequence of
  genuine in-situ star formation combined with secular dynamical
  evolution (due to internal processes induced by a galaxy's spiral
  arms, bars, galactic winds, and dark matter) and enviromental
  effects.

\item The HeB stars are mainly distributed in clumps, with the
  youngest being the most clumped. Clumps of different age can be
  concentrated in the same place or be completely detached from each
  other. From a dynamical point of view, our HeB stars are younger
  than the estimated dynamical age of the large-scale interaction
  between the host galaxy and its neighbors, suggesting that these HeB
  clumps are not simply detached from the main body of the host galaxy
  but instead formed in situ.

\item With respect to the RGB population, the centroid and dispersion
  of the HeB spatial distributions varies from galaxy to galaxy. There
  are galaxies where the HeB clumps coincide with the central
  concentration of RGB stars, and others where the HeB clumps are
  completely detached with or without a significant RGB population
  counterpart. In most cases the HeB stars are far more concentrated
  than the RGB stars, whereas in a few cases they are equally
  extended.
\end{itemize}

In order to qualitatively classify these features we assign a short
code of three capital letters to our SFDs. Depending on how the HeB
clumps are arranged with respect to one another, we assign a first
letter ``S'' (similar distributions) or ``O'' (if an offset is
present). The second and third letters describe how the overall HeB
population is arranged relatively to the RGB population. If the
centroids' distance of the two species is shorter than 250 pc (in
order to avoid completeness biases, HeB stars older than 40 Myr were
not included while calculating the HeB centroid), we assign a letter
``S'' (similar centroids), otherwise we assign a letter ``O''
(offset). The third letter refers to the spatial dispersion of the
clumps relatively to the dispersion of the RGB population. If clumps
are far more concentrated than the RGB stars we assign a letter ``C''
(concentrated), otherwise we assign a letter ``D'' (diffuse). Finally,
we add the suffix ``t'' if the HeB distributions are twisted with
respect to the RGB one.

Figure \ref{maps1} shows, from top to bottom, the spatial maps for the
galaxies IC4247, NGC3738, NGC5253 and NGC5474.

{\bf IC4247}: As listed in NED\footnote{The NASA Extragalactic
  Database.}, IC4247 is likely a small spiral, part of the Centaurus A
group (see e.g. \citealt{banks99}). Following our notation, IC4247 can
be classified as SSC, since most of the HeB clumps are nested and
mainly located in the central parts of the RGB distribution. However,
the oldest HeB stars shows also a hint of wings that are reminiscent
of a spiral structure. Such structures are not visible in the RGB
stars, suggesting that they may not be density waves but star forming
regions stretched by differential rotation.

{\bf NGC3738}: Listed as Im in NED, it is roughly located in the Canes
Venatici I group (\citealt{kara03}). According to \cite{he04} the
closest companion is NGC4068, at a distance of 490 kpc. Despite this
isolation, the HI component of NGC 3738 is morphologically and
kinematically disturbed, with a kinematically distinct gas cloud in
the line of sight of the HI disk. They suggest that NGC3738 is the
result of an advanced merger or ram pressure stripping. From this
point of view, the lack of any recent burst in our SFH for NGC3738 may
indicate that the merging event is older than 100 Myr. In our notation
NGC3738 can be classified as OSC, since most of its HeB stars are in
the central part of the RGB distribution, with no significant offset
in their centroids, and the HeB stars systematically progress towards
the North-West (by about half a kpc) as their age
decreases. \cite{hunter98} found a similar dichotomy, with roughly
half of the inner part of the optical galaxy undergoing an intense
star formation episode, with pressure and gas density enhanced by
30\%$–$70\% with respect to the other side. They also found that the
HI velocity fields exhibit significant deviations from ordered
rotation and there are large regions of high-velocity dispersion,
suggesting that such larger-scale conditions could have resulted from
the merger of two SFDs.

{\bf NGC5253}: Listed as Im in NED, NGC5253 is a member of the
Centaurus A group. The H$\alpha$ images
(\citealt{martin98,cal04,meu06}) show multiple filamentary and
bubble-like structures perpendicular to the optical major axis and
extending beyond the stellar distribution, in contrast with the outer
optical isophotes that resemble an elliptical galaxy. The behaviour of
the neutral gas within NGC 5253 is also very complex. \cite{lelli14}
modelled the HI emission with a disk dominated by radial motions and
derived an inflow/outflow timescale of $\sim 100-200$ Myr (consistent
with the starburst time scale inferred by \citealt{mq10}). In addition,
shadowing of the diffuse X-ray emission by the cooler disk gas
(\citealt{ott05}) may suggest that the radial motions are an inflow. A
similar conclusion was reached by \cite{miura18} analyzing CO(2–1)
observations taken with ALMA. \cite{ls12} analyzed deep HI data
proposing that the very peculiar HI morphology and kinematics of NGC
5253 could be explained by an interaction scenario.

In our classification NGC5253 is a SSCt, although a small fraction of
HeB stars are also located very far from the central region. Centroids
of HeB stars coincide with the centroid of the RGB population. In
general, the HeB and RGB distributions share elliptical shapes, but
the major axes of the HeB stars are clearly twisted compared to the
major axis of the RGB stars. Clearly, most of the last 50 Myr activity
of NGC5253 is taking place in the central part of the galaxy, possibly
triggered by gas falling in.

{\bf NGC 5474}: Classified as SAcd in NED, this galaxy is part of the
M101 Group, 90 kpc to the south of the grand design spiral M101. Deep
21-cm mapping shows gas between the two galaxies at intermediate
velocities (\citealt{mihos12}), a possible feature of a close passage
of the two. The most prominent feature of NGC 5474 is its bulge, which
exhibits a significant offset from the disk of the galaxy
(\citealt{hh79,korn98}). At odds with this asymmetry, the HI velocity
field indicates normal differential rotation (\citealt{rd94}),
with a distribution following that of the optical disk (and
significantly warped beyond it).

In our classification it is an SOD. Indeed, all young generations are
similarly distributed, significantly offset, and similarly diffuse
compared to the RGB distribution. In addition, young stars have a
complex distribution characterised by a major central concentration and
a diffuse component. Interestingly, the central component shows three
large substructures where the SF process has been continuous in the
last 60 Myr (or more).

Figure \ref{maps2} shows from top to bottom the galaxies NGC1705,
UGC7242, NGC5238 and UGCA281. 

{\bf NGC1705}: Classified as SA0/BCG in NED, this galaxy is rather
isolated, with its closest neighbour at more than 500 kpc (LSBG
F157-089). Its HI disk (\citealt{meurer98}) is warped and
significantly offset compared to the stellar component. Moreover,
\cite{meurer92} studied the H$\alpha$ emission line kinematics,
detecting the presence of a kpc-scale expanding super-shell of ionised
gas centered on the central nucleus with a blue-shifted emission
component at 540 km s$^{-1}$, most likely a galactic wind powered by
SN explosions from its super star cluster \citep[see
also][]{annibali03}.

As already discussed in Paper I, NGC 1705 shows a slightly declining
activity for the past 100 Myr. This behavior changed drastically 10
Myr ago, when the SFR increased by a factor of two or more over the
100 Myr averaged SFR. After the peak, the SFR remained constant and no
drop is detected.

NGC1705 can be classified as OSC. Most of the recent activity is
confined to the inner part of the RGB ellipse, 100-200 pc offset from
the RGB centroid, with a striking pattern of SF progressing from East
to West. Most of the oldest HeB stars are in a filamentary structure
near the RGB centroid, whereas the youngest HeBs are highly
concentrated in the super star cluster to the West
of the RGB centroid. Despite this recent concentration, some SF is
also continuing in the previously most active region.

{\bf UGC7242}: Classified as Scd in NED, this galaxy is located in the
eastern part of the M81 group. The distribution of HeB stars exhibits
two major concentrations: a prominent central clump and a minor spur
in the South direction. The latter feature tends to disappear moving
at young ages, but low-number statistics prevent us from robustly
exploring this effect. Overall HeB stars cover a large fraction of the
RGB ellipse and no significant offset is found. In our scheme it can
be classified as SOC.

{\bf NGC5238}: Classified as SABdm in NED, this galaxy belongs to the
Canes Venatici I group. From the HI perspective, \cite{cannon16} found
that the HI disk is asymmetric in the outer regions, and the HI
surface density maximum is not coincident with the central optical
peak, but rather is offset to the North-East by $\sim 300$ pc. Our
maps suggest that the iso-contours of the young population are also
twisted with respect to the RGB ellipse. All our HeB samples have
their centroids offset compared to the RGB one and cover a significant
fraction of the RGB ellipse. In our classification NGC5238 is SSCt,
although a South-East spur that is clearly visible among the older HeB
stars tends to disappear in the youngest HeBs.

{\bf UGCA281}: Classified as Sm in NED, this galaxy is isolated and
located in the Canes Venatici I group. \cite{vt83} explored the HI
distribution and discovered a core-halo structure, with the visible
star-forming regions located near that core, but slightly shifted with
respect to the peak in HI. From Fig. \ref{maps2} it is visible that
the HeB distributions show a major clump and a significant diffuse
component, with the former slightly offset compared to the RGB
centroid (note that the lack of RGB stars near the HeB major
concentration is due to photometric incompleteness). Overall, UGCA281
can be classified as SSC.

Figure \ref{maps3} shows from top to bottom the galaxies UGC5340,
ESO486, UGC685, and UGC695. 

{\bf UGC5340}: Classified as Im in NED and located in the Lynx Cancer
void (\citealt{pt11}), this galaxy exhibits a very distorted
morphology, with a cometary tail (pointing to the South in our maps)
populated by stars of all ages (\citealt{tik14a,sacchi16}) and
particularly rich in H II regions, which extend from the main body of
the galaxy for a projected length of 5 kpc, and is most likely an
accreted secondary body (\citealt{tik14a,sacchi16, annibali16}). This
galaxy is clearly interacting. Indeed, combining HST/ACS and LBT/LBC
images, \citet{sacchi16} and \citet{annibali16} discovered another
substructure/satellite. Being at 12.7 Mpc, UGC5340 is among the most
distant galaxies in our survey, hence the oldest HeB bin is highly
incomplete (as well as the central portion of the RGB spatial
distribution). Although the centroids of HeB stars have similar
coordinates, it is clear that 20-40 Myr ago the galaxy was much more
active in the main body than in the cometary tail, while in the last
20 Myr the main body and the cometary tail appear similarly
populated. Compared to RGBs, HeBs centroids are significantly offset
and their overall distribution much more elongated. In particular, no
significant RGB counterpart is visible in the tail, whereas the
northern side shows the opposite. A significant fraction of the 68\%
ellipse is occupied by HeB stars, suggesting that current SF is
globally involving this galaxy. In conclusion, we suggest a type OOD
for UGC5340.

{\bf ESO486-G021}: This galaxy is classified as a likely spiral in
NED. In our map the HeBs follow an elongated structure extending from
the SE to the NW, with no significant differences in the the three age
bins. Avoiding the oldest bin of HeBs, because poorly populated,
centroids of the younger HeBs match well the RGB's centroid. HeB stars
tend to fill the RGB 68\% ellipse with several of them being located
outside. In conclusion, our classification is SSD.

{\bf UGC685}: Classified as SAm in NED, UGC685 is quite isolated in
space. Located at 1.3 Mpc from the centroid of the 17+6 Association
(\citealt{tully06}), it hosts a few HII regions that are all
concentrated to the South-East of the center of the galaxy
(\citealt{hopp99}).  Using NIR imaging data, \cite{hopp99} found
little signs of irregularities, while the HI distribution is much more
extended than the optical (stellar) light distribution (see also
\citealt{roy11}).

In our HeB maps, UGC685 appears as an arc whose centroid is only
slightly offset from the RGB one. We also see a clear progression of
SF, with the main overdensity of HeB stars moving from East to
West. RGBs cover a larger field of view than HeBs. We classify this
galaxy as OSC.

{\bf UGC695}: UGC695 is an Sc type galaxy in the Eridanus Void. Most
of its current SF shows no trends as a function of time and is
concentrated in a single clump, roughly coincident with the RGB
centroid. The RGB stars are far more diffuse. Our classification for
this system is therefore SSC.

Figure \ref{maps4} shows from top to bottom the galaxies IC559,
NGC5477, NGC4485, and UGC4459.

{\bf IC559}: Classified as Sc, this galaxy is located in the
Lynx-Cancer Void. The paucity of HeB stars makes our classification
difficult. Overall, the HeB centroids coincide with that of the RGB
stars and no SF trend is visible. The RGB distribution is far more
extended than that of the HeBs. We suggest a classification SSC.

{\bf NGC5477}: Classified as SAm in NED, NGC 5477 belongs to the M101
group. Our maps indicate that HeBs extend across much of the RGB 68\%
ellipse, without showing any age-related change in their overall
distribution. Compared to the RGB centroid, the HeB centroids are
slightly offset towards the North-East, hence opposite to the
direction of M101, where, instead, a possible connecting bridge seems
to exist.  All together, these features suggest a classification SOD.

{\bf NGC 4485}: Classified as IBm in NED, this galaxy forms with
NGC4490 an isolated bright pair of interacting galaxies.  With stellar
masses of, respectively, $0.82\times 10^9$ vs $7.2\times 10^9$
M$_{\odot}$ (\citealt{pearson18}), and a projected distance of 7.7 kpc
(\citealt{elme98}), they represent a more massive analogue of the
Magellanic Clouds. Both galaxies show signs of tidal disruption, as
suggested by the presence of an extended HI envelope (about 50 kpc in
projection; \citealt{huc80}) surrounding the system, a dense bridge of
gas connecting the pair, and most of the ongoing star formation taking
place primarily between the two galaxies (as indicated by H$\alpha$
imaging; \citealt{thr89}). \cite{pearson18} modelled the system using
N-body and test-particle simulations, simultaneously reproducing the
observed present day morphology and kinematics. According to this
study, the tidal forces from NGC 4490 alone are sufficient to match
the observed gas properties as long as NGC4485's spin is prograde with
respect to its orbit and has a high inclination orbit with respect to
NGC 4490. The $\approx$ 50 kpc (projected) envelope would consist of
material from NGC 4485, lost during the first pericentric encounter of
the two galaxies ($\approx$ 1.4 Gyr ago). This suggests that tidal
interactions between two low mass galaxies can push the gas to large
distances, producing a massive envelope of neutral HI, without
invoking stellar feedback (see \citealt{clemens98}) or perturbations
from a massive host (like the Milky Way for the Magellanic Clouds).

In our classification, NGC4485 is OOD. Compared to the RGB stars the
current SF is shifted to the West, and appears to be progressing to
the SW (in the direction of the companion NGC4490), with the most
recent SF event taking place in the periphery of the galaxy, at a
projected distance larger than 1.5 kpc from the 20-40 Myr old
population. This reminds one of the activity of the 30 Doradus region,
the most active place in the LMC, located just north of the eastern
tip of the LMC bar. The activity between NGC4485 and NGC4490 is
probably caused by NGC4490, which is pulling the gas, promoting
SF. Interestingly, our recovered SFH (see Fig. \ref{SFHs2}) does not
show a significant recent SF enhancement, corroborating the conclusion
of \cite{thr89} that the interaction between NGC 4490 and NGC 4485 may
have rearranged the ISM in the two galaxies, causing regions of active
star formation to be re-distributed, but without affecting too much
the rate of star formation averaged over the entire galaxy.

{\bf UGC4459}: Classified as Im in NED, UGC 4459 is a quite isolated
member of the M81 group, with its nearest neighbour UGC 4483 at a
projected distance of about 223 kpc. From the point of view of the gas
component, the HI distribution shows two major peaks
(\citealt{walter07,begum06}), with a large scale velocity gradient
aligned along the line connecting the two
(\citealt{begum06}). Moreover, the south-eastern half of the galaxy is
receding with a rapid change in velocity with galacto-centric
distance, while the approaching (northwestern) half of the galaxy
shows a smoother gradient. \cite{begum06} reject as unlikely the
hypothesis that the peculiar kinematics of the galaxy is due to ram
pressure.

In our maps the RGB stars show a rather asymmetric distribution, with
a major lobe extending in the East direction. Centroids of RGB and HeB
stars are similar, but the highest concentration of HeB stars is
elongated and aligned with the line connecting the HI peaks (along the
direction North West-South East). In our classification UGC4459 is SSC.

As discussed in \cite{weisz08}, asymmetries of RGB stars and isolation
of UGC4459 may indicate a late-stage merger between two fainter
dwarfs. However, the disturbed kinematics and the evidence that the
young populations do not follow the distribution of RGB stars maybe
also point to a recent gravitational disturbance with a minor
satellite that has not been detected yet.


Figure \ref{maps5} shows from top to bottom the galaxies UGC1249,
UGC5139, UGC4305, and NGC3274. 

{\bf UGC1249}: Classified as SBm in NED, UGC1249 is a dwarf spiral
galaxy that is in interaction with the barred spiral NGC 672, located
at a linear projected separation of 16.3 kpc (\citealt{heald11}). The
HI distribution shows a clear connection between the two galaxies,
with the outskirts and interface region more disturbed than optical
region. Moreover, a tidal arm trailing behind NGC 672 also suggests an
ongoing tidal interaction. The pair seems to be in an early stage of
interaction (\citealt{heald11}). \citealt{tik14} studied the
luminosity functions of red supergiants and peripheral AGB stars in
both galaxies, suggesting that simultaneous enhancements of star
formation in the two galaxies occurred in the intervals 20–30 and
450–700 Myr ago.

In our maps the RGB stars show an asymmetric distribution, with a
major concentration (incomplete in its center) and a diffuse
component. The HeB stars are offset with respect to the RGB stars,
with at least three major sub-structures: an elongated structure
extending from North-West to South-East, visible at all ages, two
southern minor clumps, one of which (the southeast one, located
partially outside the 68\% RGB ellipse) is mostly visible in the range
of ages 20-40 Myr. These findings suggest a strong off-center activity
at these epochs, corroborating the result of \cite{tik14}. Overall,
the HeB stars fill much of the RGB ellipse, although the higher
concentration is on one side (south-west). Interestingly, this side
points in the opposite direction with respect to NGC672. Using our
notation, UGC1249 is OOD.

{\bf UGC5139}: Classified as IABm in NED, this galaxy is a member of
the M81 group. \cite{ott01} combined HI maps with UBV(RI)$_c$ and
H$\alpha$ observations, revealing the presence of a supergiant shell
with a diameter of 1.7 kpc covering half of the optical extent. They
suggested that in the past this galaxy could have been a BCD,
and an intense episode of SF and subsequent SN explosions may have
blown the supergiant HI shell. H$\alpha$ emission is predominantly
found in the south–east side of the galaxy.

The paucity of HeB stars and the little spatial coverage of our
observations prevent a firm classification for this galaxy. HeBs
occupy most of the 68\% RGB ellipse. A tentative classification could
be OOD, since the HeB stars are significantly offset compared to the
RGB distribution, and a mild spatial trend with age is also visible
(younger HeBs are progressively shifted to the south-west). Interestingly,
the south–eastern region of UGC5139 points to the M81 triplet (M81,
M82, and NGC3077) and shows a steep gradient in the HI distribution,
whereas the opposite side (north–west) shows systematically higher
velocity dispersions (\citealt{ott01}).  All of these asymmetries may
indicate that ram pressure stripping and/or tidal forces may be at
work in this galaxy.

{\bf UGC4305}: Classified as Im in NED, UGC4305 belongs to the M81
group and is rather isolated. Despite its isolation, the HI
distribution is characterised by numerous holes (\citealt{puche92}),
whose origin has been associated with stellar feedback from massive
stars (\citealt{puche92}) or stellar feedback from multiple
generations of SF spread out over tens or hundreds of Myr
(\citealt{weisz09}). The SFH solutions from our work previously
presented in Paper I are consistent with an almost constant SF
activity over the last 180 Myr, with mild enhancements whose intensity
is at most 1.5 times higher than the 100 Myr averaged SFR.

Compared to the ACS pointings, our WFC3 observations cover only part
of the galaxy, making it difficult to compare RGB and HeB
distributions. However, it is clear from the maps that in the East
direction HeB stars extend well beyond the 68\% RGB ellipse, with no
obvious trend with age. This may suggest a classification of
SOD. Using deep H$\alpha$ images, \cite{egorov16} performed a detailed
analysis of the star forming regions in the supergiant H\textsc{ i}
shell also targeted by our WFC3 field; on the eastern side, where we
find the centroids of HeB stars, they measured a less dense medium
that is allowing a faster expansion of the ionized gas in the center
of the complex. They indeed found 3 faint expanding ionized
superbubbles there, which possibly interact with each other, and 3
superbubbles located at the periphery of the complex.

{\bf NGC3274}: NGC3274 is classified as SABd in NED. To date, no
studies have been performed on the stellar population of this
galaxy. Our maps do not suggest a clear age dependence of the HeB
spatial distributions. The overall structure of the HeB stars younger
than 40 Myr is rather elongated, resembling a bar or a spiral
arm. Compared to RGB stars, centroids of HeB stars are offset,
especially the oldest ones. Our tentative classification is SOD.

Figure \ref{maps6} shows from top to bottom the galaxies NGC4656,
NGC4449 and NGC4248.

{\bf NGC4656}: Classified as SBm in NED, NGC4656 is a distorted
edge-on spiral galaxy in the NGC4631 group. The first studies (see,
e.g., \citealt{dev64,nilson73}) of this galaxy concluded that the
bright ``hook'' shaped northeast end of NGC4656 is a separate galaxy
from NGC4656, hence, some papers still refer to this system as NGC
4656/4657. Moreover, NGC 4656 was thought to be interacting with its
closer neighbour, NGC 4631, an edge-on spiral galaxy at a projected
distance of 71 kpc, as suggested by a bridge of neutral hydrogen
(\citealt{rob68}) extending from NGC4656 to NGC 4631. However, a later
study focusing on surface brightness profiles (\citealt{stay83})
concluded that NGC 4656 and NGC 4657 are part of the same galaxy, with
no obvious feature indicating an interaction with NGC 4631. In this
direction, \cite{md15} discovered a giant stellar tidal stream in the
halo of NGC 4631, partially extending between NGC 4631 and NGC
4656. However, orbital considerations made these authors suggest that
the stream around NGC 4631 is likely due to interactions between this
galaxy and its dwarf satellites. \cite{rand94} noted disturbances in
the HI kinematics, suggesting that NGC 4656 could be a system with two
loosely wrapped tidal arms, or a ring galaxy of the Cartwheel type
seen almost edge-on (which may suggest a collision with a companion
near the center, causing a radially propagating density wave). More
recently, \cite{sch12} discovered a candidate tidal dwarf in the NGC
4656 system, which was dubbed NGC 4656UV. They compared the spectral
energy distribution of all regions of NGC 4656 and NGC 4656UV with a
set of stellar population synthesis models, concluding that the
metallicity of NGC 4656UV is too low ($\sim$ 10 times lower than
that of NGC 4656) to be a tidal dwarf, which are typically metal rich
objects (see, e.g., \citealt{weil03}), and proposed that NGC4656UV
formed out of gas stripped from the outskirts of NGC 4656 during an
encounter between NGC 4656 and NGC 4631. From dynamical arguments,
they estimate the age of this event about 200-300 Myrs
ago. \cite{dem12} found eight UV sources younger than 100 Myr in the
area of NGC4656UV, corroborating the tidal dwarf hypothesis. Finally,
by using Fabry-Perot and GMOS multi-slit data, \cite{mun18} found that
NGC 4656UV has a low metallicity (12+log(O/H)$\sim 8.2$, very similar
to that of NGC 4656), which places this galaxy on the mass metallicity
relation for normal dwarf galaxies (contrary to what is expected for
tidal dwarfs). Moreover, through the analysis of the radial velocity
profiles and by fitting a kinematic model to the observed velocity
field, they confirmed that NGC 4656 is consistent with one single body
instead of two objects (NGC 4656 and NGC 4657). Based on these
results, \cite{mun18} suggested that NGC 4656 and NCG 4656UV are a
pair of interacting galaxies, with NGC4656UV being a normal dwarf not
of tidal origin. Interestingly, their H$\alpha$ SFRs also indicate an
enhancement in the star formation at the north region (the ``hook''),
which is the closest to NGC 4656UV. NGC4656UV is beyond our field of
view, in the north-east direction.

In our maps, the oldest HeB bin is very incomplete, while the younger
HeBs are statistically robust and show interesting features. First, we
clearly see two different concentrations. A clustered component, whose
centroid is closer to the RGB centroid, and a very elongated tail,
extending from the clustered component to the north-east (the
``hook''). Moreover, in the last 0-20 Myr the clustered component has
become more active with respect to the tail, suggesting that the SF is
progressing toward the south-west, in apparent contrast with the
result of \cite{mun18}. However, their finding is that the very recent
SFR is more intense in the north-east region, compared to the central
and south-west one. Our result is that the north-east region is older
than the central clustered component. In other words, if the hook
region has been caused by the interaction with NGC4656UV, this
interaction may be not happening right now. From these
characteristics, a tentative classification for NGC4656 is OOD.

{\bf NGC 4449}: Classified as SBm in NED, NGC4449 is located in the
Canes Venatici I group. It is similar in size and mass to the Large
Magellanic Cloud, with ongoing and intense star formation distributed
along a bar-like structure (\citealt{hill94}). Its closest neighbours
are the gas rich dwarf galaxy DDO 125, at a projected distance of 40
kpc, and a tidally distorted, old and a very low surface brightness
dwarf, NGC4449B (\citealt{md15,rich12}), located at a projected
distance of 9 kpc from NGC4449.

The gas component shows a variety of morphological features like
filaments, arcs, and loops extending for several kpc
(\citealt{hunter97}), well beyond the optical galaxy. \cite{hunter98}
found a strong condensation of gas centered on the optical galaxy with
a diameter of about 9 kpc, in turn embedded in an elongated ellipse of
lower column density with a major axis of 35 kpc. Beyond the ellipse
the gas distribution is dominated by streamers, with the longest
originating from the eastern side of the ellipse and extending
south-west for 25 kpc toward a large HI concentration. Kinematically,
the inner and outer parts of the neutral gas form two separate systems
that are counter-rotating (\citealt{hunter98}), which generally
signals the recent accretion of gas.

Despite these asymmetries, it is not clear whether the observed
streamers are tidal tails caused by dynamical interaction
(\citealt{hunter98}). \cite{toloba16} studied the internal stellar
kinematics and metallicities of NGC4449 and NGC4449B, finding similar
radial velocities to within measurement uncertainties, consistent with
what may be expected if NGC4449B were gravitationally bound to NGC
4449.

NGC4449 was a challenging galaxy to model in both Paper I and
\citet{sacchi17}, where we analyzed the optical (F555W/F814W)
counterpart of the LEGUS catalog for this galaxy. Its high levels of
differential reddening (the highest we find in the 23 dwarfs analyzed
here) make the CMD harder to interpret, smearing its features
exacerbating the separation between MS and post-MS stars. In the UV
case, we also needed an IMF flatter than a Salpeter one ($s = -2.0$)
to provide a better fit to the data, even though the degeneracy
between mass and age for very massive stars prevents us from making
strong claims in this regard.

In our spatial maps, the young component of NGC4449 shows a prominent
bar, with two streams stemming from its extremities. Compared to the
RGB distribution, these features occupy a modest fraction of the
ellipse.  The stream extending to north is also connected to another
stellar overdensity. Moving to younger ages, the bar becomes thinner
and the ratio between the number of HeB stars in the overdensity and
in the bar grows. All centroids are significantly offset. We classify
this galaxy as OOC.

{\bf NGC4248}: Classified as a likely spiral in NED, NGC4248 is a
member of the Canes Venatici II Group (\citealt{fou92}), whose
brightest member, M106, can be considered a Milky Way analogue. Using
deep, wide-field surface photometry, \cite{watkins16} found
extremely boxy and offset outer isophotes, a clear evidence of tidal
distortion. Our HeBs maps show an elongated structure which moves to
north-east with time. Compared to the RGB distribution, HeB stars
occupy a tiny region and are clearly offset with respect to the RGB
centroid. Following our scheme, NGC4248 can be classified as OOC.

\section{Discussion and conclusions}

Following our previous work on the star formation history of NGC4449,
NGC1705, and NGC 4305 (Holmberg II), presented in Paper I, we have
investigated the stellar populations, CMDs, SFHs, and SF patterns of
the full sample of 23 LEGUS star forming dwarf galaxies.

The aim of this study was to characterize starbursts, both in terms of
duration and spatial distribution of the star formation process in
these active systems. Comparing different galaxies in the most
homogeneous approach possible is also key to connecting the SF
activity with the environment in which galaxies live, and to find
possible mechanisms triggering the formation of stars in this kind of
galaxies.

To this purpose, we took advantage of the UV bands of \textit{HST},
thanks to the data obtained within the LEGUS survey from both new and
archival observations with the WFC3 and ACS. By studying the U vs
U$-$V CMDs of our targets, we were able to infer detailed SFHs in the
last $\sim 100$ Myr with unprecedented time resolution, and to compare
our results with FUV and H$\alpha$ integrated information. While the
SFHs exhibit different shapes from galaxy to galaxy, particularly at
the youngest epochs, the general trend is a relatively flat SF as a
function of time, with no major bursts, and the strongest SF
enhancements exceeding the 100-Myr average only by a factor of
2-3. These results hold irrespective of: 1) the stellar evolution
library used to reconstruct the SFH, despite the different assumption
in the underlying stellar physics; 2) the mass fraction of atomic gas
(HI). Indeed, when the whole sample is divided into two sub-groups,
gas-rich and gas-poor SFDs, no systematic difference is observed in
their average SFHs. 

In a few cases (NGC4449, NGC5253, NGC3738, and NGC4656) the levels of
stellar crowding and/or differential reddening found in the fields
challenged our ability to reproduce the star-counts and color spread
of the observational CMDs. This is due to large completeness
variations on the scale of fractions of a kpc, very difficult to
account for even using our artificial star procedure, specifically
designed to this purpose. Also our extinction modeling, based on two
parameters, might be too simplistic when applied to extreme cases like
these. On the other hand, stellar evolution models need to be taken
with caution when studying these shorter wavelengths, especially in
the color transition from MS to post-MS stars; indeed, the synthetic
CMDs of NGC4449 and NGC3738 present a gap (much larger than
photometric uncertainties) between the MS and the post-MS phase that
is completely absent in the observed CMDs. A similar gap was already
found by \cite{tang14} in several dwarf galaxies, and can be overcome
by extending the stellar mechanism of the overshooting at the base of
the convective envelope.

We found very interesting results when comparing our derived SFRs with
integrated FUV and H$\alpha$ rates. For half of the sample, our
60-Myr-average CMD-based SFR is more than two times the FUV SFR
(corrected for extinction), whereas the CMD-based 10 Myr-average SFRs
are consistent with the H$\alpha$ estimates. This corroborates the
result of \cite{mq15b}, extending the sample of galaxies where the
discrepancy is found.

Finally, using HeB stars in different age intervals, we have studied
how the star formation has been spatially progressing in the last 60
Myr. We found a large variance of spatial configurations, both in
terms of concentration with respect to the RGB stars distribution and
localization of different HeBs generations. About 44\% of our galaxies
shows a spatial progression of the SF in the last 60 Myr. Compared to
RGB stars, only a few galaxies show very centralized concentrations of
young stars, while the typical distribution is neither clearly
centralized nor clearly diffuse. About 52\% of the sample is
caracterized by offsets larger than 250 pc between RGB and HeB spatial
centroids. In terms of spatial coverage, about 44\% of galaxies is
caracterized by a population of HeB stars that are spread out as much
as the RGB stars. This result overall confirms the
\citeapos{mq12}{mq12} findings, who studied the concentration of blue
HeB stars (from optical CMDs) over the last 100–250 Myr, suggesting
that the starburst phenomenon is not necessarily a SF event taking
place in the very center of the system.\\

We can briefly summarize our main results here.

\begin{itemize}

\item In the last 100 Myr the dwarfs of the LEGUS sample show
  different SFH, but no significant bursts (larger than 2-3
  times the 100-Myr-average) are detected. On average, the activity in
  the last 100 Myr is consistent with a constant trend, irrespective
  of the mass fraction of atomic gas in the galaxies;

\item Overall the synthetic CMDs reproduce well qualitatively and
  quantitatively the observations, except in a few very crowded and/or
  extincted galaxies;

\item The starburst phenomenon is not exclusively a concentrated and
  instantaneous SF event. A large fraction of our galaxies show: 1) very
  recent diffuse (compared to RGB stars) SF; 2) off-center recent or
  ongoing SF; 3) spatial progression of SF;

\item Most of the galaxies in the sample show a CMD-based average SFR
  in the last 60 Myr higher than the FUV-based SFR (based on existing
  scaling relations). The CMD-based average SFR in the last 10 Myr is
  instead in good agreement with the H${\alpha}$-based SFR.

\end{itemize}
  
\acknowledgments 

MC and MT kindly acknowledge partial funding from INAF PRIN-SKA-2017, program 1.05.01.88.04, and INAF Main Stream SSH, program 1.05.01.86.28.


\begin{thebibliography}{}

\bibitem[Annibali et al.(2003)]{annibali03} Annibali, F., Greggio, L., Tosi, M., Aloisi, A., \& Leitherer, C.\ 2003, \aj, 126, 2752 

\bibitem[Annibali et al.(2008)]{annibali08} Annibali, F., Aloisi, A.,
  Mack, J., et al.\ 2008, \aj, 135, 1900 

\bibitem[Annibali et al.(2009)]{annibali09} Annibali, F., Tosi, M., Monelli, M., et al.\ 2009, \aj, 138, 169 


\bibitem[Annibali et al.(2012)]{annibali12} Annibali, F., Tosi, M., Aloisi, A., van der Marel, R.~P., \& Martinez-Delgado, D.\ 2012, \apjl, 745, L1 


\bibitem[Annibali et al.(2015)]{annibali15} Annibali, F., Tosi, M., Pasquali, A., et al.\ 2015, \aj, 150, 143 

\bibitem[Annibali et al.(2016)]{annibali16} Annibali, F., Nipoti, C., Ciotti, L., et al.\ 2016, \apjl, 826, L27 

\bibitem[Ashley et al.(2014)]{ashley15} Ashley, T., Elmegreen, B.~G., Johnson, M., et al.\ 2014, \aj, 148, 130

\bibitem[Banks et al.(1999)]{banks99} Banks, G.~D., Disney, M.~J., Knezek, P.~M., et al.\ 1999, \apj, 524, 612

\bibitem[Begum et al.(2006)]{begum06} Begum, A., Chengalur, J.~N., Karachentsev, I.~D., et al.\ 2006, \mnras, 365, 1220

\bibitem[Bellazzini et al.(2001)]{bellazzini01} Bellazzini, M., Ferraro, F.~R., \& Pancino, E.\ 2001, \apj, 556, 635 

\bibitem[Bekki \& Couch(2003)]{bekki03} Bekki, K., \& Couch, W.~J.\ 2003, \apjl, 596, L13 

\bibitem[Bekki \& Freeman(2002)]{bf02} Bekki, K., \& Freeman, K.~C.\ 2002, \apjl, 574, L21 


\bibitem[Berg et al.(2012)]{berg12} Berg, D.~A., Skillman, E.~D., Marble, A.~R., et al.\ 2012, \apj, 754, 98 



\bibitem[Bernard et al.(2012)]{bernard12} Bernard, E.~J., Ferguson, A.~M.~N., Barker, M.~K., et al.\ 2012, \mnras, 426, 3490 



\bibitem[Bertelli et al.(1994)]{bertelli94} Bertelli, G., Bressan, A.,
  Chiosi, C., Fagotto, F., \& Nasi, E.\ 1994, \aaps, 106,

\bibitem[Bressan et al.(2012)]{bressan12} Bressan, A., Marigo, 
P., Girardi, L., et al.\ 2012, \mnras, 427, 127 


\bibitem[Bressan et al.(2015)]{bressan15} Bressan, A., Girardi, L., Marigo, P., Rosenfield, P., \& Tang, J.\ 2015, Asteroseismology of Stellar Populations in the Milky Way, 39, 25 



\bibitem[Bureau \& Carignan(2002)]{BC02} Bureau, M., \& Carignan, C.\
  2002, \aj, 123, 1316 

\bibitem[Calzetti et al.(2004)]{cal04} Calzetti, D., Harris, J., Gallagher, J.~S., et al.\ 2004, \aj, 127, 1405


\bibitem[Calzetti et al.(2007)]{calzetti07} Calzetti, D., Kennicutt, R.~C., Engelbracht, C.~W., et al.\ 2007, \apj, 666, 870


\bibitem[Calzetti et al.(2015)]{calzetti15} Calzetti, D., Lee, J.~C., Sabbi, E., et al.\ 2015, \aj, 149, 51 

\bibitem[Cannon et al.(2016)]{cannon16} Cannon, J.~M., McNichols, A.~T., Teich, Y.~G., et al.\ 2016, \aj, 152, 202


\bibitem[Caffau et al.(2011)]{caffau11} Caffau, E., Ludwig, H.-G., Steffen, M., Freytag, B., \& Bonifacio, P.\ 2011, \solphys, 268, 255 

\bibitem[Cardelli et al.(1989)]{cardelli89} Cardelli, J.~A., Clayton,
  G.~C., \& Mathis, J.~S.\ 1989, \apj, 345, 245 

\bibitem[Choi et al.(2016)]{choi16} Choi, J., Dotter, A., Conroy, C., et al.\ 2016, \apj, 823, 102 


\bibitem[Cignoni et al.(2015)]{cigno15} Cignoni, M., Sabbi, E., 
van der Marel, R.~P., et al.\ 2015, \apj, 811, 76 

\bibitem[Cignoni et al.(2016)]{cigno16} Cignoni, M., Sabbi, E., van
  der Marel, R.~P., et al.\ 2016, \apj, 833, 154 

\bibitem[Cignoni et al.(2018)]{cigno18} Cignoni, M., Sacchi, E., Aloisi, A., et al.\ 2018, \apj, 856, 62 

\bibitem[Clemens et al.(1998)]{clemens98} Clemens, M.~S., Alexander, P., \& Green, D.~A.\ 1998, \mnras, 297, 1015


\bibitem[Croxall et al.(2009)]{crox09} Croxall, K.~V., van Zee, L., Lee, H., et al.\ 2009, \apj, 705, 723-738 

\bibitem[Dalcanton et al.(2012)]{dalcanton12} Dalcanton, J.~J., Williams, B.~F., Melbourne, J.~L., et al.\ 2012, \apjs, 198, 6 

\bibitem[Dale et al.(2009)]{dale09} Dale, D.~A., Cohen, S.~A., Johnson, L.~C., et al.\ 2009, \apj, 703, 517

\bibitem[De Marchi et al.(2016)]{demarchi16} De Marchi, G., Panagia, N., Sabbi, E., et al.\ 2016, \mnras, 455, 4373

\bibitem[Egorov et al.(2016)]{egorov16} Egorov, O.~V., Lozinskaya, T.~A., \& Moiseev, A.~V.\ 2016, The Interplay Between Local and Global Processes in Galaxies,, 30

\bibitem[Firth et al.(2006)]{firth06} Firth, P., Evstigneeva, E.~A.,
  Jones, J.~B., et al.\ 2006, \mnras, 372, 1856 

\bibitem[de Mello et al.(2012)]{dem12} de Mello, D.~F., Urrutia-Viscarra, F., Mendes de Oliveira, C., et al.\ 2012, \mnras, 426, 2441

\bibitem[de Mink et al.(2012)]{demink12} de Mink, S.~E., Brott, I., Cantiello, M., et al.\ 2012, Proceedings of a Scientific Meeting in Honor of Anthony F.~J.~Moffat, 465, 65 

\bibitem[de Mink et al.(2014)]{demink14} de Mink, S.~E., Sana, H., Langer, N., et al.\ 2014, \apj, 782, 7

\bibitem[de Vaucouleurs, \& de Vaucouleurs(1964)]{dev64} de Vaucouleurs, G., \& de Vaucouleurs, A.\ 1964, \apj, 140, 1622

\bibitem[Dohm-Palmer et al.(1997)]{dp} Dohm-Palmer, R.~C., Skillman, E.~D., Saha, A., et al.\ 1997, \aj, 114, 2527 

\bibitem[Dolphin(2000)]{dolphin2000} Dolphin, A.~E.\ 2000, \pasp, 112, 1383 

\bibitem[Dotter(2016)]{dotter16} Dotter, A.\ 2016, \apjs, 222, 8 

\bibitem[Elmegreen et al.(1998)]{elme98} Elmegreen, D.~M., Chromey, F.~R., Knowles, B.~D., et al.\ 1998, \aj, 115, 1433

\bibitem[Elmegreen et al.(2012)]{elme12} Elmegreen, B.~G., Zhang, H.-X., \& Hunter, D.~A.\ 2012, \apj, 747, 105

\bibitem[Elmegreen \& Scalo(2006)]{es06} Elmegreen, B.~G., \& Scalo, J.\ 2006, \apj, 636, 149 

\bibitem[Evstigneeva et al.(2007)]{evsti07} Evstigneeva, E.~A., Drinkwater, M.~J., Jurek, R., et al.\ 2007, \mnras, 378, 1036 

\bibitem[Fouque et al.(1992)]{fou92} Fouque, P., Gourgoulhon, E., Chamaraux, P., et al.\ 1992, \aaps, 93, 211

\bibitem[Gerola et al.(1980)]{gerola80} Gerola, H., Seiden, P.~E., \& Schulman, L.~S.\ 1980, \apj, 242, 517 

\bibitem[Gallart et al.(2015)]{gallart15} Gallart, C., Monelli, M., Mayer, L., et al.\ 2015, \apjl, 811, L18 

\bibitem[Hao et al.(2011)]{hao11} Hao, C.-N., Kennicutt, R.~C., Johnson, B.~D., et al.\ 2011, \apj, 741, 124

\bibitem[Harris et al.(1997)]{harris97} Harris, J., Zaritsky, D., \&
  Thompson, I.\ 1997, \aj, 114, 1933 

\bibitem[Heald, \& HALOGAS Team(2011)]{heald11} Heald, G.~H., \& HALOGAS Team\ 2011, American Astronomical Society Meeting Abstracts \#217 217, 246.19
\bibitem[Heckman \& Leitherer(1997)]{heck97} Heckman, T.~M., \& Leitherer, C.\ 1997, \aj, 114, 69 

\bibitem[Heckman et al.(2001)]{heck01} Heckman, T.~M., Sembach, K.~R., Meurer, G.~R., et al.\ 2001, \apj, 554, 1021 

\bibitem[Hill et al.(1994)]{hill94} Hill, R.~S., Home, A.~T., Smith, A.~M., et al.\ 1994, \apj, 430, 568 


\bibitem[Hopp(1999)]{hopp99} Hopp, U.\ 1999, \aaps, 134, 317

\bibitem[Huchtmeier et al.(1980)]{huc80} Huchtmeier, W.~K., Seiradakis, J.~H., \& Materne, J.\ 1980, \aap, 91, 341


\bibitem[Hunter, \& Gallagher(1997)]{hunter97} Hunter, D.~A., \& Gallagher, J.~S.\ 1997, \apj, 475, 65


\bibitem[Hunter et al.(1998)]{hunter98} Hunter, D.~A., Wilcots, E.~M., van Woerden, H., Gallagher, J.~S., \& Kohle, S.\ 1998, \apjl, 495, L47 

\bibitem[Hunter et al.(1999)]{hunter99} Hunter, D.~A., van Woerden,
  H., \& Gallagher, J.~S.\ 1999, \aj, 118, 2184 

\bibitem[Hunter \& Elmegreen(2004)]{he04} Hunter, D.~A., \& Elmegreen, B.~G.\ 2004, \aj, 128, 2170 


\bibitem[Huchtmeier \& Richter(1989)]{hr89} Huchtmeier, W.~K., \& Richter, O.~G.\ 1989, Science, 246, 943 

\bibitem[Johnson et al.(2003)]{j03} Johnson, K.~E., Indebetouw, R., \&
  Pisano, D.~J.\ 2003, \aj, 126, 101 

\bibitem[Johnson et al.(2013)]{joh13} Johnson, B.~D., Weisz, D.~R., Dalcanton, J.~J., et al.\ 2013, \apj, 772, 8

\bibitem[Karachentsev et al.(2003)]{kara03} Karachentsev, I.~D., Sharina, M.~E., Dolphin, A.~E., et al.\ 2003, \aap, 398, 467

\bibitem[Karczewski et al.(2013)]{kar13} Karczewski, O.~{\L}., Barlow, M.~J., Page, M.~J., et al.\ 2013, \mnras, 431, 2493 

\bibitem[Kennicutt(1998)]{k98} Kennicutt, R.~C.\ 1998, \araa, 36, 189

\bibitem[Kennicutt et al.(2009)]{k09} Kennicutt, R.~C., Hao, C.-N., Calzetti, D., et al.\ 2009, \apj, 703, 1672

\bibitem[Kere{\v s} et al.(2005)]{keres15} Kere{\v s}, D., Katz, N., Weinberg, D.~H., \& Dav{\'e}, R.\ 2005, \mnras, 363, 2 

\bibitem[Koleva et al.(2014)]{koleva14} Koleva, M., De Rijcke, S., Zeilinger, W.~W., et al.\ 2014, \mnras, 441, 452 

\bibitem[Kornreich et al.(1998)]{korn98} Kornreich, D.~A., Haynes, M.~P., \& Lovelace, R.~V.~E.\ 1998, \aj, 116, 2154

\bibitem[Kroupa(2001)]{kroupa01} Kroupa, P.\ 2001, \mnras, 322, 
231 

\bibitem[Lah{\'e}n et al.(2019)]{lah19} Lah{\'e}n, N., Naab, T., Johansson, P.~H., et al.\ 2019, \apjl, 879, L18

\bibitem[Lee et al.(2009)]{lee09} Lee, J.~C., Gil de Paz, A., Tremonti, C., et al.\ 2009, \apj, 706, 599-613 

\bibitem[Lee et al.(2011)]{lee11} Lee, J.~C., Gil de Paz, A., Kennicutt, R.~C., et al.\ 2011, \apjs, 192, 6

\bibitem[Lelli et al.(2014)]{lelli14} Lelli, F., Verheijen, M., \&
  Fraternali, F.\ 2014, \mnras, 445, 1694 

\bibitem[Leroy et al.(2013)]{leroy13} Leroy, A.~K., Walter, F., Sandstrom, K., et al.\ 2013, \aj, 146, 19

\bibitem[L{\'o}pez-S{\'a}nchez et al.(2012)]{ls12} L{\'o}pez-S{\'a}nchez, {\'A}. R., Koribalski, B.~S., van Eymeren, J., et al.\ 2012, \mnras, 419, 1051

\bibitem[Marigo et al.(2017)]{marigo17} Marigo, P., Girardi, L., Bressan, A., et al.\ 2017, \apj, 835, 77 


\bibitem[Martin(1998)]{martin98} Martin, C.~L.\ 1998, \apj, 506, 222

\bibitem[Mart{\'{\i}}nez-Delgado et al.(2012)]{md12} Mart{\'{\i}}nez-Delgado, D., Romanowsky, A.~J., Gabany, R.~J., et al.\ 2012, \apjl, 748, L24 

\bibitem[Mart{\'\i}nez-Delgado et al.(2015)]{md15} Mart{\'\i}nez-Delgado, D., D'Onghia, E., Chonis, T.~S., et al.\ 2015, \aj, 150, 116

\bibitem[Martins et al.(2012)]{martin12} Martins, F., F{\"o}rster Schreiber, N.~M., Eisenhauer, F., \& Lutz, D.\ 2012, \aap, 547, A17 




\bibitem[Mihos et al.(2013)]{mihos12} Mihos, J.~C., Harding, P., Spengler, C.~E., et al.\ 2013, \apj, 762, 82




\bibitem[McQuinn et al.(2010)]{mq10} McQuinn, K.~B.~W., Skillman,
  E.~D., Cannon, J.~M., et al.\ 2010, \apj, 724, 49 
\bibitem[McQuinn et al.(2011)]{mq11} McQuinn, K.~B.~W., Skillman, E.~D., Dalcanton, J.~J., et al.\ 2011, \apj, 740, 48 

\bibitem[McQuinn et al.(2012)]{mq12} McQuinn, K.~B.~W., Skillman, E.~D., Dalcanton, J.~J., et al.\ 2012, \apj, 759, 77 

\bibitem[McQuinn et al.(2015)-a]{mq15a} McQuinn, K.~B.~W., Mitchell, N.~P., \& Skillman, E.~D.\ 2015, \apjs, 218, 29 
\bibitem[McQuinn et al.(2015)-b]{mq15b} McQuinn, K.~B.~W., Skillman, E.~D., Dolphin, A.~E., et al.\ 2015, \apj, 808, 109


\bibitem[Meynet \& Maeder(1997)]{mm97} Meynet, G., \& Maeder, A.\ 1997, \aap, 321, 465 



\bibitem[Melisse \& Israel(1994)]{mi94} Melisse, J.~P.~M., \& Israel, F.~P.\ 1994, \aap, 285, 51 


\bibitem[Meurer et al.(1992)]{meurer92} Meurer, G.~R., Freeman, K.~C., Dopita, M.~A., \& Cacciari, C.\ 1992, \aj, 103, 60 

\bibitem[Meurer et al.(1998)]{meurer98} Meurer, G.~R., Staveley-Smith, L., \& Killeen, N.~E.~B.\ 1998, \mnras, 300, 705 

\bibitem[Meurer et al.(2006)]{meu06} Meurer, G.~R., Hanish, D.~J., Ferguson, H.~C., et al.\ 2006, \apjs, 165, 307

\bibitem[Miura et al.(2018)]{miura18} Miura, R.~E., Espada, D.,
  Hirota, A., et al.\ 2018, \apj, 864, 120

\bibitem[Moustakas et al.(2010)]{mou10} Moustakas, J., Kennicutt, R.~C., Jr., Tremonti, C.~A., et al.\ 2010, \apjs, 190, 233-266 

\bibitem[M{\"u}hle et al.(2005)]{mule05} M{\"u}hle, S., Klein, U., Wilcots, E.~M., \& H{\"u}ttemeister, S.\ 2005, \aj, 130, 524 

\bibitem[Murphy et al.(2011)]{murphy11} Murphy, E.~J., Condon, J.~J., Schinnerer, E., et al.\ 2011, \apj, 737, 67

\bibitem[Mu{\~n}oz-Elgueta et al.(2018)]{mun18} Mu{\~n}oz-Elgueta, N., Torres-Flores, S., Amram, P., et al.\ 2018, \mnras, 480, 3257

\bibitem[Nilson(1973)]{nilson73} Nilson, P.\ 1973, Acta Universitatis Upsaliensis. Nova Acta Regiae Societatis Scientiarum Upsaliensis - Uppsala Astronomiska Observatoriums Annaler

\bibitem[Ott et al.(2001)]{ott01} Ott, J., Walter, F., Brinks, E., et al.\ 2001, \aj, 122, 3070

\bibitem[Ott et al.(2005)]{ott05} Ott, J., Walter, F., \& Brinks, E.\ 2005, \mnras, 358, 1423 


\bibitem[Patton et al.(2013)]{patton13} Patton, D.~R., Torrey, P., Ellison, S.~L., Mendel, J.~T., \& Scudder, J.~M.\ 2013, \mnras, 433, L59 

\bibitem[Paxton et al.(2011)]{paxton11} Paxton, B., Bildsten, L., Dotter, A., et al.\ 2011, \apjs, 192, 3 

\bibitem[Paxton et al.(2013)]{paxton13} Paxton, B., Cantiello, M., Arras, P., et al.\ 2013, \apjs, 208, 4 

\bibitem[Pearson et al.(2018)]{pearson18} Pearson, S., Privon, G.~C., Besla, G., et al.\ 2018, \mnras, 480, 3069

\bibitem[Paxton et al.(2015)]{paxton15} Paxton, B., Marchant, P., Schwab, J., et al.\ 2015, \apjs, 220, 15 

\bibitem[Puche et al.(1992)]{puche92} Puche, D., Westpfahl, D.,
  Brinks, E., \& Roy, J.-R.\ 1992, \aj, 103, 1841 

\bibitem[Pustilnik, \& Tepliakova(2011)]{pt11} Pustilnik, S.~A., \& Tepliakova, A.~L.\ 2011, \mnras, 415, 1188

\bibitem[Ramachandran et al.(2019)]{rama19} Ramachandran, V., Hamann, W.-R., Oskinova, L.~M., et al.\ 2019, \aap, 625, A104

\bibitem[Rand(1994)]{rand94} Rand, R.~J.\ 1994, \aap, 285, 833

\bibitem[Rich et al.(2012)]{rich12} Rich, R.~M., Collins, M.~L.~M., Black, C.~M., et al.\ 2012, \nat, 482, 192 

\bibitem[Roberts(1968)]{rob68} Roberts, M.~S.\ 1968, \apj, 151, 117

\bibitem[Roychowdhury et al.(2011)]{roy11} Roychowdhury, S., Chengalur, J.~N., Kaisin, S.~S., et al.\ 2011, \mnras, 414, L55

\bibitem[Rownd et al.(1994)]{rd94} Rownd, B.~K., Dickey, J.~M., \& Helou, G.\ 1994, \aj, 108, 1638

\bibitem[Sabbi et al.(2012)]{sabbi12} Sabbi, E., Lennon, D.~J., Gieles, M., et al.\ 2012, \apjl, 754, LL37 


\bibitem[Sabbi et al.(2015)]{sabbi15}  Sabbi, E., Lennon, D.~J.,
  Anderson, J., et al.\ 2016, \apjs, 222, 11 

\bibitem[Sabbi et al.(2018)]{sabbi18} Sabbi, E., Calzetti, D., Ubeda, L., et al.\ 2018, \apjs, 235, 23 

\bibitem[Sacchi et al.(2016)]{sacchi16} Sacchi, E., Annibali, F.,
  Cignoni, M., et al.\ 2016, \apj, 830, 3 

\bibitem[Sacchi et al.(2017)]{sacchi17} Sacchi, E., Cignoni, M., A. Aloisi, et al., 2018, \apj, 857, 63S

\bibitem[Sana et al.(2013)]{sana13} Sana, H., de Koter, A., de Mink, S.~E., et al.\ 2013, \aap, 550, A107 

\bibitem[Sana \& Evans(2011)]{sana11} Sana, H., \& Evans, C.~J.\ 2011, Active OB Stars: Structure, Evolution, Mass Loss, and Critical Limits, 272, 474 




\bibitem[Schaerer et 
al.(1993)]{schaerer93} Schaerer, D., Meynet, G., Maeder, A., \& Schaller, G.\ 1993, \aaps, 98, 523 

\bibitem[Schechtman-Rook, \& Hess(2012)]{sch12} Schechtman-Rook, A., \& Hess, K.~M.\ 2012, \apj, 750, 171

\bibitem[Schneider et al.(2018)]{sch18} Schneider, F.~R.~N., Sana, H., Evans, C.~J., et al.\ 2018, Science, 359, 69

\bibitem[Stayton et al.(1983)]{stay83} Stayton, L.~C., Angione, R.~J., \& Talbert, F.~D.\ 1983, \aj, 88, 602

\bibitem[Starkenburg et al.(2016)]{stark16} Starkenburg, T.~K., Helmi, A., \& Sales, L.~V.\ 2016, \aap, 595, A56 

\bibitem[Stewart et al.(2000)]{stewart00} Stewart, S.~G., Fanelli, M.~N., Byrd, G.~G., et al.\ 2000, \apj, 529, 201 

\bibitem[Stierwalt et al.(2015)]{stier15} Stierwalt, S., Besla, G., Patton, D., et al.\ 2015, \apj, 805, 2 

\bibitem[Stierwalt et al.(2017)]{stier17} Stierwalt, S., Liss, S.~E., Johnson, K.~E., et al.\ 2017, arXiv:1701.01731 

\bibitem[Stil \& Israel(2002)]{stil02} Stil, J.~M., \& Israel, F.~P.\ 2002, \aap, 392, 473 



\bibitem[Stinson et al.(2007)]{stinson07} Stinson, G.~S., Dalcanton, J.~J., Quinn, T., Kaufmann, T., \& Wadsley, J.\ 2007, \apj, 667, 170 

\bibitem[Tang et al.(2014)]{tang14} Tang, J., Bressan, A., Rosenfield,
  P., et al.\ 2014, \mnras, 445, 4287

\bibitem[Theis(1999)]{theis99} Theis, C.\ 1999, Reviews in Modern Astronomy, 12, 309 

\bibitem[Thronson et al.(1989)]{thr89} Thronson, H.~A., Hunter, D.~A.,
  Casey, S., et al.\ 1989, \apj, 339, 803

\bibitem[Tikhonov et al.(2014)-a]{tik14a} Tikhonov, N.~A., Galazutdinova, O.~A., \& Lebedev, V.~S.\ 2014, Astronomy Letters, 40, 1

\bibitem[Tikhonov et al.(2014)-b]{tik14} Tikhonov, N.~A., Galazutdinova, O.~A., \& Lebedev, V.~S.\ 2014, Astronomy Letters, 40, 11

\bibitem[Tognelli et al.(2011)]{tognelli11} Tognelli, E., Prada Moroni, P.~G., \& Degl'Innocenti, S.\ 2011, \aap, 533, A109 

\bibitem[Toloba et al.(2016)]{toloba16} Toloba, E., Guhathakurta, P.,
  Romanowsky, A.~J., et al.\ 2016, \apj, 824, 35

\bibitem[Tolstoy et al.(2009)]{THT09} Tolstoy, E., Hill, V., \& Tosi, M.\ 2009, \araa, 47, 371 

\bibitem[Tosi et al.(1991)]{tosi91} Tosi, M., Greggio, L., Marconi, G., et al.\ 1991, \aj, 102, 951

\bibitem[Tosi et al.(2001)]{tosi01} Tosi, M., Sabbi, E., Bellazzini, M., et al.\ 2001, \aj, 122, 1271 

\bibitem[Tully(1988)]{tully88} Tully, R.~B.\ 1988, \aj, 96, 73 

\bibitem[Tully et al.(2006)]{tully06} Tully, R.~B., Rizzi, L., Dolphin, A.~E., et al.\ 2006, \aj, 132, 729

\bibitem[Turner et al.(2015)]{turner15} Turner, J.~L., Beck, S.~C., Benford, D.~J., et al.\ 2015, \nat, 519, 331

\bibitem[van Woerden et al.(1975)]{vanw75} van Woerden, H., Bosma, A., \& Mebold, U.\ 1975, La Dynamique des galaxies spirales, 241, 483 

\bibitem[van der Hulst, \& Huchtmeier(1979)]{hh79} van der Hulst, J.~M., \& Huchtmeier, W.~K.\ 1979, \aap, 78, 82

\bibitem[Viallefond, \& Thuan(1983)]{vt83} Viallefond, F., \& Thuan, T.~X.\ 1983, \apj, 269, 444

\bibitem[Walter et al.(2007)]{walter07} Walter, F., Cannon, J.~M., Roussel, H., et al.\ 2007, \apj, 661, 102

\bibitem[Watkins et al.(2016)]{watkins16} Watkins, A.~E., Mihos, J.~C., \& Harding, P.\ 2016, \apj, 826, 59


\bibitem[Weidner, \& Kroupa(2005)]{wk05} Weidner, C., \& Kroupa, P.\ 2005, \apj, 625, 754

\bibitem[Weilbacher et al.(2003)]{weil03} Weilbacher, P.~M., Duc, P.-A., \& Fritze-v. Alvensleben, U.\ 2003, \aap, 397, 545


\bibitem[Weisz et al.(2008)]{weisz08} Weisz, D.~R., Skillman, E.~D., Cannon, J.~M., et al.\ 2008, \apj, 689, 160-183 

\bibitem[Weisz et al.(2009)]{weisz09} Weisz, D.~R., Skillman, E.~D., Cannon, J.~M., et al.\ 2009, \apj, 704, 1538-1569 

\bibitem[Weisz et al.(2011)]{weisz11} Weisz, D.~R., Dalcanton, J.~J., Williams, B.~F., et al.\ 2011, \apj, 739, 5 

\bibitem[Weisz et al.(2012)]{weisz12} Weisz, D.~R., Johnson, B.~D., Johnson, L.~C., et al.\ 2012, \apj, 744, 44 



\end{thebibliography}
\end{document}